\documentclass[amsfonts,showpacs, longbibliography,nofootinbib,amsmath,amssymb,twocolumn]{aastex631}

\usepackage{subfigure}
\usepackage{changepage}
\usepackage{bm,graphics,graphicx,epsfig,soul,latexsym,hyperref,float,mathrsfs,multirow,comment}
\let\splitbox\relax
\usepackage{adjustbox}
\usepackage{color}
\usepackage{amsmath}
\usepackage[normalem]{ulem}
\usepackage{afterpage}
\usepackage[usenames]{xcolor}
\usepackage{enumitem}
\usepackage{orcidlink}

\graphicspath{{figure/}{./}}
\definecolor{darkgreen}{rgb}{0,0.5,0}

\hypersetup{
    bookmarks=true,         
    unicode=false,          
    pdftoolbar=true,        
    pdfmenubar=true,        
    pdffitwindow=false,     
    pdfstartview={FitH},    
    pdftitle={Aligned_spins_phasing},    
    pdfauthor={Author},     
    pdfsubject={Subject},   
    pdfcreator={Creator},   
    pdfproducer={Producer}, 
    pdfkeywords={keyword1} {key2} {key3}, 
    pdfnewwindow=true,      
    colorlinks=true,       
    linkcolor=red,          
    citecolor=blue,        
    filecolor=magenta,      
    urlcolor=darkgreen,           
    linktocpage=true
}

\allowdisplaybreaks
\begin{document}

\title{Characterizing Binary Black Hole Subpopulations in GWTC-4 with Binned Gaussian Processes: On the Origins of the $35M_{\odot}$ Peak.}
\author{Omkar Sridhar\orcidlink{0009-0008-3375-1077
}}\email{omkarnm1401@gmail.com} \affiliation{Department of Physics and Astronomy, Northwestern University, 2145 Sheridan Road, Evanston, IL 60208, USA}
\affiliation{Center for Interdisciplinary Exploration and Research in Astrophysics (CIERA), Northwestern University, 1800 Sherman Ave,
Evanston, IL 60201, USA}
\author{Anarya Ray\orcidlink{0000-0002-7322-4748}} \email{anarya.ray@northwestern.edu} \affiliation{Center for Interdisciplinary Exploration and Research in Astrophysics (CIERA), Northwestern University, 1800 Sherman Ave,
Evanston, IL 60201, USA}
\affiliation{NSF-Simons AI Institute for the Sky (SkAI), 172 E. Chestnut Street, Chicago, IL 60611, USA}
\author{Vicky Kalogera\orcidlink{0000-0001-9236-5469}}\affiliation{Department of Physics and Astronomy, Northwestern University, 2145 Sheridan Road, Evanston, IL 60208, USA}
\affiliation{Center for Interdisciplinary Exploration and Research in Astrophysics (CIERA), Northwestern University, 1800 Sherman Ave,
Evanston, IL 60201, USA}
\affiliation{NSF-Simons AI Institute for the Sky (SkAI), 172 E. Chestnut Street, Chicago, IL 60611, USA}

\begin{abstract}
Understanding the astrophysical origins of binary black holes requires accurate and flexible modeling of multi-dimensional population properties. In this \textit{Letter}, using a data-driven framework based on binned Gaussian processes, we characterize the joint distribution of BBH primary masses, mass ratios, and effective inspiral spins. We identify three distinct subpopulations in the GWTC-4 sample of observations and investigate their astrophysical origins. We find that only one of the three subpopulations exhibits the $35M_{\odot}$ peak, which is characterized by a strong preference for equal mass systems and isotropic spin orientations. Our inferred distributions are consistent with a predominantly dynamical origin of this feature. By comparing with theoretical simulations, we further show that the subpopulation that exhibits the $35M_{\sun}$ peak can exclusively comprise dynamically assembled systems in globular clusters, specifically if black hole birth spins are in the range~$(0.1-0.2)$, whereas the other two subpopulations require substantial contributions from alternative formation channels. We constrain the \textit{lower bound} on the merger rate of BBHs in globular clusters to be $0.69^{+0.23}_{-0.33} \rm{Gpc}^{-3}\rm{yr}^{-1}$, which is consistent with most theoretical predictions(that can range from $0.2-57\rm{Gpc}^{-3}\rm{yr}^{-1}$ depending on modeling assumptions). We conclude that dynamical formation in globular clusters remains a strong candidate for the origin of this excess near $30-40M_{\odot}$ and that more data and targeted parametric models are necessary to rigorously establish this interpretation.
\end{abstract}
\date{\today}


\section{Introduction} \label{sec:intro}

    The first part of the fourth observing run of the LIGO-Virgo-KAGRA detector network \citep[LVK;][]{LIGOScientific:2025slb, LIGOScientific:2014pky,Acernese2015, Akutsu2021} concluded in 2024, and 153 confidently identified binary black hole (BBH) signals were released as part of the fourth gravitational wave (GW) transient catalog \citep[GWTC-4;][]{Abac_data_2025}. With a growing number of GW observations by the LVK, studying the population properties of BBH parameters and the correlations between them provides a unique opportunity to probe the astrophysical phenomena that led to the formation of these GW sources. Currently proposed mechanisms for BBH formation include isolated binary evolution in galactic fields \citep{PortegiesZwart:1997ugk,Belczynski:2001uc, Stevenson:2017tfq}; dynamical formation in dense environments \citep{Kulkarni:1993fr, Sigurdsson:1993zrm, PortegiesZwart:1999nm, Ziosi:2014sra}; hierarchical mergers in stellar clusters \citep{Fishbach2017, Gerosa2017, Rodriguez:2019huv, Doctor2020, Kimball:2020qyd} and disks of active galactic nuclei \citep{Stone:2016wzz,Bartos:2016dgn,McKernan:2019beu}; and evolution in hierarchical triple systems \citep{Thompson_2011, Silsbee:2016djf, Antonini:2017ash, Antognini:2015ima, Stegmann:2021jen, Chattopadhyay:2022buz, Trani:2021tan, Stegmann:2025zkb}. In principle, one can elucidate the partially unconstrained physical phenomena underlying these evolutionary pathways by identifying their imprints as specific population features in the observed sample of BBHs.

As the detection sample continues to grow, two key features in the BBH mass spectrum have emerged as robust characteristics of the underlying population. A sharp peak near $10M_{\odot}$ and an over-abundance of systems near $35M_{\odot}$ have been reported in GWTC-4 using a large number of both strongly parametrized and flexible population models, all of which broadly agree on their nature and existence~\citep{Abac_pop_2025}. Upon its discovery in the first catalog \citep[GWTC-1;][] {LIGOScientific:2018mvr,Talbot:2018cva}, the $35M_{\odot}$ peak was claimed to be a manifestation of the pile-up due to (pulsational) pair-instability supernovae \citep[PPISNe;][]{Woosley:2016hmi,Spera:2017fyx, Farmer:2019jed, Farmer:2020xne, Ziegler:2020klg, Hendriks:2023yrw}. The location of this pile-up in the mass spectrum has strong implications for stellar evolution theory \citep{Belczynski:2017gds,Stevenson:2019rcw,Golomb:2023vxm}. 

Previous studies have shown that a (P)PISNe peak near $30-40M_{\odot}$ is in tension with simulations of massive star evolution and that current understanding of the physical processes inside massive stellar cores places it at $50M_{\odot}$ or higher (see, e.g., \cite{Hendriks:2023yrw} and references therein). This has motivated alternative formation scenarios that can explain the $35M_{\odot}$ feature, such as stable mass transfer and chemically homogeneous evolution in isolated stellar binaries \citep{deMink:2016vkw,Marchant:2016wow,Briel:2022cfl}, BBHs from population III stars \citep{Kinugawa:2014zha, Kinugawa:2020xws, Kinugawa:2021qee}, dynamical formation in globular clusters \citep[GCs;][]{Antonini2023, Ray2024, Wong:2020ise}, and hierarchical mergers~\citep{Tiwari:2020otp, Mahapatra:2022ngs} as well as stellar mergers and accretion~\citep{Kiroglu2025a, Kiroglu:2025vqy, Kremer:2020wtp} in GCs or other dense environments. Each of these scenarios can lead to unique distributions of other BBH parameters such as spins, mass ratios, and redshifts. In particular, coevolution of systems in isolated stellar binaries can lead to preferentially aligned spin components, whereas dynamical assembly leads to isotropic spin orientations~\citep{Mapelli:2021gyv, Chattopadhyay:2023pil, Rodriguez:2019huv, Rodriguez_2022} relative to the orbit~\citep[see also,][who show that BBH+star collisions can lead to a fraction of aligned mergers in clusters as well]{Kiroglu:2025bbp}. Hence, to constrain the contributions of these individual formation pathways to the astrophysical population and thereby establish the origin of the $35M_{\odot}$ feature, it is necessary to identify and characterize subpopulations in the multi-dimensional BBH parameter space, including masses and spins or combinations thereof. 

Several studies have investigated population-level correlations in the BBH detection sample, using both strongly modeled and non-parametric approaches. In GWTC-4, the high mass subpopulation $(\gtrsim40M_{\odot})$ has been thoroughly investigated with both strongly modeled~\citep[e.g.,][]{Ray2025_PISN, Tong:2025wpz} and semi-parametric~\citep[e.g.,][]{Antonini:2025zzw, Wang:2025_PISN} approaches. However, these studies have not explored in detail the astrophysical origins of the $35M_{\odot}$ peak. On the other hand, \cite{Banagiri2025} have parametrized the population distribution of BBH masses, component spin magnitudes, and orientations, identifying three distinct subpopulations. They find that the subpopulation in the $20-40M_{\odot}$ mass range has a unique mass-ratio distribution that prefers equal mass systems as compared to the other two. However, they cannot distinguish the spin distributions of this subpopulation from those of the low mass one. This can result from the low measurability of individual spin magnitudes and orientations compared to effective spin parameters or the strongly parametrized nature of their population inference. 

In an earlier study, \cite{Ray2024} probed the existence of subpopulations associated with the $35M_{\odot}$ feature using a data-driven approach. By analyzing GWTC-3 BBHs using a non-parametric model for the joint distribution of BBH component masses and effective inspiral spins\footnote{Effective inspiral spin is the mass-weighted sum of spins aligned with the orbital angular momentum}, they found evidence of a subpopulation associated with the $30-40M_{\sun}$ feature, which corresponds to a symmetric effective spin distribution peaking near zero. Although effective inspiral spins are less interpretable astrophysically than component spin magnitudes and tilt distributions, they are measured with greater precision. Moreover, they can inform on whether the population is dominated by systems with spins preferentially aligned with the orbit or by systems with isotropic spin orientations. Given that a symmetric effective spin distribution about zero implies preferentially isotropic spin orientations, \cite{Ray2024} claimed, for the first time, hints of a dynamical origin of the $35M_{\odot}$ peak in GWTC-3. For a comprehensive review of which formation scenarios are likely (or unlikely) to be responsible for the $35M_{\odot}$ feature given GWTC-3 data, see \cite{Roy:2025ktr}.

Repeating the analysis of \cite{Ray2024} on GWTC-4, \cite{Abac_pop_2025} reported that the trends in the effective spin distribution of $30-40M_{\odot}$ BBHs have persisted as a robust feature in the updated detection sample, which comprises twice as many observations. However, \textit{it was not explored whether the $35M_{\odot}$ feature originates exclusively from a dynamically formed subpopulation}. BBHs forming through dynamical encounters in GCs can be expected to peak in the $30-40M_{\odot}$ range, correspond to a symmetric effective spin distribution peaking at zero, and strongly prefer equal mass systems \citep{Rodriguez:2016vmx,Farr:2017uvj,Antonini2023}. Therefore, to ascertain whether or not BBHs in the $30-40M_{\odot}$ range are predominantly formed in GCs, it is necessary to search for such a subpopulation in the space of BBH primary~(heavier component) masses, mass ratios, and effective inspiral spins, preferably with a non-parametric framework to avoid model-induced biases \citep{Callister:2022qwb, Callister_Farr_2024,Alvarez-Lopez:2025ltt}. Owing to its high flexibility and construction in the space of BBH \textit{component masses} and effective spins, the models of \cite{Ray2024} are unable to directly constrain a subpopulation that is consistent with the effective spin and \textit{ mass-ratio} distributions predicted by BBH formation in GCs. For the same reasons, such an analysis could not assure whether the inferred mass-spin correlations are, in fact, not manifested due to an underlying mass-ratio-effective spin correlation being marginalized while reconstructing the conditional mass-spin distributions.

In this \textit{Letter}, using a non-parametric function for the joint distribution of BBH primary masses, mass ratios, and effective inspiral spins, constructed from binned Gaussian Processes~\citep[BGPs;][]{Mohite2022, Ray:2023, Ray2024} we characterize BBH subpopulations in GWTC-4 and constrain the astrophysical origins of the three robust features in the mass distribution, namely the $10M_{\odot}$ peak, the $35M_{\odot}$ feature and the high mass~$(>45M_{\odot})$ cliff.  We find evidence for three distinct subpopulations in the space of BBH primary masses, mass ratios and effective inspiral spins spanning the entire detection sample. In particular, we show that only one subpopulation, consistent with preferentially equal mass-systems and a symmetric effective spin distribution peaking narrowly close to zero, contributes to the $35M_{\odot}$ peak, whereas the other two demonstrate a monotonic fall-off in the $30-40M_{\odot}$ range. We find that the mass-spectrum of this subpopulation differs from the other two by more than $90\%$ significance in the $30-40M_{\sun}$ range. By comparing with theoretical simulations of dynamical BBH formation in GCs, we show that only this subpopulation can comprise such systems predominantly, whereas the other two show substantial contributions from alternate channels. We conclude that, given GWTC-4, dynamical assembly in GCs likely dominates the $30-40M_{\odot}$ range and hence is a strong candidate for the origins of the $35M_{\odot}$ feature in the BBH mass-spectrum. We further constrain the range of BH birth-spins in GCs that are consistent with our inferred distributions.

 Additionally, we find that for the lower and higher mass subpopulations, the inferred distributions are consistent with the conclusions of previous studies by \cite{Godfrey:2023oxb,Sadiq_2024} and \cite{MaganaHernandez:2024qkz, MaganaHernandez:2025fkm,Ray2025_PISN}, respectively. In particular, we find that lower masses below $\sim 30 M_{\odot}$ are associated with a mass-ratio distribution peaked near $0.6-0.8$. For masses higher than $\sim 45 M_{\odot}$, we find that the inference can be prior driven and that flexible priors lead to no clear evidence of a PISN mass gap in agreement with the findings of \cite{Ray2025_PISN}. Furthermore, we do not find strong evidence of a mass-ratio-effective spin correlation in any range of primary mass. We validate our inference by exploring variations in bin resolutions and analyzing a large simulated catalog, demonstrating that our inference is not prior-driven.

The rest of this \textit{Letter} is organized as follows. We begin with a description of the model that we employed, in Sec. \ref{sec:methods}. In Sec. \ref{sec:results}, we outline the main results we obtain from inference on GWTC-4 data. We discuss the astrophysical implications of our inference results in Sec. \ref{sec:astro imps}, concluding with a brief synopsis of the results and future plans for improving and expanding our inference model in light of upcoming releases by the LVK from the later parts of the fourth observing run.

\section{Methods}\label{sec:methods}



For examining potential subpopulations in GWTC-4 data, the quantity of interest in our analysis is the merger rate density as a function of chosen parameters, i.e. the number of events for a given set of primary mass, mass ratio and effective spin of the binary. 
We model the rate density as a piecewise binned function in three dimensions, such that it is treated as a hyper-parameter to be inferred in each bin \citep{Mohite2022}. Specifically, we define:

\begin{equation} \label{eq: rate density def}
    n^{\gamma} = \frac{d N^{\gamma}}{d \log{m_1} dq d \chi_{\text{\rm eff}} d V_c dt_s},
\end{equation}

where $n^\gamma$ is the merger rate density per co-moving volume ($V_c$), source-frame time ($t_s$), log primary mass ($m_1$), mass-ratio ($q$) and effective spin ($\chi_{\rm eff}$) in the $\gamma$th bin.



Our population model permits a broad range of correlations in the ($m_1, q, \chi_{\rm eff}$) space, and is not informed a priori by astrophysically motivated functional forms. The joint distribution takes the following form:

\begin{widetext}

\begin{equation} \label{eq: Rate density prob dist rel}
\begin{split}
    \frac{d N}{d m_1 dq  d z d \chi_{\text{\rm eff}}} (m_1, q, z, \chi_{\rm eff}| \vec{n}, \kappa)  &= T_r\sum_\gamma \frac{n^\gamma}{m_1} \frac{dV_c}{dz} (1+z)^{\kappa-1} \times  \begin{cases} 
      1 & (m_1,q,\chi_{\text{\rm eff}}) \in \gamma^{th} ~\text{bin} \\
      0 & \text{otherwise} 
   \end{cases} 
\end{split}
\end{equation}

\end{widetext}

where the left-hand side represents the number of merger events per primary mass, mass ratio, effective spin and redshift; $T_r$ is the observation time in detector frame, and $\kappa$ is the redshift evolution parameter \citep{Fishbach:2018}. 
Note that the model in Eq.(\ref{eq: Rate density prob dist rel}) does not assume any particular form for the mass or spin distributions, and is flexible enough to, in principle, infer any possible correlations between parameters of interest, up to the chosen bin resolution. In addition, unphysical bins in the $(m_1,q)$ space, i.e., bins with $q < m_{\rm min}/m_1$, where $m_{\rm min}$ is the lowest $m_1$ bin, are removed to ensure that our model covers the same range of masses for both components.

The merger rate density within each bin is inferred using Bayesian hierarchical inference, with the occurrence of BBH events modeled as an inhomogeneous Poisson process  \citep{Messenger:2012jy,Mandel:2019, Wysocki2019, Vitale:2020,Essick:2025zed}. We account for Malmquist biases in the inferred distribution using a large set of simulated signals drawn from a fiducial population. Under these considerations, we construct our rate density likelihood using observed GW data.
Like previous implementations \citep{Mohite2022,Ray:2023, Ray2024}, the ingredients involved in constructing the rate-density likelihood have the advantage of being precomputable for a given $\kappa$ value, 
thereby contributing to the computational efficiency of this inference framework. 

We note, here, that in highly flexible models, the smoothing prior can influence the distribution non-negligibly in regions of sparse data~\citep{Heinzel:2024, Heinzel:2023hlb}, such as the very low~$(m_1<7M_{\odot}, \chi_{\rm eff}<-0.5)$ and very high~$(m_1>90M_{\odot}, \chi_{\rm eff}>0.5)$ mass and effective spin ends. Furthermore, as an inherently three-dimensional model, our framework fixes the distribution of other BBH parameters to restrictive functional form. However, given the low measurability of the ignored parameters and a lack of evidence in favour of any strong correlation between them and the parameters that are flexibly modeled, it is unlikely that our astrophysical conclusions regarding $m_1,q,\chi_{\rm eff}$ subpopulations in the data-rich regions of parameter space are affected by such limitations.

The logarithmic rate density is modeled using a Gaussian process (GP) prior, which regularizes the population distribution in regions of parameter space with sparse data \citep{Foremanmackey2014, Mandel2017}. We use an exponential quadratic kernel to express the covariance matrix of the GP. The means, amplitude of the covariances, and correlation length scales of the GP are hyperparameters we infer simultaneously with the rate densities, and are in turn modeled using standard normal, standard half-normal, and log-normal prior distributions, respectively \citep{Mohite2022, Ray:2023, Ray2024}. The prior mean and variance of the length scales are chosen such that the furthest and nearest bin centers are within $4\sigma$ of the prior distribution.

To infer the joint posterior distribution of the merger rate densities and the hyperparameters describing the GP, we employ Hamiltonian Monte Carlo (HMC) sampling with the No U-Turn Sampler algorithm \citep{Brooks_2011, Hoffman2014}. We extend the publicly available inference pipeline \texttt{gppop}\footnote{https://github.com/AnaryaRay1/gppop/tree/main} \citep{Ray:2023, Ray2024} to implement our inference. This code in turn relies on the \texttt{PyMC} \citep{Oriol_2023} library to conduct the HMC sampling.

While constructing the population likelihood, the Monte Carlo sums used to approximate the integrals over single-event posterior samples and detectable simulations can be prone to uncertainties \citep{Farr2019, Essick2022, Talbot:2023pex, Heinzel:2025ogf}, as a result of which there is a possibility of biases arising from non-converged point estimates to uncertain integrals. We monitor the variance in our likelihood function arising from that of the Monte Carlo integrals and penalize it accordingly \textit{during sampling}, using a variance cut, as proposed by \cite{Talbot:2023pex}. This was not implemented in the previous BGP analyses of \cite{Ray:2023, Ray2024, MaganaHernandez:2024}, where coarse bin resolutions allowed for this check to be carried out in post-processing. We find that for higher resolutions, fewer samples are rejected in post-processing leading to a higher effective sample size at the same computational cost. Further details are discussed in Appendix \ref{app: MC convergence}.

\section{Results: Subpopulations in GWTC-4} \label{sec:results}

We analyze GWTC-4 data \citep{Abac_data_2025} which comprises of 153 confidently observed\footnote{i.e., with a false alarm rate of one per year or lower.} BBH events \citep{Abbott_2020, Abbott_2021, Abbott2023, Abac_pop_2025, Essick2022apJ}, excluding clear population outliers. Due to cosmological expansion, the observed frequency/time-evolution and flux of the signal measured by the detectors get redshifted, and detector frame mass and luminosity distance samples need to be converted to source frame through a cosmological model, which we assume to be Planck2015 \citep{Planck:2015fie}. We choose bins uniform in log mass and a specific set of $\chi_{\rm eff}$ bins, both identical to the BGP inference of \cite{Abac_pop_2025}. For mass-ratios, we choose uniform bins between 0.1 and 1. The exact locations of bin edges can be found in Table~\ref{tab: Bin choices}. To ensure that the inferred conclusions are not sensitive to binning choices, we provide inference results for an alternate bin choice with different primary mass bins in Appendix \ref{app: Case_28_bins}. Those results are fully consistent with the ones presented here.



\begin{table}[H] 
\begin{adjustwidth}{0cm}{}
\begin{tabular}{cc}
\hline
\hline
Parameter & Bin edges \\
\hline
$m_1 (M_{\odot})$           & \hspace{1em} log-uniform(5, 200, 23)                                                                              \\

\hline
$\chi_{\rm eff}$ & \hspace{-3em} \begin{tabular}[c]{@{}c@{}} -0.7, -0.6, -0.4, -0.3, -0.2,   \\ -0.1, -0.05, 0.0, 0.05, 0.1, 0.15,  \\  0.2, 0.3, 0.4, 0.6, 0.7
       \end{tabular}                                                                              \\
\hline
$q$   & \hspace{-2em} \begin{tabular}[c]{@{}c@{}} 0.1, 0.2, 0.3, 0.4, 0.5,  \\ 0.6, 0.7, 0.8, 0.9, 1.0 \end{tabular}  \\
\hline
\end{tabular}
\end{adjustwidth}
\caption{The bin choice used in this section.}\label{tab: Bin choices}
\end{table} 

\begin{figure*}
\centering

\includegraphics[width = 0.48\linewidth]{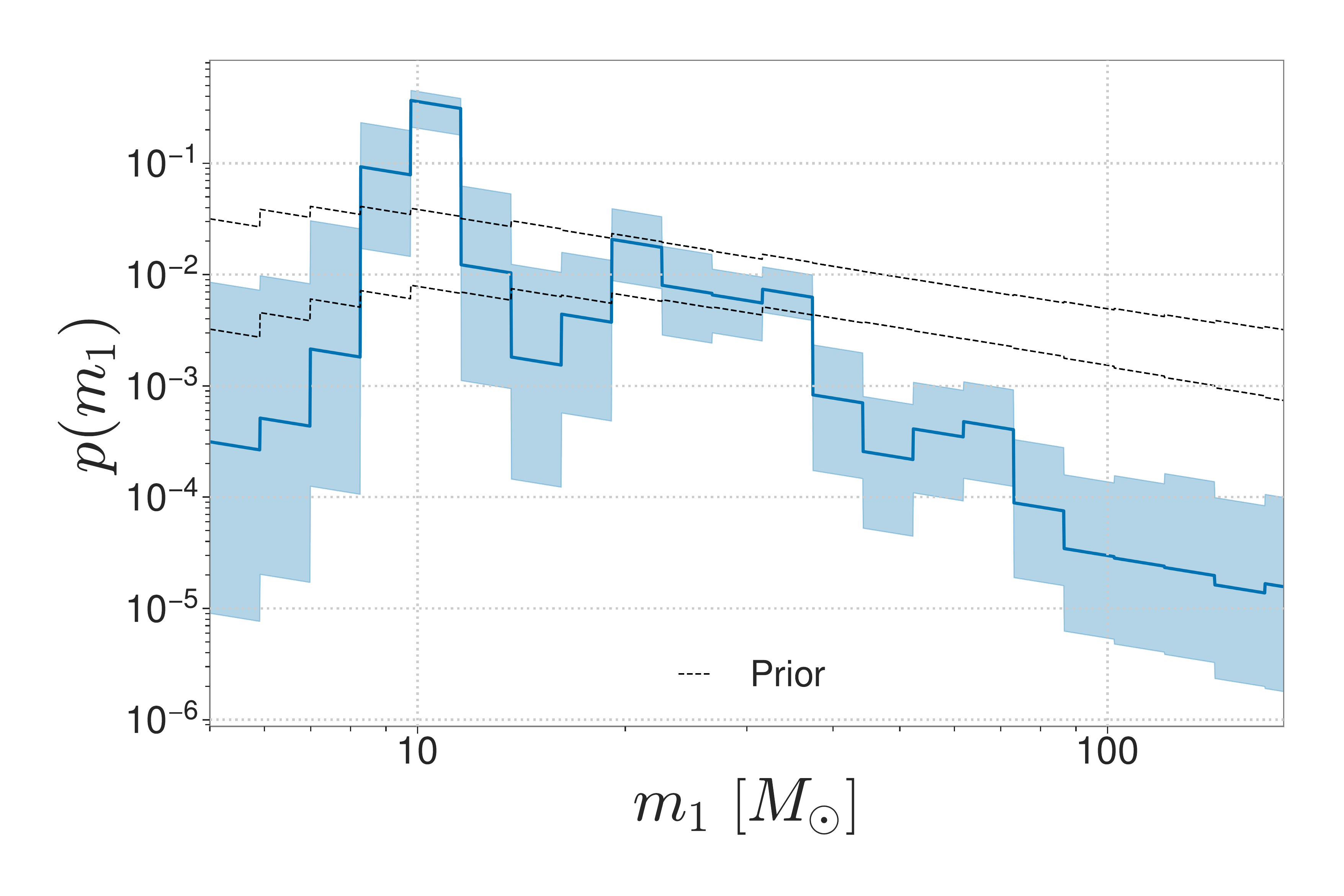}
\includegraphics[width = 0.48\linewidth]{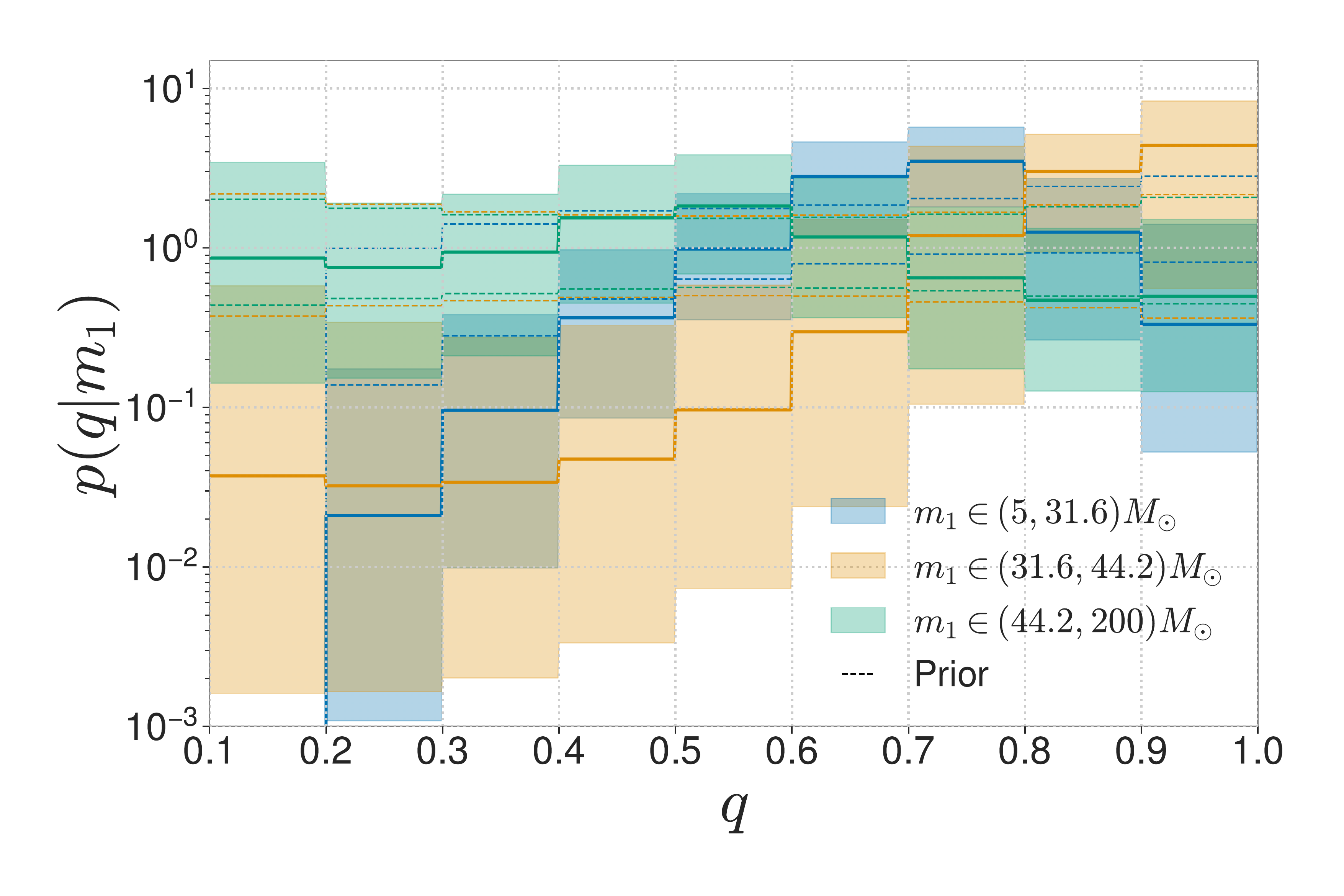}

\includegraphics[width = 0.48\linewidth]{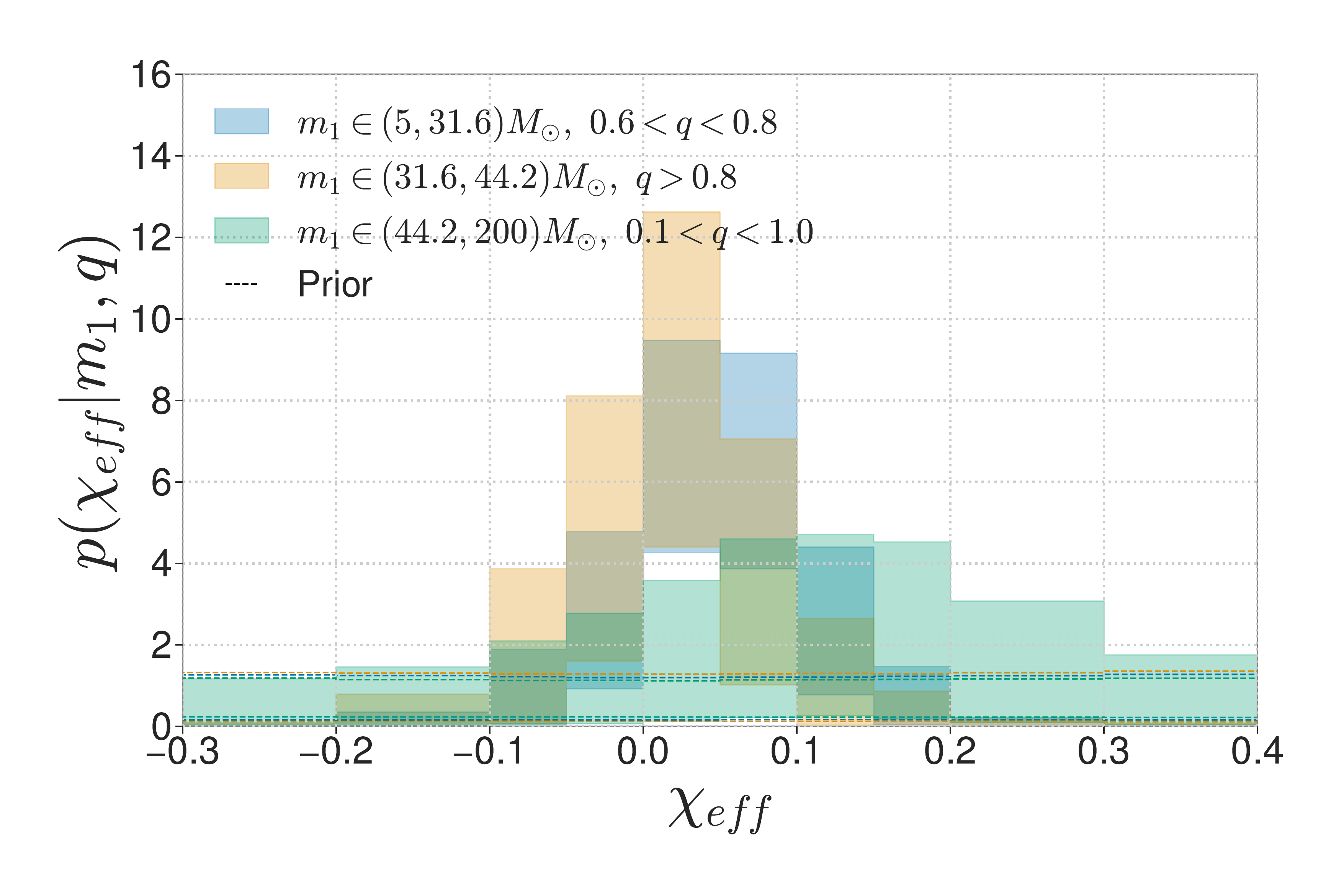}
\includegraphics[width = 0.48\linewidth]{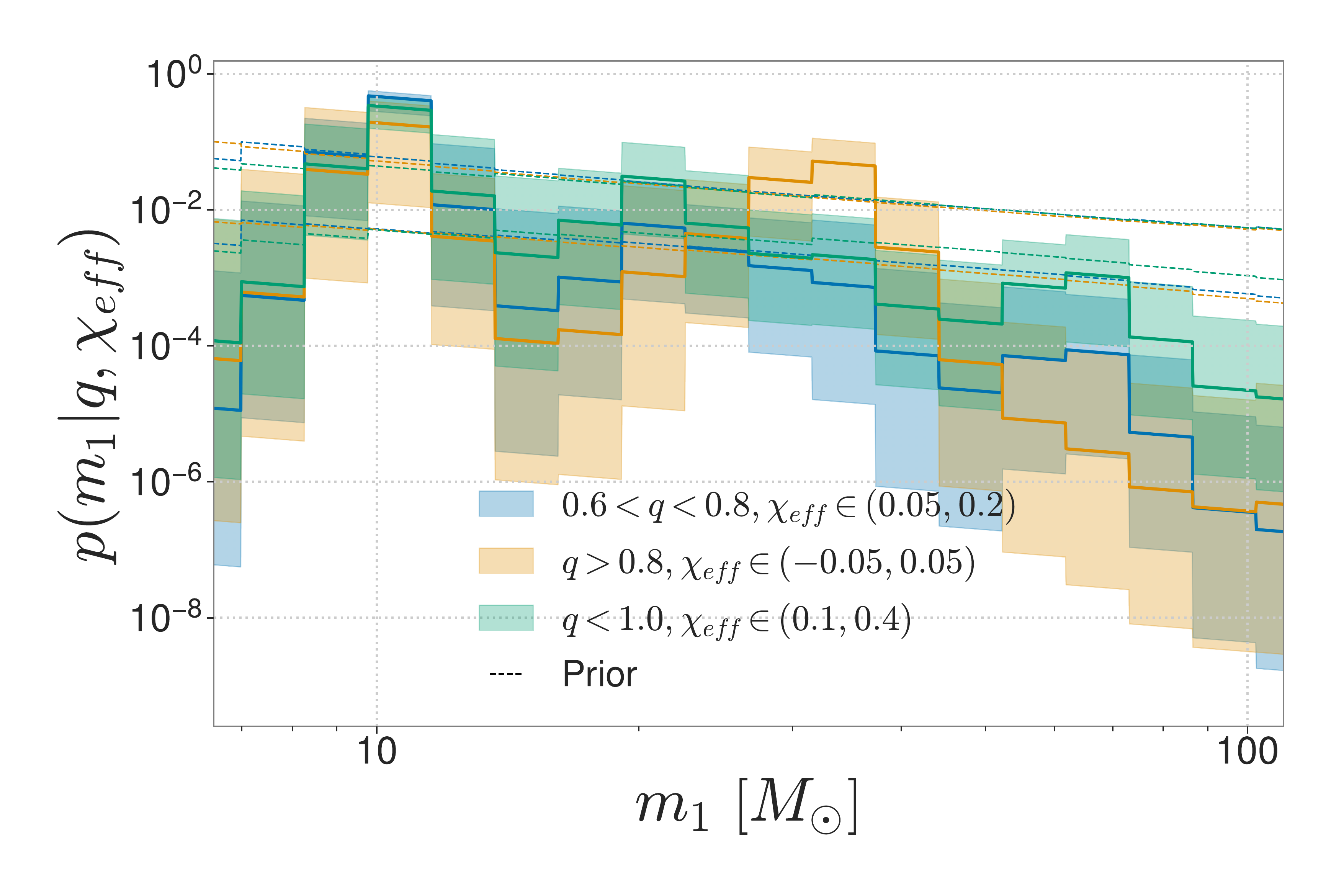}

\caption{
Marginal and conditional distributions of primary mass, mass-ratio and effective spin inferred from GWTC-4 for the bin choice described in Table \ref{tab: Bin choices}. The top panels show the marginal primary mass distribution on the left and the mass ratio distribution conditioned on three primary mass ranges to the right. The bottom panels show the effective spin distribution for three different primary mass and mass-ratio ranges to the left and the primary mass distribution conditioned on three different mass ratio and effective spin ranges on the right.
}
\label{fig: Subpop_case_9}
\end{figure*}

\subsection{Three distinct subpopulations} \label{subsec: Subpopulations}

From our inferred joint distribution of BBH primary masses, mass-ratios, and effective inspiral spins, we find evidence for three distinct subpopulations. Guided by the over-densities in the marginal mass-distribution, we reconstruct the conditional distribution of BBH mass-ratios given different ranges of primary mass, namely:
\begin{enumerate}
    \item{Subpopulation 1: $m_1\in (5,31.6)M_{\odot}$},
    \item{Subpopulation 2: $m_1\in (31.6, 44.2)M_{\odot}$}, and
    \item{Subpopulation 3: $m_1>44.2M_{\odot}$}.
\end{enumerate}
We then construct the inferred conditional distributions of $\chi_{\rm eff}$ given particular ranges of primary-mass and mass-ratio. Finally, we reconstruct the conditional distribution of primary masses given mass ratio and effective spin ranges to confirm whether various features in the mass distribution indeed correspond to distinct mass ratio and effective spin distributions. 

Our results, displayed in Figure~\ref{fig: Subpop_case_9}, indicate the existence of three distinct BBH subpopulations in the GWTC-4 sample, which we present below.



\begin{itemize}
    \item {Subpopulation 1 comprises predominantly low-mass systems in the $m_1 \in (5, 31.6) M_{\odot}$ range displaying \textit{a sharp peak in the mass-distribution near $10M_{\odot}$}. It is characterized by \textit{a peak in the mass-ratio distribution} in the ranges ($0.6 < q < 0.8$) and \textit{a positively skewed effective spin distribution} peaked away from $\chi_{\rm eff}=0$, indicative of preferential alignment and non-negligible spin magnitudes. This subpopulation does not exhibit a peak in the mass distribution near $30-40M_{\odot}$.}
    \item {Subpopulation 2 consists of binaries with intermediate $m_1 \in (31.6, 44.2) M_{\odot}$. \textit{It contributes to the $\sim35M_{\odot}$ peak in the mass distribution} and is characterized by preferentially equal mass systems \textit{with the mass-ratio distribution railing against $q=1$}. It has an \textit{effective spin distribution peaked near and symmetric about $\chi_{\rm eff}=0$} indicating isotropic spin orientations and small spin magnitudes.}
    \item {Subpopulation 3 \textit{contributes dominantly to the higher end of the mass spectrum at} $m_1 \gtrsim  44.2 M_{\odot}$, has a broad, \textit{nearly flat mass-ratio distribution} with support for both equal mass and asymmetric mass systems, and \textit{a broad, positively skewed effective spin distribution} preferring spin alignment and large spin magnitudes. The effective spin and mass-ratio distributions are clearly distinct from the rest of the ensemble in multiple bins with more than $90\%$ confidence. \textit{It also displays a hint of a bump near $60-70M_{\odot}$} \citep{MaganaHernandez:2024qkz,MaganaHernandez:2025fkm, Wang:2025_PISN} and a fall off in merger rate density above $70M_{\odot}$. We note here, that due to the scarcity of observations, this mass-range is partially informed by the prior unlike other two subpopulations}
\end{itemize}
From the reconstructed conditional mass distributions (see bottom right panel of Figure \ref{fig: Subpop_case_9}), only Subpopulation 2 exhibits the $35M_{\odot}$ peak, whereas the other two demonstrate a monotonic fall-off in the $30-40M_{\odot}$ mass range.

\begin{figure*}
\centering
    \includegraphics[width=0.48\textwidth]{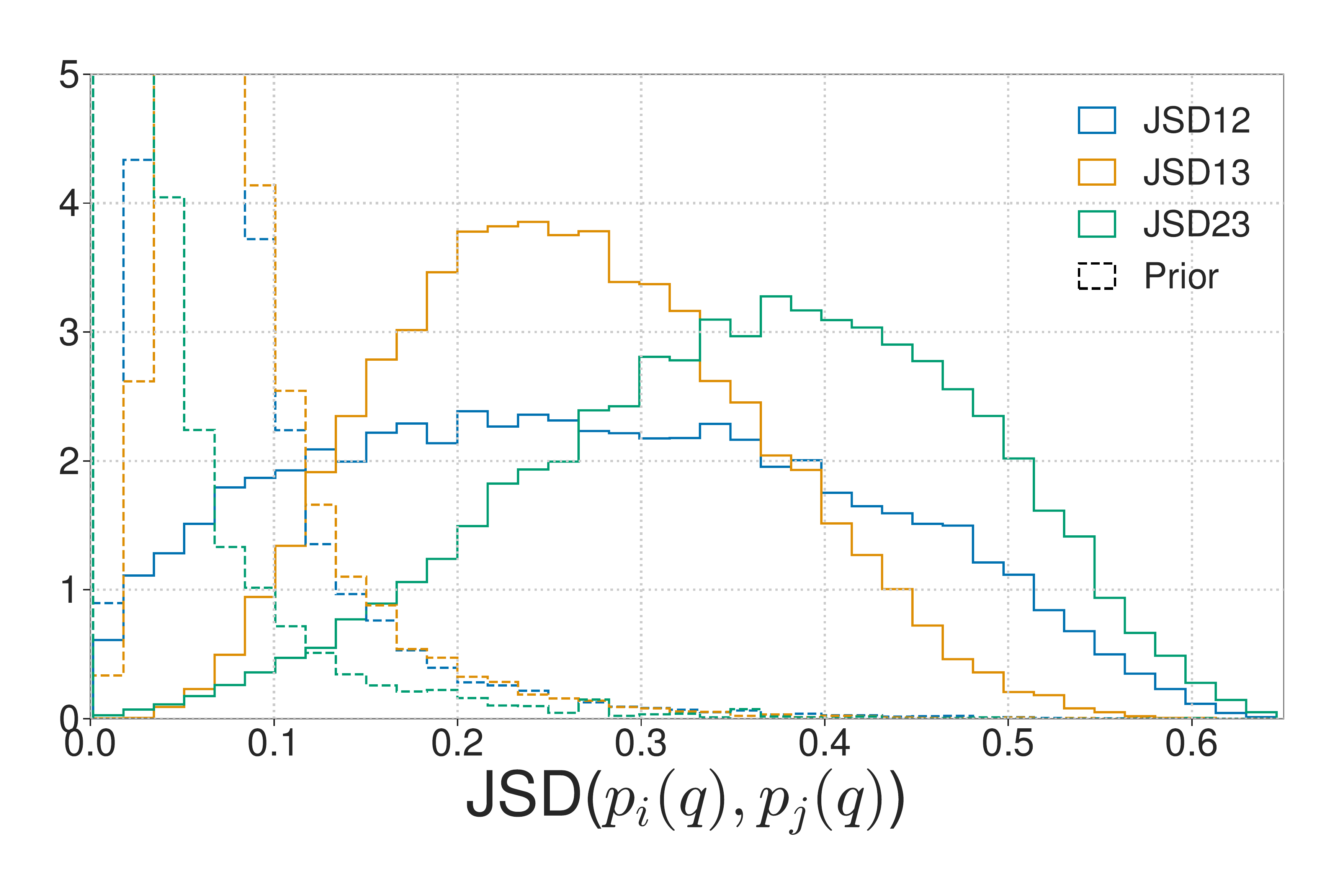}
    \includegraphics[width=0.48\textwidth]{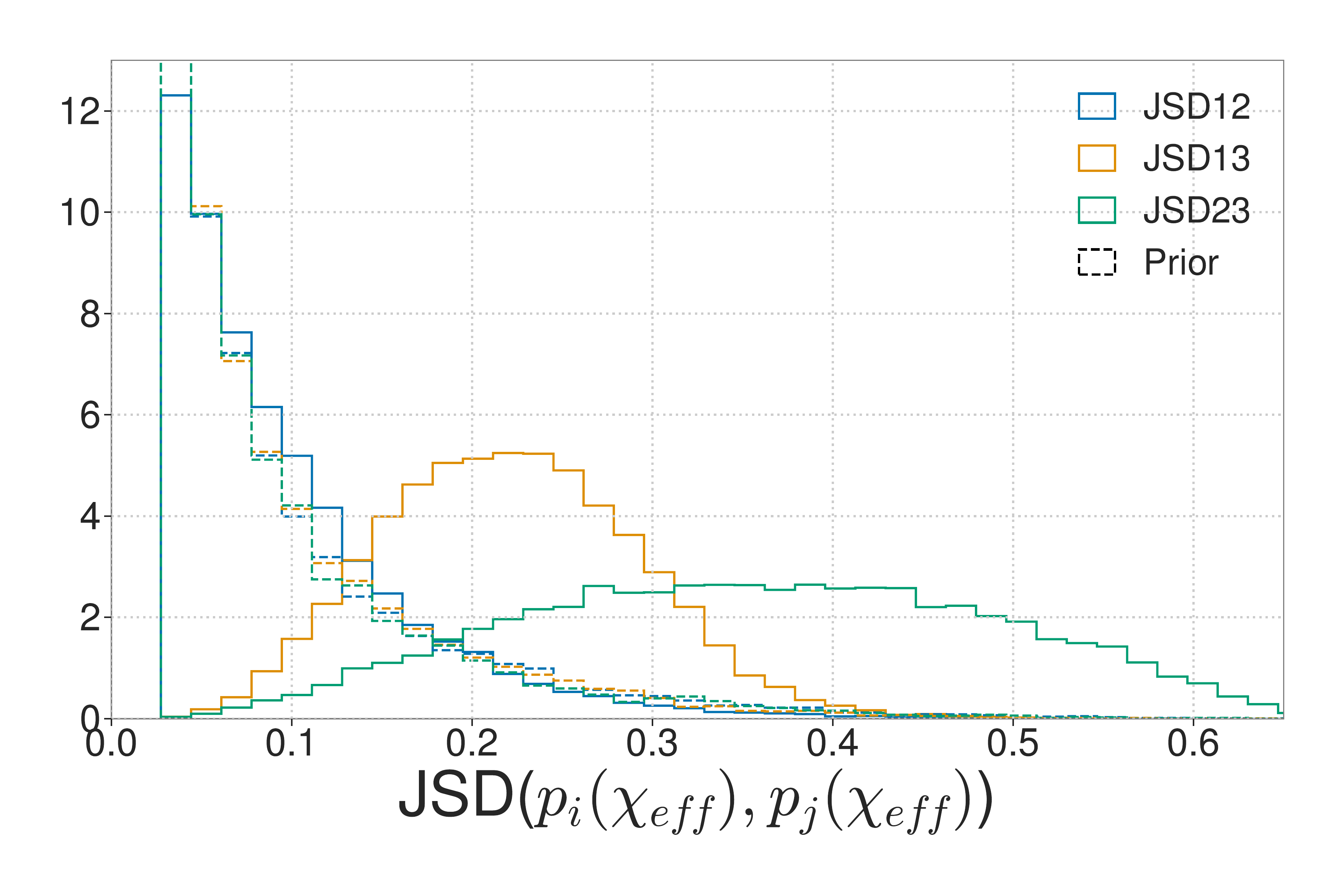}
    \caption{\label{fig:JSD}JS divergence between conditional mass-ratio~(\textit{left}) and $\chi_{eff}$~(\textit{right}). Here the subscripts 1,2, and 3 represent the three subpopulations/ mass ranges (low-mass, $30-40M_{\odot}$ and high mass) respectively.}    
\end{figure*}

To quantitatively assess the number of subpopulations consistent with the data and establish their existence, we compute the Jensen-Shannon~(JS) divergence between the corresponding conditional distributions of $q$ and $\chi_{eff}$. In Figure~\ref{fig:JSD}, we show the posterior and prior of the JS divergences for each pair of subpopulation and show that the data prefer distinct~($0<JSD\leq 1$) $q$ and $\chi_{eff}$ distributions over ones that are identical~($JSD\sim0$). Note that for $p(\chi_{eff})$, the JS divergence between subpopulations 1 and 2 are barely more informative than the prior and more data might be necessary to quantitatively establish whether or not these two distributions are confidently distinct. However, as we show later on in Section~\ref{sec:astro imps}, the distributions are distinct enough to establish that certain BBH formation scenarios predict spin distributions that are inconsistent with Subpopulation 1 but not with Subpopulation 2. In Appendix~\ref{sec:app-mass-boundaries}, we further explore the sensitivity of our conclusions to the mass-ranges chosen to demarcate the various subpopulations by computing JS divergences between $q$ and $\chi_{eff}$ distributions conditioned on different mass-ranges than the ones considered in this section. We find that perturbations in the edges of the mass-range do not change our conclusions quantitatively.

We note that a highly flexible BGP model can uncover previously unexplored trends in the population, but struggles to completely disentangle the relative abundances of specific subpopulations. Merger rate densities in a specific set of bins can only be interpreted as lower or upper bounds when compared with the theoretical predictions of a formation channel whose distributions are expected to peak in the respective ranges of the BBH parameter space. Nevertheless, the distinct features of these subpopulations can still inform on the specifics of evolutionary pathways that contribute dominantly to BBH formation in these regions of the parameter space. We discuss such astrophysical implications of our results in the context of theoretical predictions from proposed formation scenarios, as well as the results of existing subpopulation studies with GWTC-4, in Section~\ref{sec:astro imps}.

\subsection{Mass-ratio - effective spin correlation} \label{subsec: q-chieff}

\begin{figure}
\centering
\includegraphics[width = 0.48\textwidth]{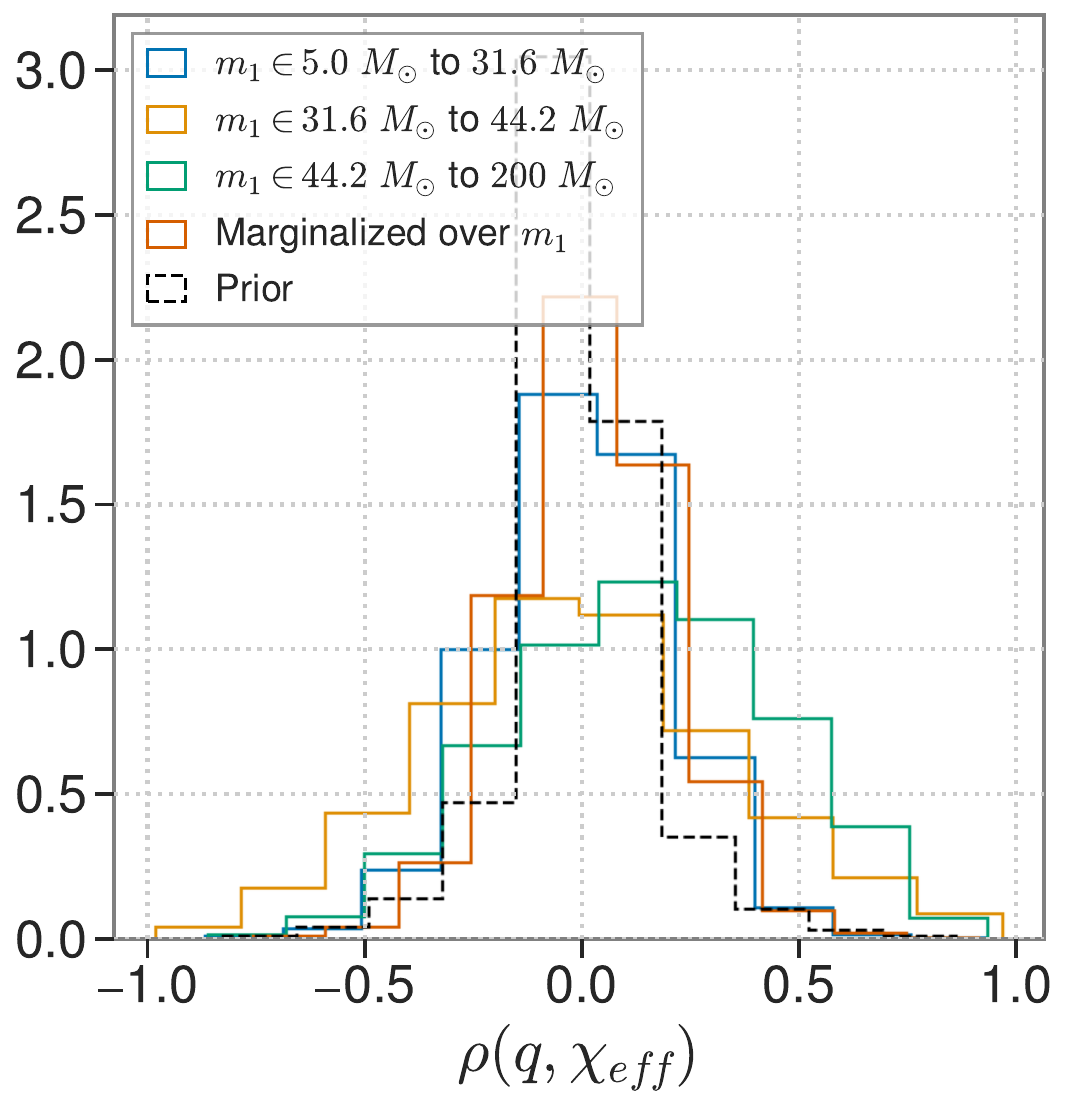}
\caption{
The Pearson correlation coefficient posteriors between effective spin and mass-ratio for three different mass ranges, namely $m_1 \in (5.0 M_{\odot}, 31.6 M_{\odot})$, $m_1 \in (31.6 M_{\odot}, 44.2 M_{\odot})$ and $m_1 \in (44.2 M_{\odot}, 200 M_{\odot})$, along with the same marginalized across the entire primary mass space and for the prior distribution.
    }
\label{fig:PCC_case_9}
\end{figure}


Next we investigate the $q-\chi_{\rm eff}$ correlation that has been reported by numerous previous works including \cite{Callister2021, Adamcewicz2023, Heinzel:2023hlb, Heinzel:2024hva, Abac_pop_2025}. We search for this correlation not only in the entire BBH population but also in specific ranges of primary mass, which is possible from our data-driven constraints on the joint distribution of $m_1,q,\chi_{\rm eff}$. In order to quantify the existence of this correlation, we compute the posterior distribution of the Pearson correlation coefficient \citep{Pearson_1896} between effective spin and mass ratio in different ranges of $m_1$, displayed in Figure \ref{fig:PCC_case_9}. 

Given that the Pearson coefficient posterior is peaked near zero as shown in Figure \ref{fig:PCC_case_9}, we conclude that there is no significant evidence for a $q-\chi_{\rm eff}$ correlation, either in specific mass ranges or in the entire BBH population marginalized over $m_1$~\citep[see also,][who find no evidence for a $q-\chi_{\rm eff}$ correlation in the high mass sub-population above $20M_{\odot}$ using GWTC-3]{Roy:2025ktr}. We also display the median merger rate density on the two-dimensional $q-\chi_{\rm eff}$ plane for different ranges of primary mass in Appendix \ref{app: q-chi-corr}, along with the conditional distributions of $\chi_{\rm eff}$ for different mass-ratio ranges, marginalized over primary mass. These corroborate that there is no significant evidence in favor of an intrinsic $q-\chi_{\rm eff}$ correlation in the underlying population. However, given our measurement uncertainties, we do not necessarily rule out a mass-ratio and effective spin correlation and find posterior support for all values of the Pearson coefficient in the range~$(-0.5, 0.5)$. Nevertheless, these results indicate that the mass-spin correlations we do find with significance are not manifesting due to marginalization over an underlying mass ratio and effective spin correlation.

\section{Discussion}\label{sec:astro imps}
\begin{figure*}
\centering

\includegraphics[width = 0.32\linewidth]{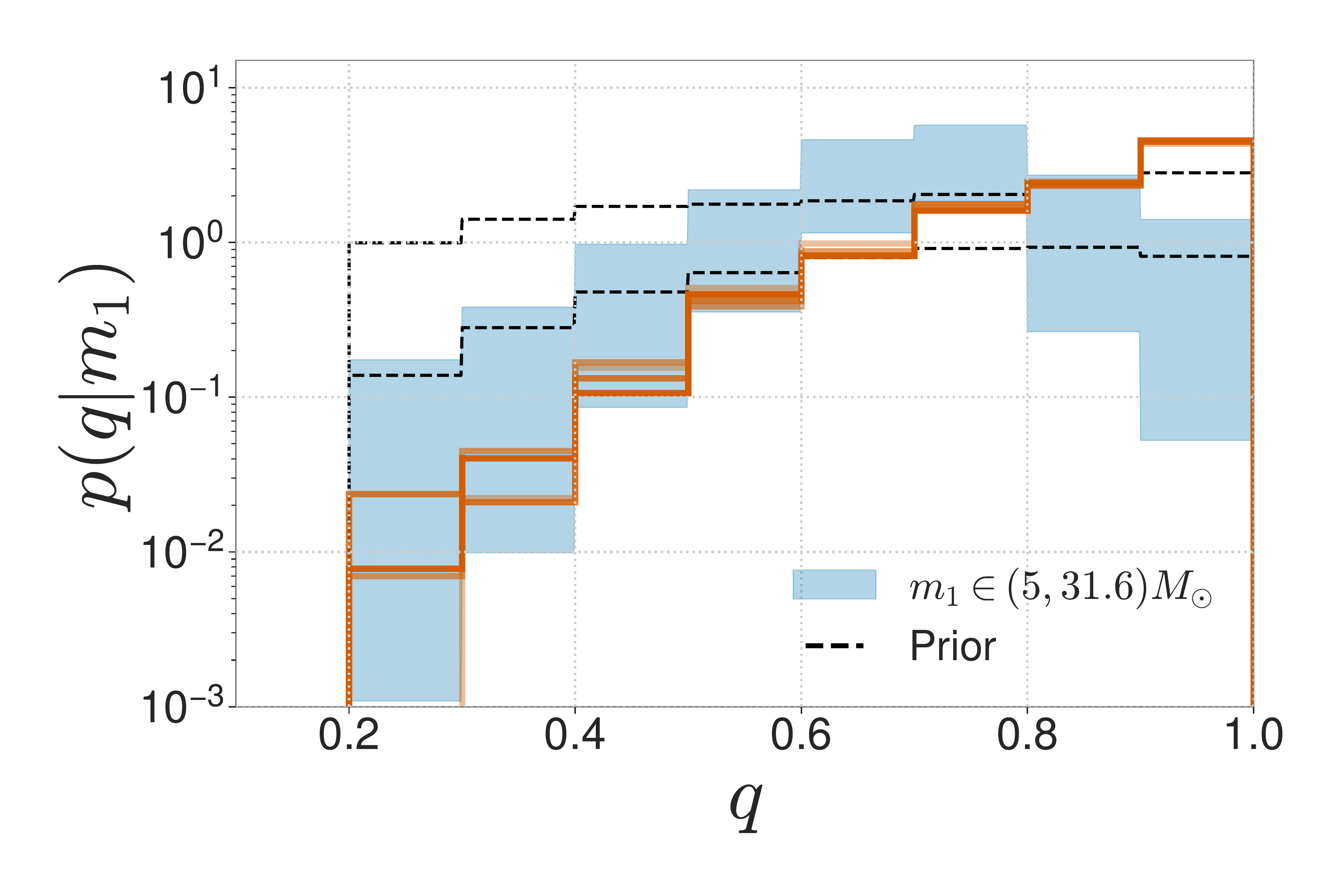}\vspace{-1mm}
\includegraphics[width = 0.32\linewidth]{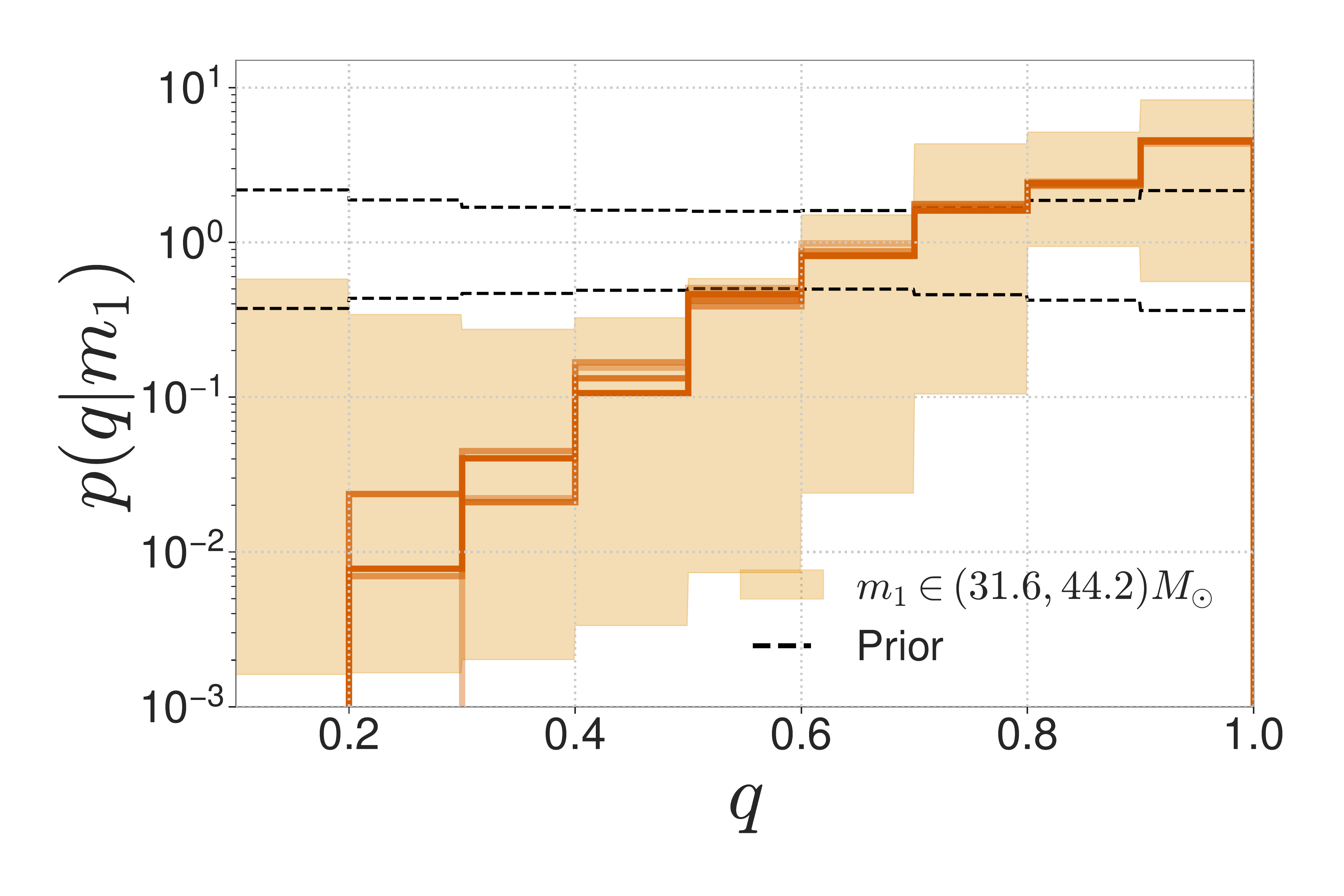}
\includegraphics[width = 0.32\linewidth]{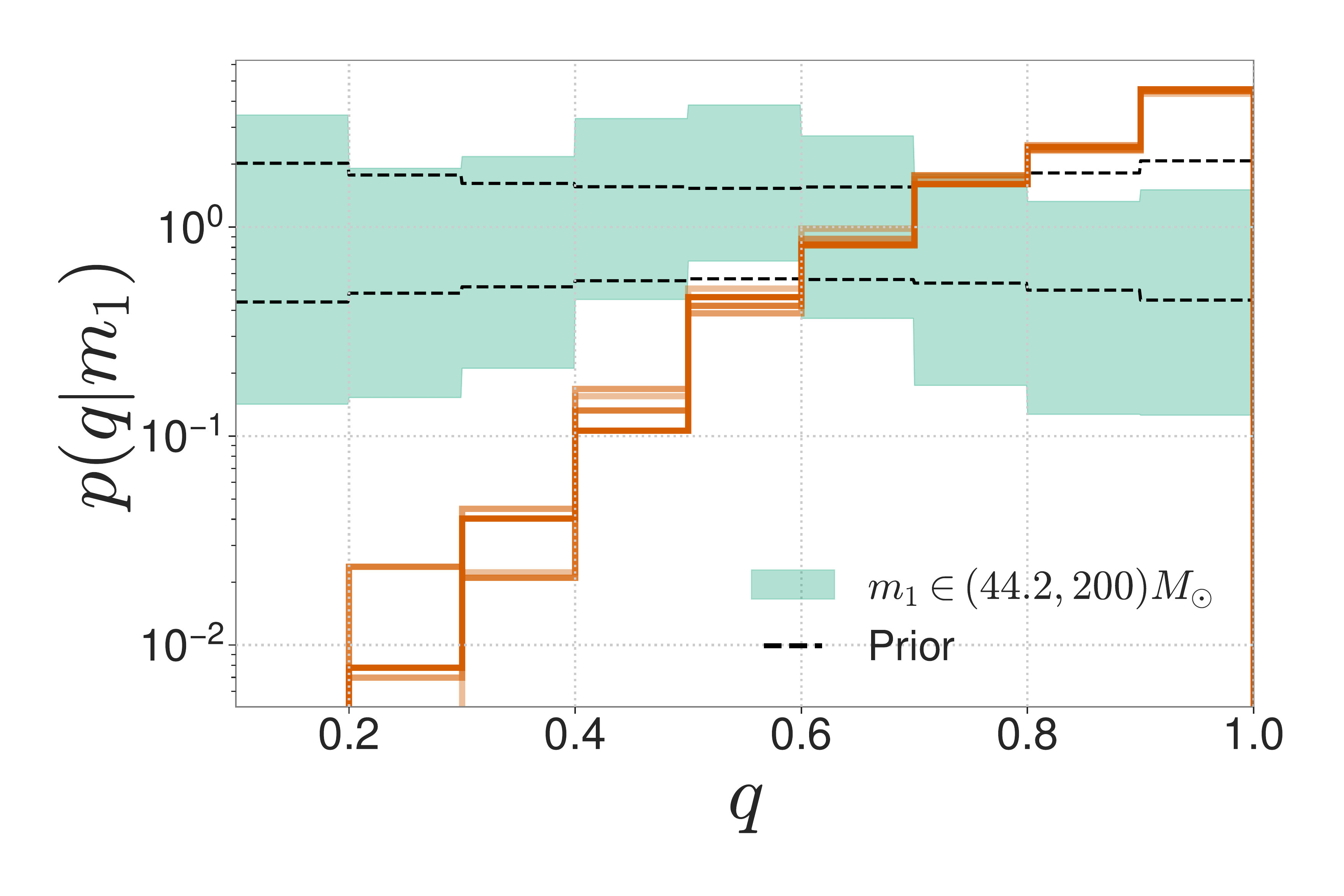}

\includegraphics[width = 0.32\linewidth]{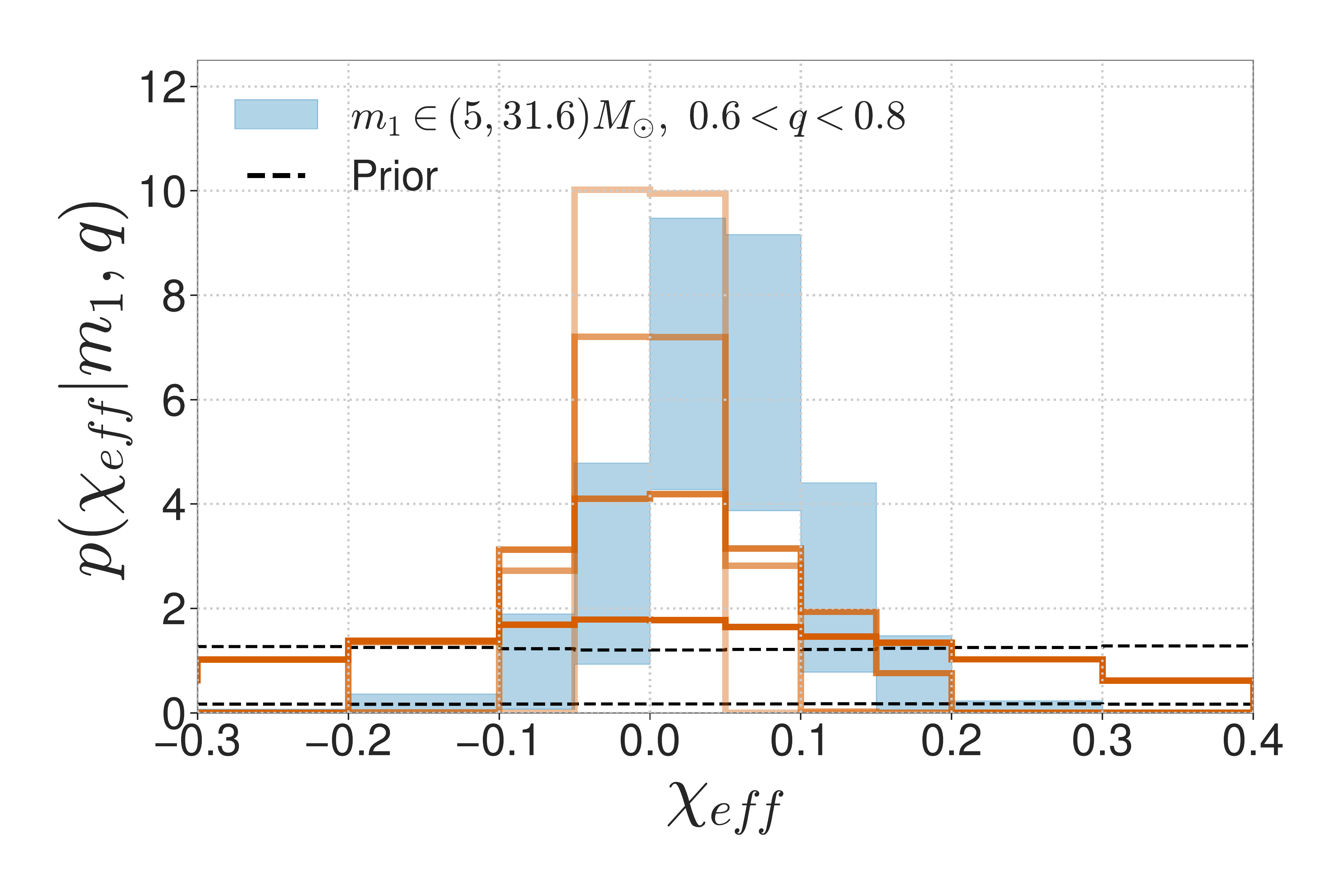}\vspace{-1mm}
\includegraphics[width = 0.32\linewidth]{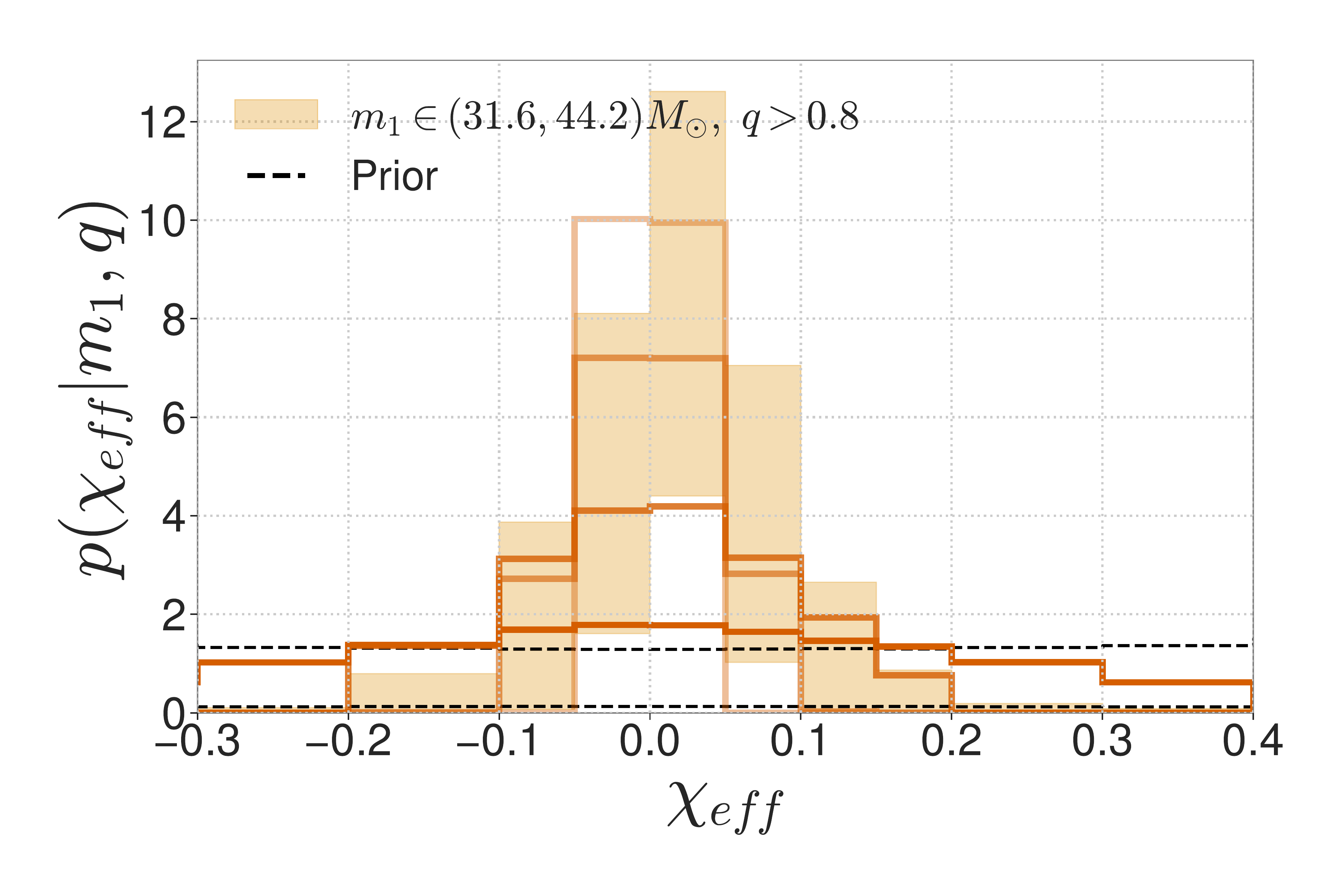}
\includegraphics[width = 0.32\linewidth]{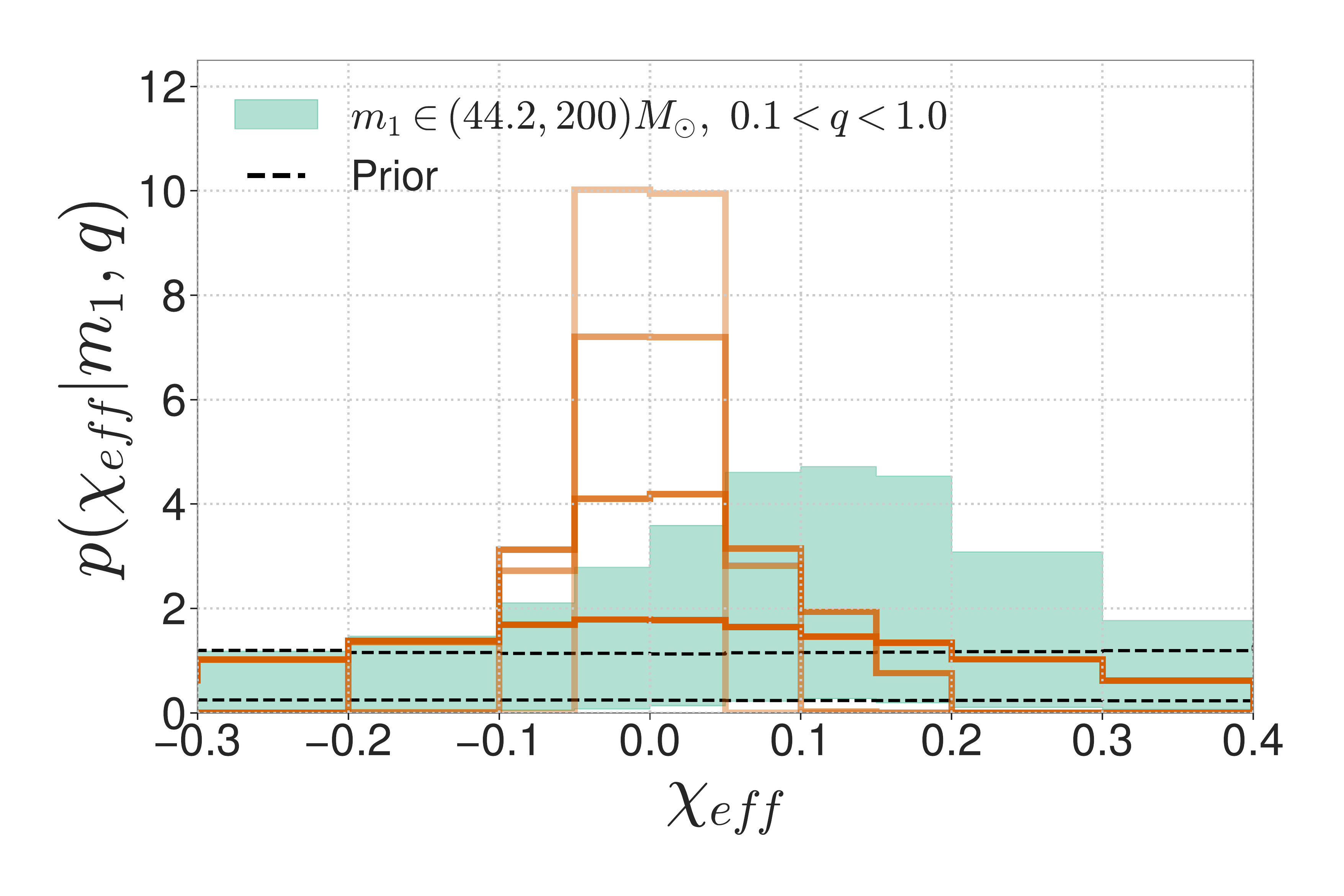}

\includegraphics[width = 0.32\linewidth]{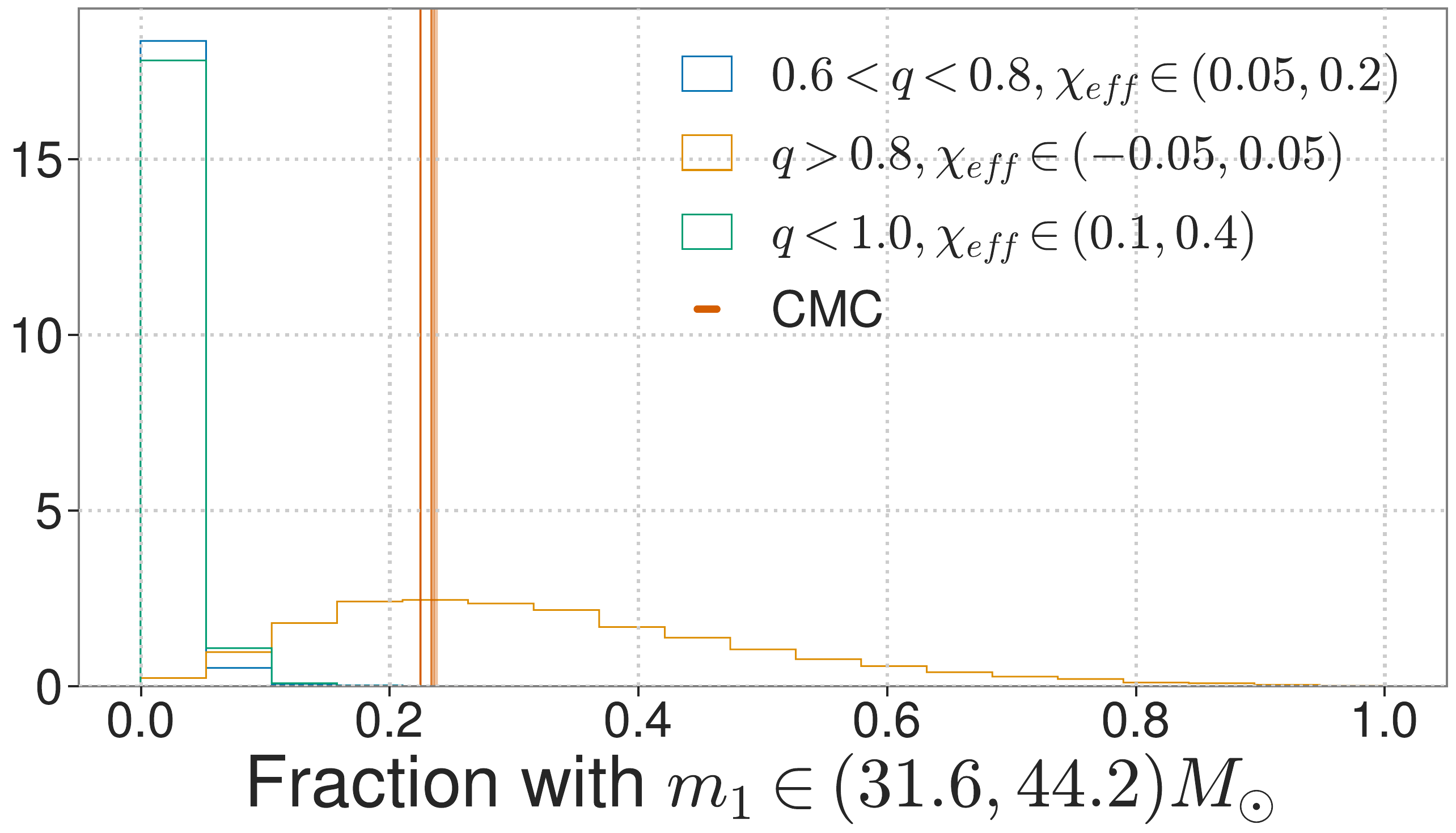} \vspace{-1mm}

\caption{
Comparison of conditional mass-ratio and effective spin distributions for the three subpopulations with the 1G+1G component of CMC simulations (which do not include hierarchical mergers) \citep{Rodriguez:2019huv}. The faintest to brightest lines correspond to birth spins of 0, 0.1, 0.2 and 0.5 respectively.
}
\label{fig: Subpop_vs_CMC_case_9}
\end{figure*}

In this work, we have resolved the joint distribution of BBH primary masses, mass ratios and effective inspiral spins using a data-driven framework for population inference based on BGP. Our examination of GWTC-4 data reveals evidence for 3 subpopulations, each corresponding to a particular feature in the primary mass distribution, that have distinct mass ratio and effective spin distributions: 
\begin{enumerate}
    \item Low primary mass ($5 M_{\odot} \lesssim m_1 \lesssim 30 M_{\odot}$) which contributes to the $\sim10 M_{\odot}$ peak, has a positively skewed effective spin distribution and a mass-ratio distribution peaking in the range $q\in(0.6,0.8)$
    \item Primary masses in $30 M_{\odot} \lesssim m_1 \lesssim 45 M_{\odot}$ which correspond to a symmetric effective spin distribution peaking near $\chi_{\rm eff} = 0$, and a mass-ratio distribution railing against $q=1$; and 
    \item High primary masses ($m_1 \gtrsim 45 M_{\odot}$) corresponding to a broad effective spin distribution with a slight preference for positive values, and a broad mass-ratio distribution with a significant fraction of binaries in the range $q\in(0.6,1)$. Due to the scarcity of events, this subpopulation is partially influenced by the prior.
\end{enumerate}
By reconstructing the primary mass distributions of each subpopulation (bottom right panel of Figure \ref{fig: Subpop_case_9}), \textit{we find that only Subpopulation~2 demonstrates the $35M_{\sun}$ peak, whereas the other two exhibit a monotonic fall off in $m_1\in(20M_{\odot},50M_{\odot})$}. On the other hand, Subpopulation 3 exhibits hints of a bump in the $60-70M_{\odot}$ range and a steady fall-off afterwards. We discuss the astrophysical implications of these correlations as follows, focusing primarily on Subpopulation 2 and the origins of the $35M_{\odot}$ peak.


For Subpopulation 2, the symmetric effective spin distribution peaking very close to zero implies relatively equal fractions of events with aligned and anti-aligned spin orientations. Hence, it is unlikely that isolated stellar binaries, wherein co-evolution of close systems tends to align their spins to the orbit \citep{Qin:2018vaa, Fuller:2019sxi, Belczynski:2017gds, Bavera:2020inc, Mandel:2020lhv}, contribute dominantly to the subpopulation that exhibits the $\simeq 35M_{\odot}$ peak. Similarly, the preference towards equal mass ratios makes it unlikely that BBH formation in the disks of active galactic nuclei, which is expected to have substantial support for both equal and unequal mass systems, contributes dominantly to this subpopulation \citep{Stone:2016wzz, Bartos:2016dgn, McKernan:2019beu}. 

Similarly, repeated mergers in dense stellar or gaseous environments are expected to comprise highly-spinning~(i.e., with dimensionless spin magnitudes of the order $0.67$) components~\citep{Berti:2008af, Hofmann:2016yih, Rodriguez:2019huv, Borchers:2025sid, Fishbach:2017dwv}, and hence exhibit a broad effective spin distribution~\citep{Antonini:2025c, Antonini:2025ilj}, which makes them inconsistent with the subpopulation 2. On the other hand, BBH mergers in triple systems, while consistent with a preference for zero effective spin due to Lidov-Kozai resonance induced by a distant perturber orienting individual spin components parallel to the orbital plane, have been shown to exhibit a positively skewed effective spin distribution~\citep{Antonini:2017ash, Rodriguez:2018jqu, Liu:2018nrf, Stegmann:2021jen, Stegmann:2025zkb} that is likely inconsistent with our inferred characteristics of Subpopulation 2. Further investigations and population models capable of inferring the distributions of effective precessing spin parameters in particular mass-ranges are necessary to verify this conclusion.

Finally, dynamically assembled systems in dense environments such as young, open, nuclear, and globular star clusters, that only comprise first generational BH components that are formed directly through stellar collapse, can be consistent with both the inferred effective spin and mass-ratio distributions for Subpopulation 2. \citep{DiCarlo:2019pmf,DiCarlo:2020lfa,Rodriguez:2019huv, Mapelli:2021gyv,Antonini:2012ad, Petrovich:2017otm, Chattopadhyay:2022}. However, only GCs can explain the abundance in the $30-40M_{\odot}$ range~\citep{Bruel:2025sdq}, whereas BBH formation in nuclear and young star clusters is expected to peak at lower and higher masses, respectively \citep{, Cheng:2023ddt, Antonini2023, Mapelli:2021gyv}. From this, we conclude that the $35M_{\odot}$ feature likely arises due to a dynamically formed subpopulation of BBHs originating in GCs. While previous studies have reported \textit{hints} of this conclusion in the data from earlier catalogs \citep{Ray2024}, our data-driven reconstruction of the BBH primary mass, mass-ratio, and effective spin distribution from GWTC-4 demonstrates stronger evidence.

However, to compare the inferred merger rates of subpopulation 2 with theoretical predictions one must consider that BBHs formed in GCs can, in principle, also contribute to the other two subpopulations. Furthermore, the predicted distributions can be susceptible to theoretical uncertainties such as the distribution of BH birth spins. In Figure \ref{fig: Subpop_vs_CMC_case_9}, we compare the subpopulations we uncovered with star cluster \citep{Rodriguez:2019huv} simulations derived from publicly available Cluster Monte Carlo \citep[CMC;][]{Pattabiraman_2013} code. The simulations were taken from the public
data release by \cite{Zevin:2020}. The simulated systems only comprise first generational mergers and correspond to the 4 specific values of BH birth spin. We find that only Subpopulation 2 is fully consistent~(at the $90$\% level of the inferred distributions) with these theoretical predictions. We further find that only the simulations that correspond to BH birth spins in the range $0.1-0.2$, are consistent, and that higher or lower values lead to an effective spin distribution that is either too broad or too narrow, respectively, as compared to the inferred one. Furthermore, the fraction of systems in the $30-40M_{\odot}$ range for the simulated populations is only consistent with that inferred from Subpopulation 2 and inconsistent with the other two. In other words, given GWTC-4, we find that \textit{dynamical mergers in GCs can contribute predominantly only in the $30-40M_{\odot}$ range}, whereas the subpopulations in other mass ranges have significant contributions from alternate formation channels. The merger rate in this mass range, therefore, serves as a robust lower bound on that of dynamically assembled BBHs in GCs, which we constrain to be $0.69^{+0.23}_{-0.33}\rm{Gpc}^{-3}\rm{yr}^{-1}$. This is consistent with most theoretical predictions~(that can range from $0.2-57\rm{Gpc}^{-3}\rm{yr}^{-1}$ depending on modeling assumptions, \cite{Mandel:2021smh}). 

Note that the mass distribution of dynamically formed BBHs in GCs need not exhibit a sharp peak near $35M_{\odot}$ itself to produce the observed feature. Even a broad mass distribution that falls off sharply above $\sim40M_{\odot}$ can explain the inferred distribution. We argue that the $35M_{\odot}$ feature can arise in the inferred distributions due to alternative formation channels (that dominate at lower masses) becoming subdominant to GCs in the $30-40M_{\odot}$ range. In such a scenario, even a nearly flat mass distribution from the GC channel that falls off rapidly above $40M_{\odot}$, can give rise to a bump in the $30-40M_{\odot}$ range relative to the rest of the mass-spectrum. Our inferred mass-ratio and effective spin distributions are fully consistent with this hypothesis. In other words, our conclusion regarding the astrophysical origin of Subpopulation 2 and the $35M_{\odot}$ feature is robust against uncertainties in the shape of the mass-spectrum of BBHs formed in GCs, as long as substantial support is expected in the $30-40M_{\odot}$ mass-range.

\begin{figure}
\centering
\includegraphics[width = 0.48\textwidth]{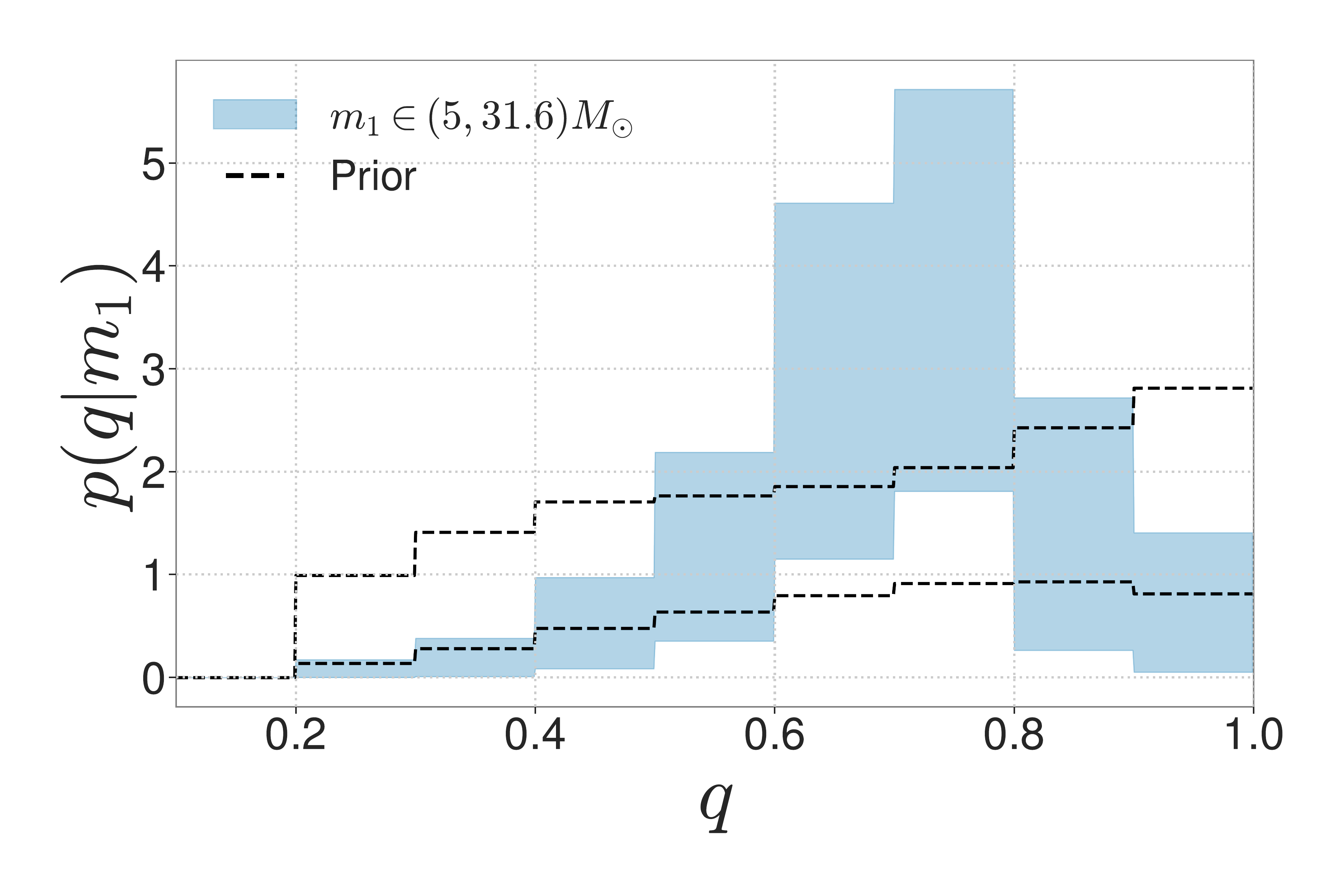}
\includegraphics[width = 0.48\textwidth]{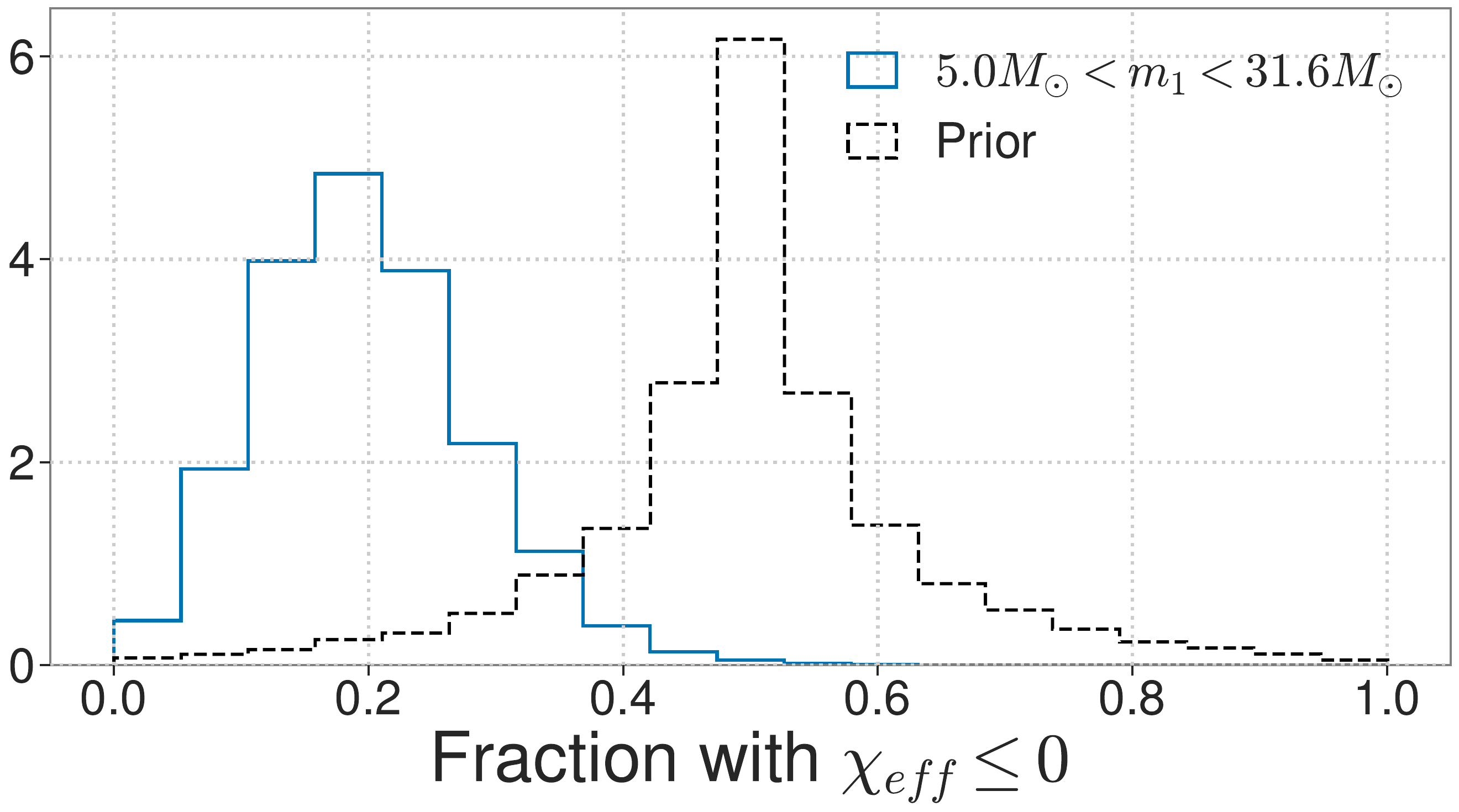}
\caption{The mass-ratio distribution~(top) and fraction of events with $\chi_{\rm eff}<0$~(bottom) for the low-mass~$(m_1<30M_{\odot})$ subpopulation.}
\label{fig:subpop1}
\end{figure}





Here we would like to note that in a recent work, \cite{Corelli:2026thw} have investigated the robustness of the $35M_{\odot}$ feature in the marginal mass-distribution by checking whether or not it is an artifact of Poisson uncertainty as opposed to a feature of the underlying distribution. By accounting for the variance due to finite catalogue size, \cite{Corelli:2026thw} have shown that the marginalized credible intervals on the mass distribution in the $30-40M_{\odot}$ region do not exhibit a statistically significant over-abundance for the (uncorrelated) population models used in their investigations. Our marginal $m_1$ distribution is consistent with the conclusion of \cite{Corelli:2026thw} and a statistically significant overabundance at $m_1\sim 35M_{\odot}$ only appears when we condition on $q>0.8,|\chi_{eff}|<0.1$. The impact of marginalizing over catalog variance on this result would necessitate repeating the analysis of \cite{Corelli:2026thw} with our fully correlated BGP model, which is left as a future exploration.

Next, we turn to Subpopulation 1, whose effective spin distribution peaking towards positive values implies preference towards aligned systems and non-negligible spin magnitudes. This is consistent with the predictions of isolated binary evolution \citep{Fuller:2019sxi,Mandel:2020lhv, Zevin:2022wrw, Briel:2022cfl, Bavera:2020inc, Bavera:2022zux}, even though, several mechanisms have been proposed for forming misaligned BBHs in isolation~\citep{Baibhav:2024rkn}. Previous investigations have shown that the $10M_{\odot}$ peak can arise from stable mass transfer in isolated stellar binaries and that it is robust against several uncertainties in theoretical modeling \citep{vanSon:2022ylf, vanSon:2022myr}. Furthermore, mass-ratio reversal during stable mass-transfer can lead to a preference towards unequal mass-ratios in the range $q\in (0.6,0.8)$ \citep{Neijssel_2019, vanSon:2021zpk, Broekgaarden:2022nst, Mould:2022xeu, Adamcewicz:2023szp, Biscoveanu:2025jpc}. The prominent peak near mass-ratios $0.6-0.8$ and a suppressed fraction of $\chi_{\rm eff}\leq 0 $ systems in our inferred distributions of Subpopulation 1~(as shown in Figure~\ref{fig:subpop1}) can therefore be indicative of substantial contributions from isolated binary evolution through stable mass transfer. However, our findings do not rule out the hypothesis that alternate pathways of isolated BBH formation contribute dominantly or substantially to Subpopulation 1.
\begin{figure}
\centering
\includegraphics[width = 0.48\textwidth]{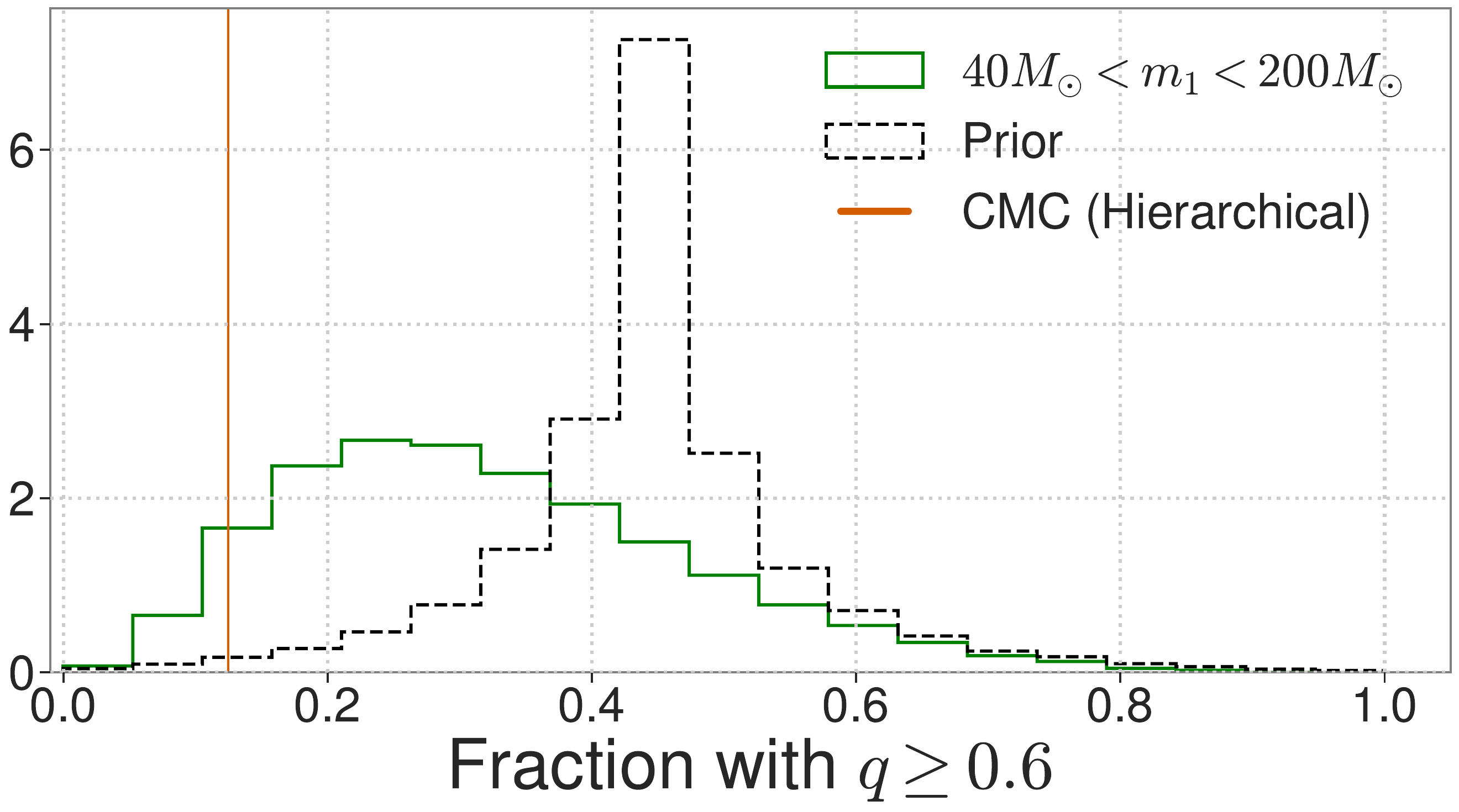}
\includegraphics[width = 0.48\textwidth]{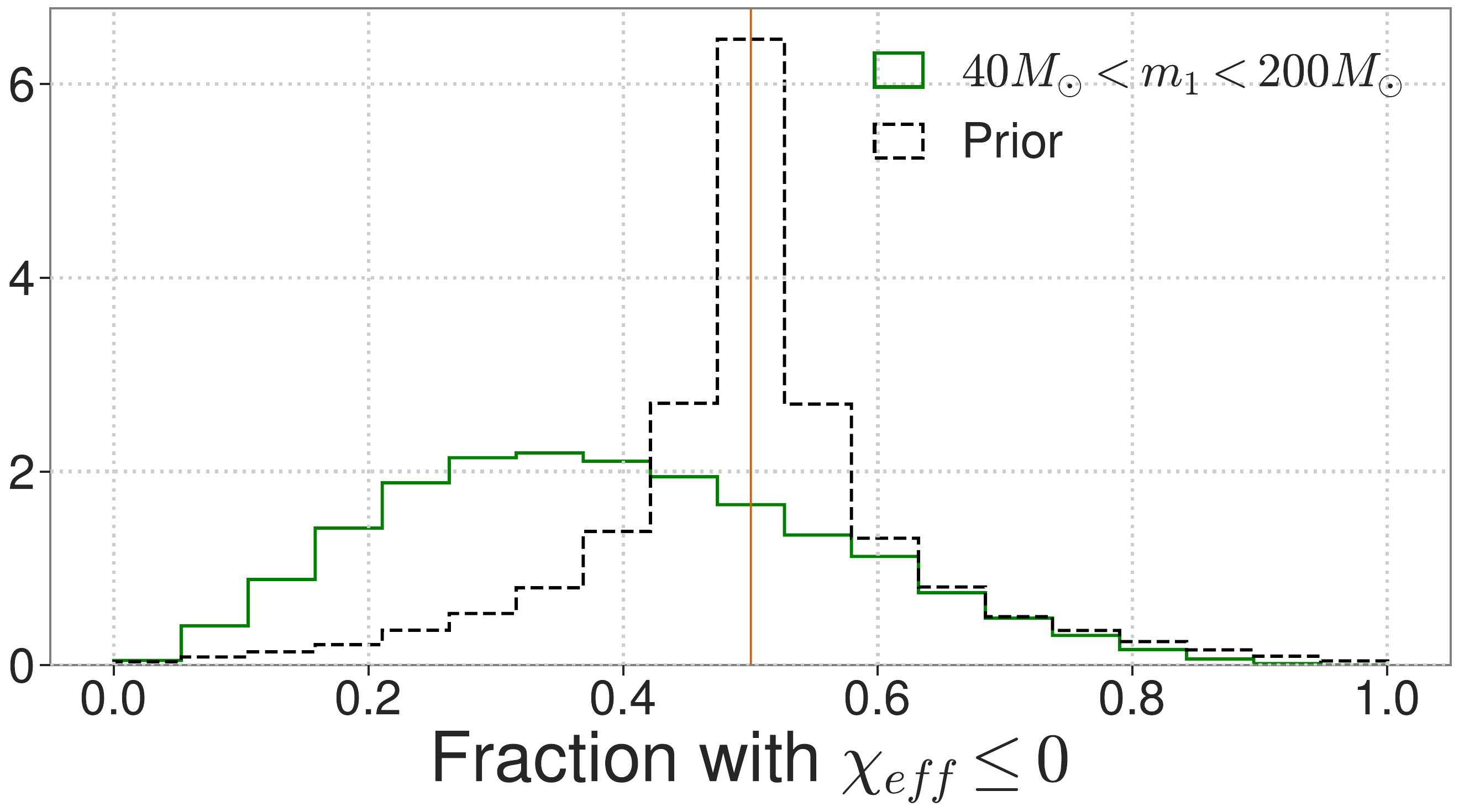}
\caption{Fraction of events with $q>0.6$~(top) and $\chi_{\rm eff}<0$~(bottom) for the high-mass~$(m_1>40M_{\odot})$ subpopulation. The orange line represents all hierarchical mergers from the simulations of \cite{Rodriguez:2019huv}, with the models corresponding to each BH birth spin combined with equal weightage.}
\label{fig:subpop3}
\end{figure}

Our findings for Subpopulation 3 are broadly consistent with previous studies that have explored trends in high-mass BBHs in GWTC-4 \citep{Banagiri2025, Wang:2025_PISN, Ray2025_PISN}. In particular, we find that for masses $\gtrsim45M_{\odot}$, there is a broadening of the $\chi_{\rm eff}$ distribution \citep{Antonini:2025zzw, Antonini:2025ilj, Antonini:2025c} and evidence of a flat mass-ratio distribution that is in alignment with the findings of \cite{Ray2025_PISN}. Figure~\ref{fig:subpop3}~(top and bottom) shows the fraction of events with $q>0.6$ and $\chi_{\rm eff}<0$ respectively, for Subpopulation 3. The inferred fractions indicate substantial contributions from alternative formation scenarios to hierarchical mergers. Potential scenarios include but are not limited to Super Eddington accretion during mass-transfer, chemically homogeneous evolution, population III stars, and accretion-driven growth after BH+Star collisions in dense environments~\citep{Briel:2022cfl, vanSon:2022ylf, deMink:2016vkw, Kinugawa:2021qee,  Kiroglu:2025vqy}. The hints of a peak in the $60-70M_{\odot}$ mass range for this subpopulation (and in the marginal $m_1$ distribution) might be indicative of the pile-up due to (P)PISNe, although, given the large error-bars, more data might be needed to confirm this conclusion. We note however that unlike Subpopulations 1 and 2, the high mass region is more strongly influenced by the prior in several parts of the $q$ and $\chi_{\rm eff}$ parameter spaces. Furthermore, flexible priors such as the one used in this work and the ones used by \cite{Ray2025_PISN} lead to inferred distributions that exhibit no clear evidence for or against a PISN mass gap.

Moving forward, the growing number of detections made by the LVK as part of their latest observing run will be vital to further tighten constraints on the nature of BBH subpopulations. To confirm a dynamical origin of the $35M_{\odot}$ peak it is necessary to simultaneously model the distribution of effective precessing spins, since the $\chi_{\rm eff}$ distributions can only inform on the fraction of alignment vs anti-alignment, and not on mis-alignment/precession, which is expected to be substantial in dynamically assembled binaries \citep{Gerosa:2021mno, Payne:2024ywe}. Similarly, data-driven reconstructions of additional substructure and focused investigations into their corresponding spin and mass-ratio subpopulations, if distinguishable, are also expected to become more feasible. A generic characterization of BBH subpopulations using higher-dimensional BGP models from future catalogs is left as an upcoming exploration. 

As noted before, a BGP-based framework, while preferable to search for previously unmodeled features and correlations in the BBH population, might not be optimal to completely isolate these clusters in the parameter space. Furthermore, the sampling techniques necessitated by such complex models do not yield Bayesian evidences, which makes it difficult to rigorously ascertain how much the data prefers the existence of these specific subpopulations. To isolate the contributions of specific subpopulations, compute merger rates that are directly comparable to theoretical simulations of specific formation channels, and perform model comparison, it is necessary to develop and constrain targeted parametrizations of the BBH population distribution guided by the trends uncovered through the non-parametric inference presented here. Such an investigation with GWTC-4 data is ongoing and will be communicated in the near future.
\section{Acknowledgements}
We are thankful to Mike Zevin and Tom Dent for insightful discussions and suggestions. A.R. was supported by the National Science Foundation~(NSF) award PHY-2207945. V.K. was supported by the D.I. Linzer Distinguished University Professorship fund. We are grateful for the computational resources provided by the LIGO laboratory and supported by National Science Foundation Grants PHY-0757058 and PHY-0823459. This material is based upon work supported by NSF’s LIGO Laboratory, which is a major facility fully funded by the National Science Foundation. This research has made use of data obtained from the gravitational Wave Open Science Center (gwosc.org), a service of LIGO Laboratory, the LIGO Scientific Collaboration, the Virgo Collaboration, and KAGRA. A.R. and V.K.  gratefully acknowledge the support of the NSF-Simons AI-Institute for the Sky (SkAI) via grants NSF AST-2421845 and Simons Foundation MPS-AI-00010513. This research was supported in part through the computational resources and staff contributions provided for the Quest high performance computing facility at Northwestern University which is jointly supported by the Office of the Provost, the Office for Research, and Northwestern University Information Technology.
\appendix

\section{Mass-ratio and effective spin distributions} \label{app: q-chi-corr}

\begin{figure*}
\centering

\includegraphics[width = 0.32\linewidth]{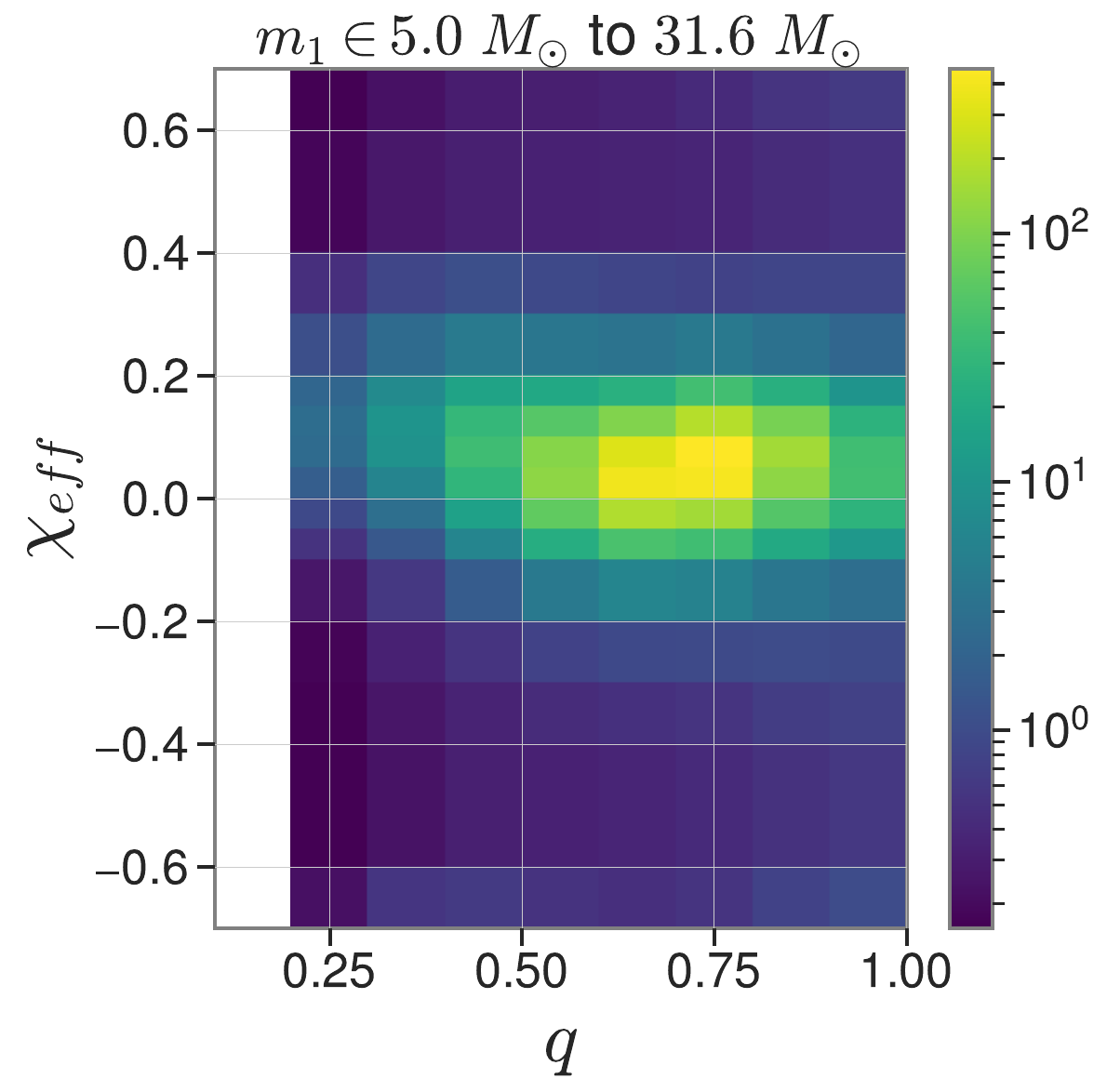}\vspace{5mm}
\includegraphics[width = 0.32\linewidth]{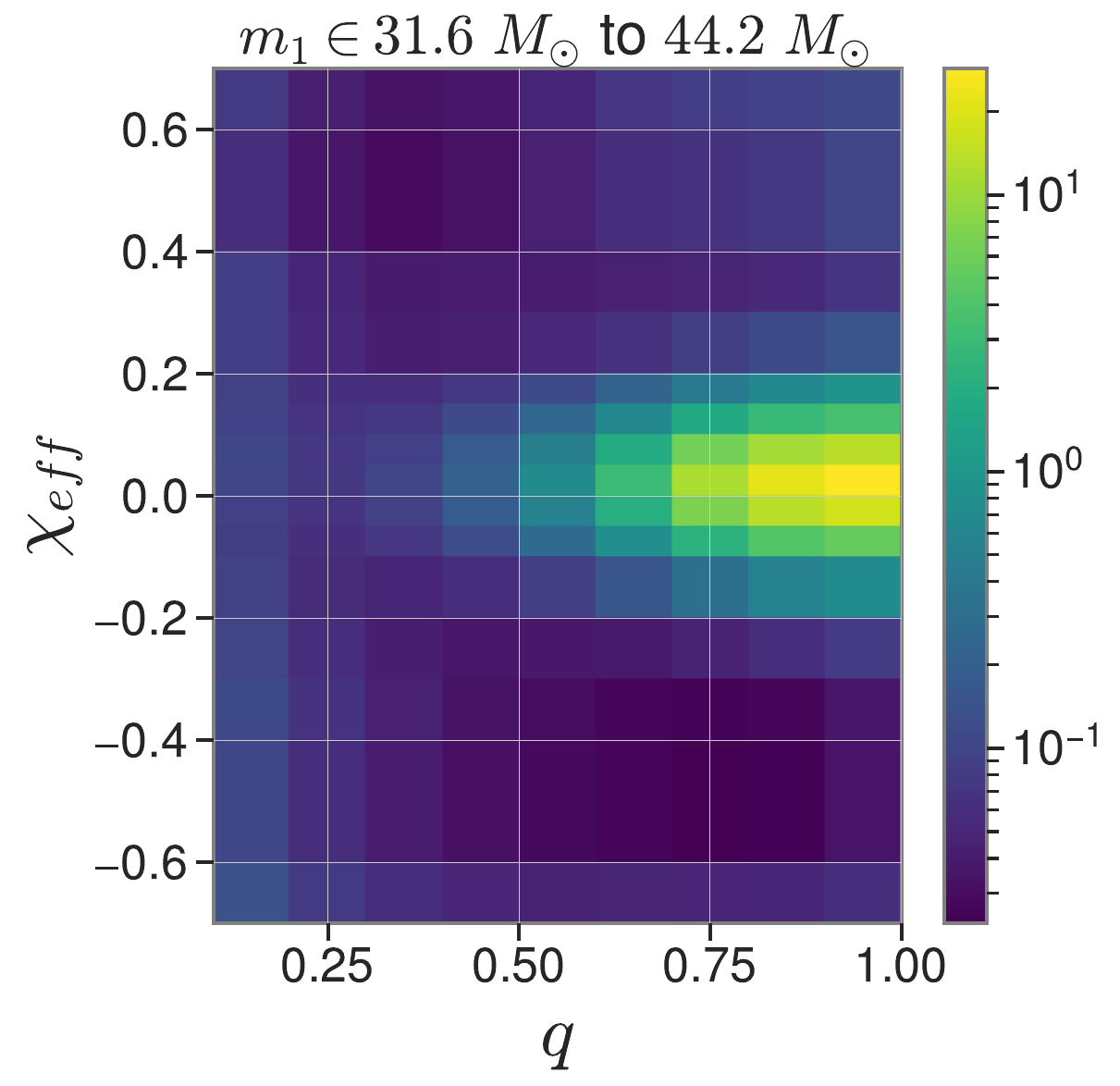}
\includegraphics[width = 0.32\linewidth]{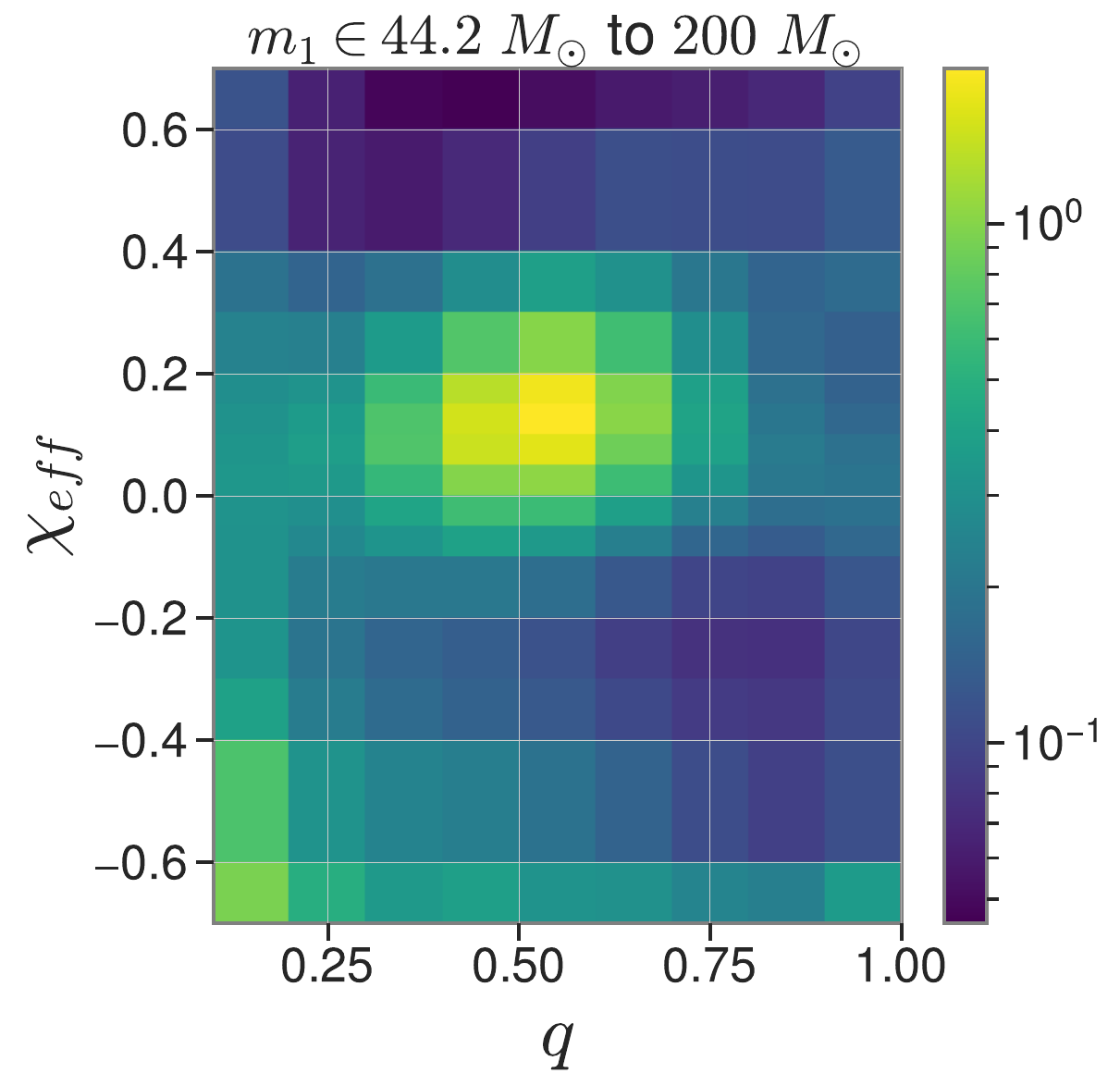}

\includegraphics[width = 0.32\linewidth]{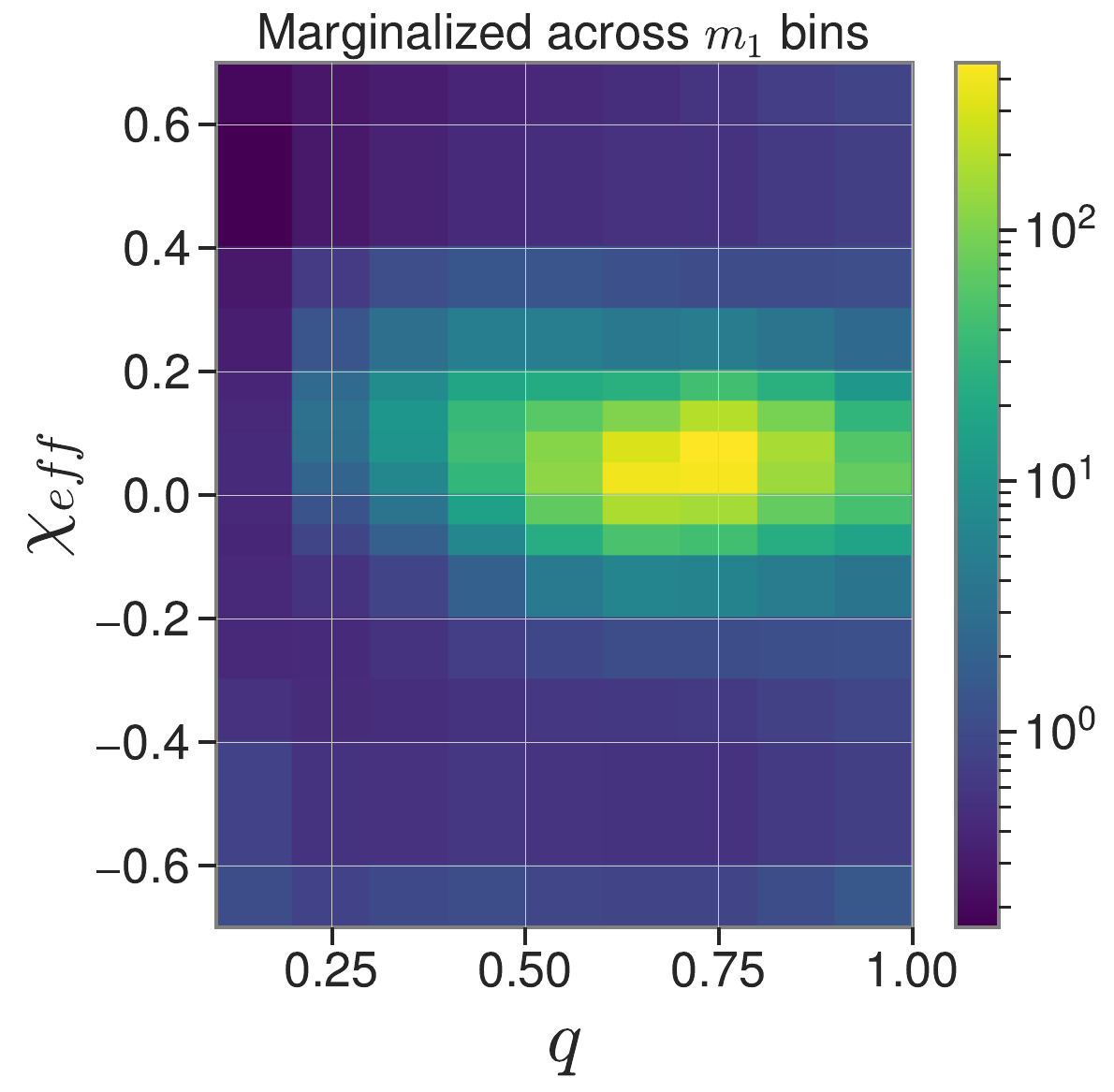}\vspace{5mm}
\includegraphics[width = 0.48\linewidth]{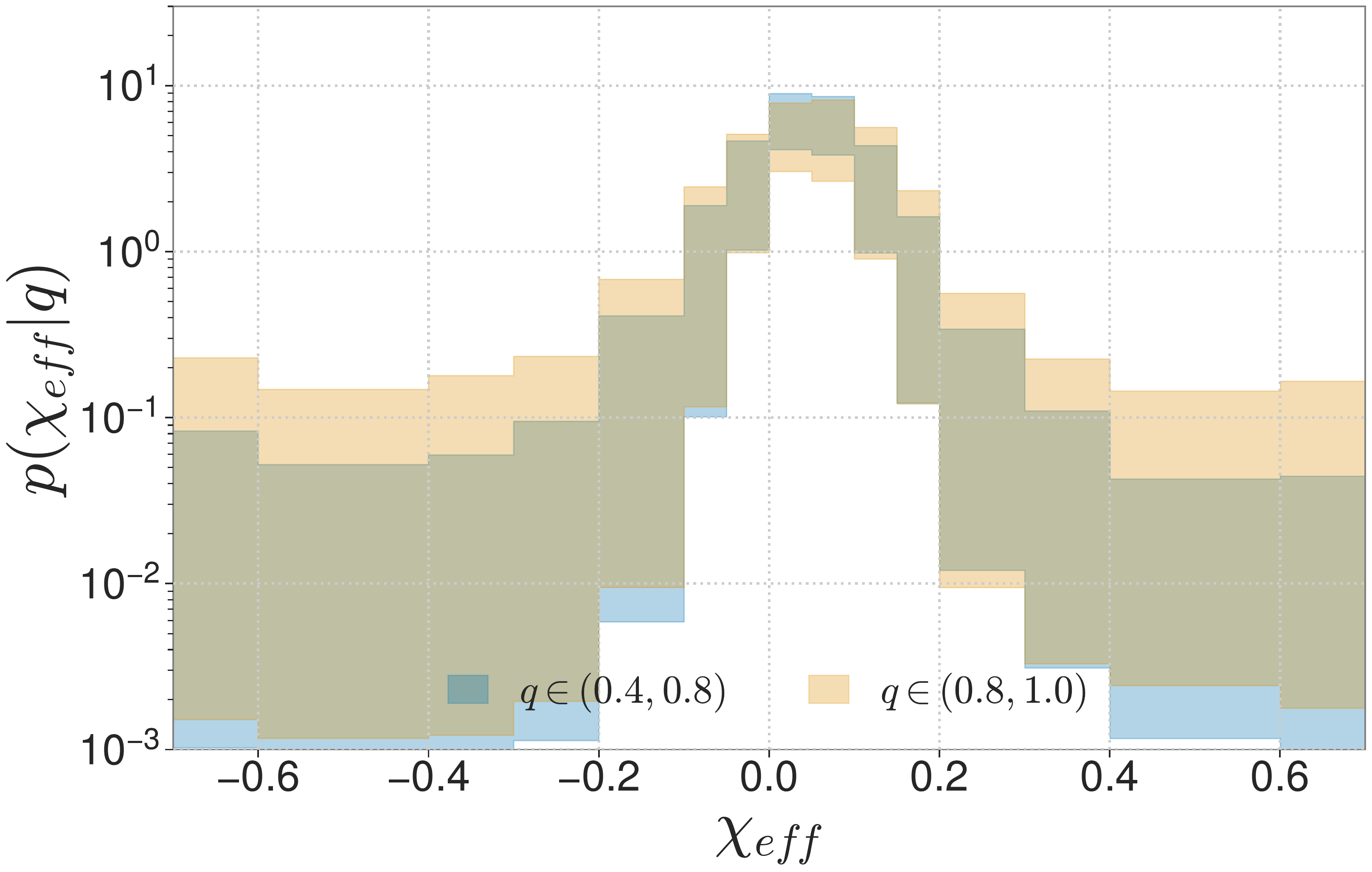}

\caption{
Inferred distributions highlighting the absence of support for a correlation between mass-ratio and effective spin. The top panels display 2d slices of the median merger rate density distribution obtained from GWTC-4, for the three primary mass ranges as in Figure \ref{fig:PCC_case_9}. The bottom left panel shows the same but for the mass-marginalized case. The bottom right panel represents the $\chi_{\rm eff}$ distribution for different ranges of mass-ratio.
    }
\label{fig:q-chi-corr}
\end{figure*}

We display the two-dimensional median merger rate density distributions in Figure \ref{fig:q-chi-corr}. They corroborate the conclusions from the Pearson posteriors in Figure \ref{fig:PCC_case_9} of an absence of $(q, \chi_{\rm eff})$ correlations in any of the primary mass ranges of interest.
\section{Variation of subpopulation mass-boundaries}
\label{sec:app-mass-boundaries}

\begin{table}[h]
\centering
\begin{tabular}{cccc}
\hline
~ & Subpopulation 1 & Subpopulation 2 & Subpopulation 3 \\
\hline
case A & $m_1\in( 5M_{\odot}, 37.4M_{\odot})$ & $m_1\in( 37.4M_{\odot}, 52.2M_{\odot})$ & $m_1\in( 52.2M_{\odot}, 200M_{\odot})$ \\
case B & $m_1\in( 5M_{\odot}, 22.6M_{\odot})$ & $m_1\in( 22.6M_{\odot}, 44.2M_{\odot})$ & $m_1\in( 44.2M_{\odot}, 200M_{\odot})$ \\
case C & $m_1\in( 5M_{\odot}, 26.7M_{\odot})$ & $m_1\in(26.7 M_{\odot}, 52.2M_{\odot})$ & $m_1\in( 52.2M_{\odot}, 200M_{\odot})$ \\
\hline
\end{tabular}
\caption{Alternate primary mass boundaries used in Figure \ref{fig:subpops-diffcase} and Figure \ref{fig:JSD-diffcase} to investigate the sensitivity of the subpopulation properties on chosen mass ranges.}
\label{tab:diffcase}
\end{table}

In this section, we investigate the sensitivity of our conclusions on the chosen mass boundaries that separate the three subpopulations. In addition to the default case used for the results presented in the main text, we explore the conditional $q$ and $\chi_{eff}$ distributions for three alternative sets of mass boundaries separating the three subpopulations, which are listed in Table~\ref{tab:diffcase}.
\begin{figure*}
    \centering
    \includegraphics[width=0.48\textwidth]{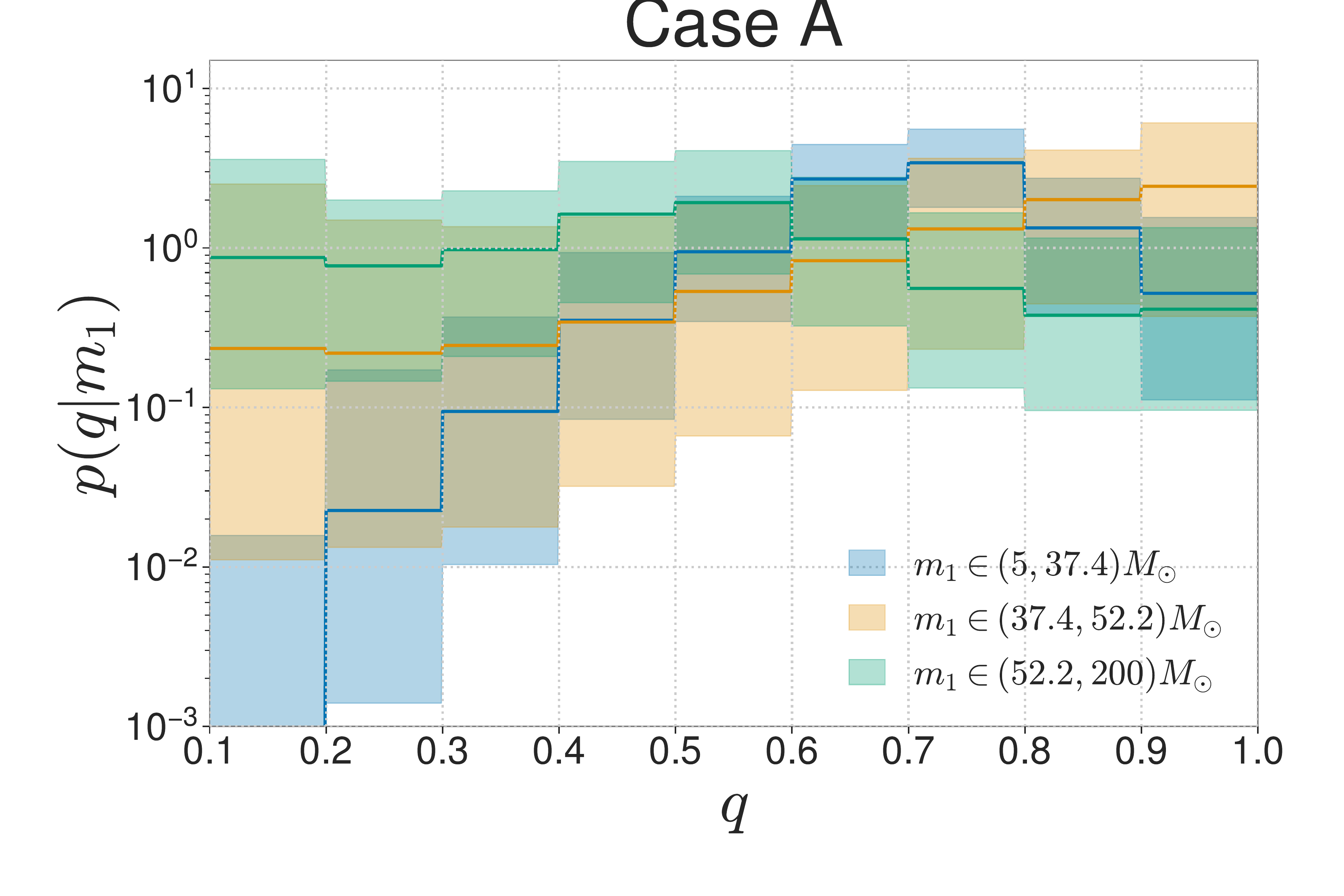}
    \includegraphics[width=0.48\textwidth]{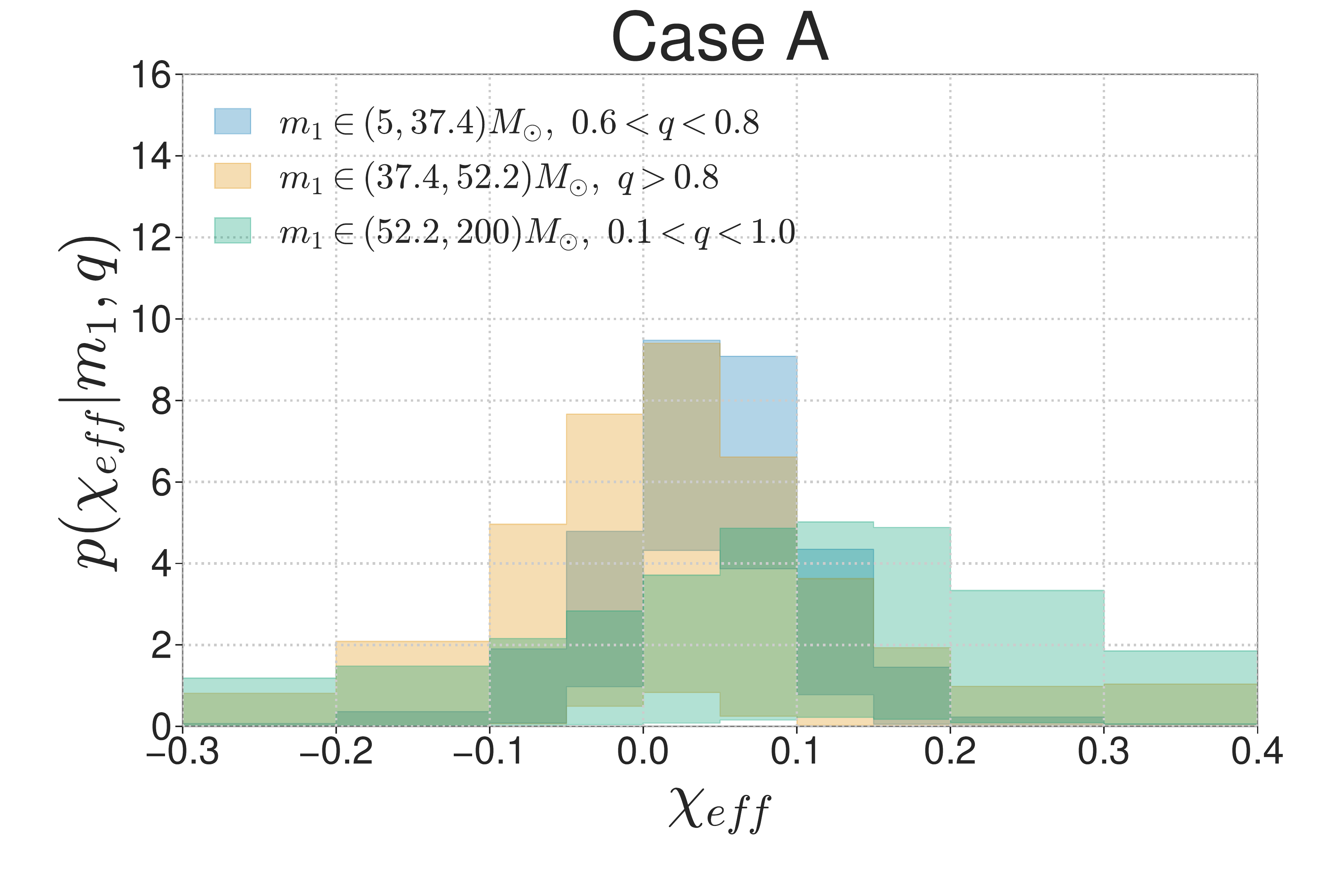}
    \includegraphics[width=0.48\textwidth]{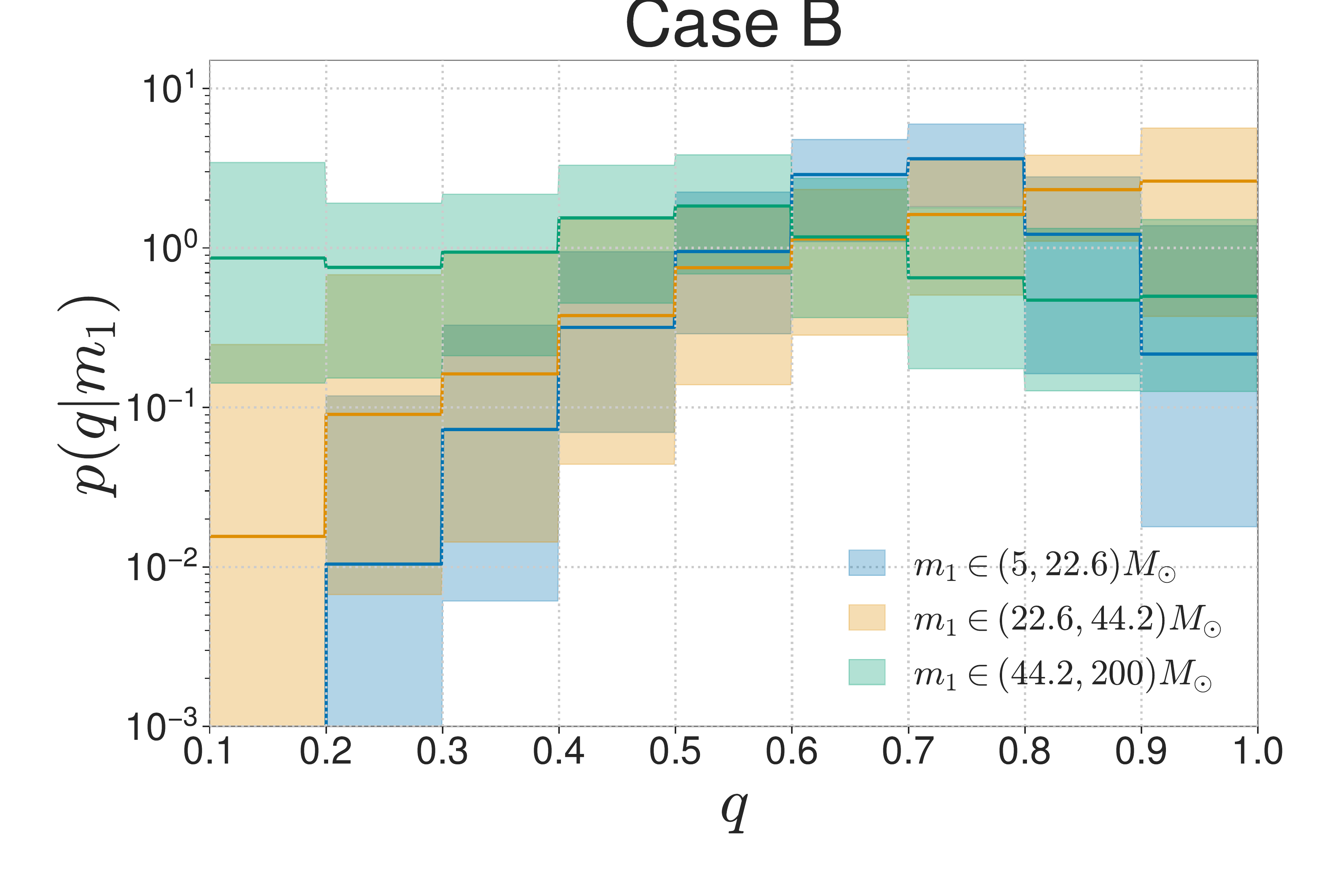}
    \includegraphics[width=0.48\textwidth]{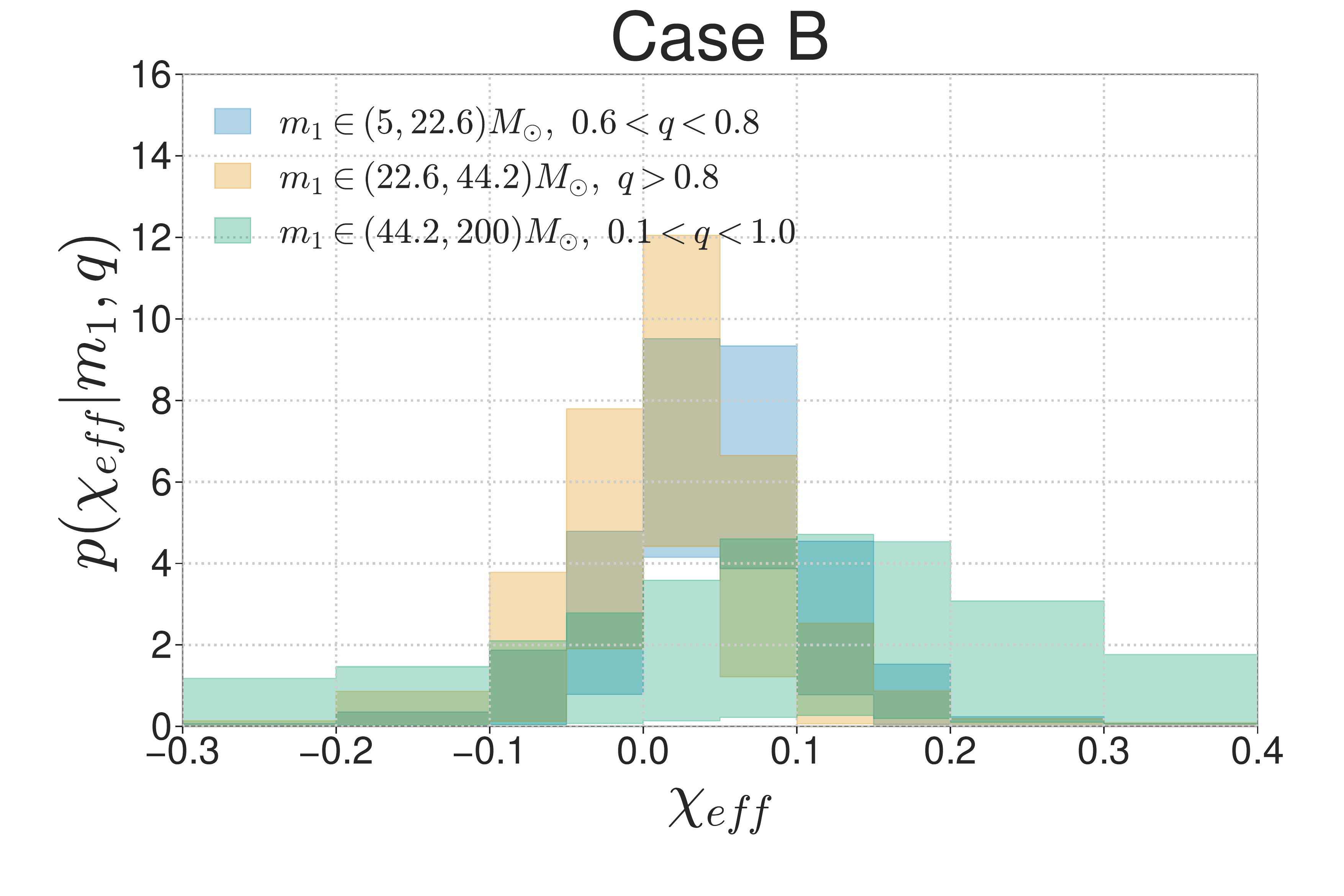}
    \includegraphics[width=0.48\textwidth]{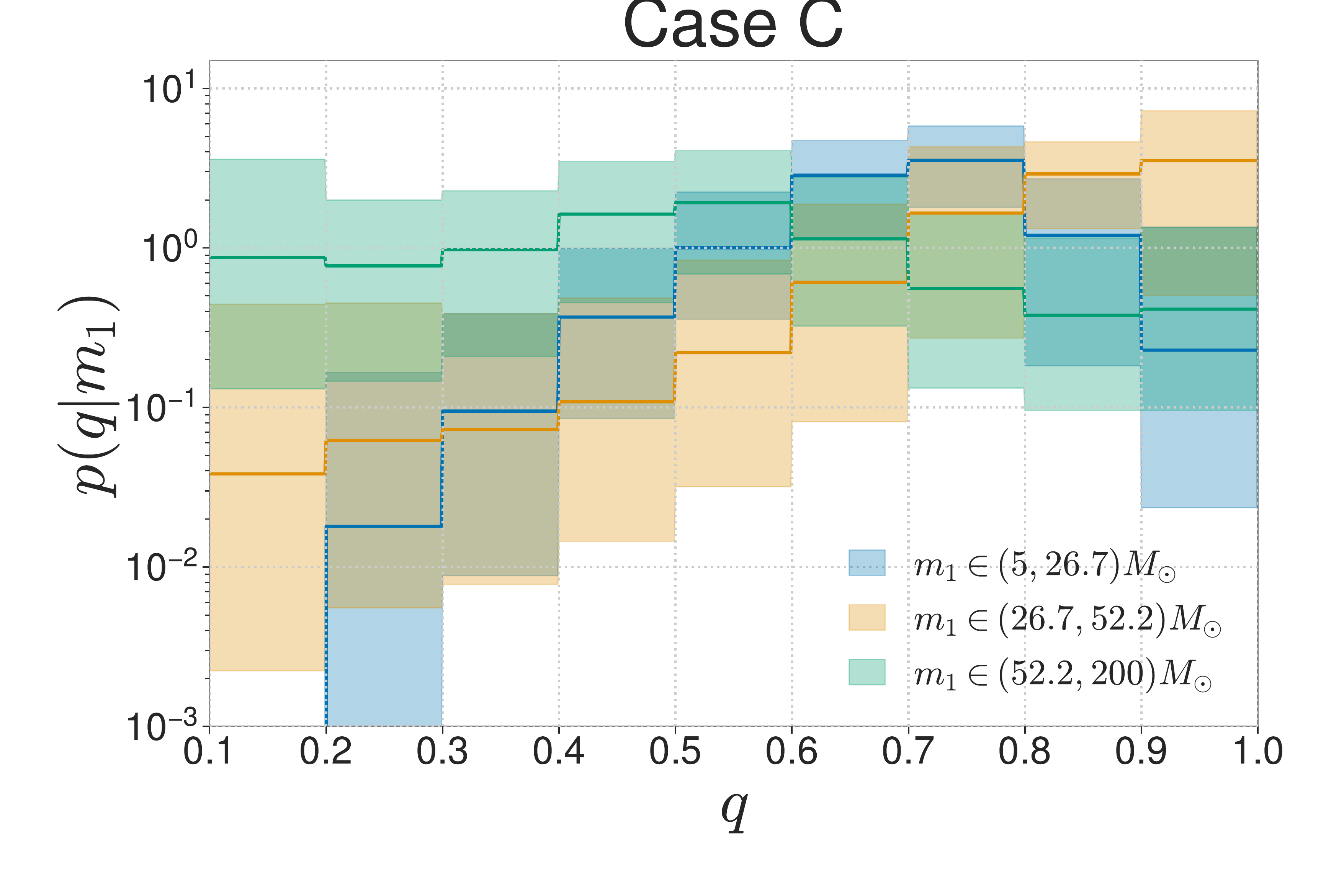}
    \includegraphics[width=0.48\textwidth]{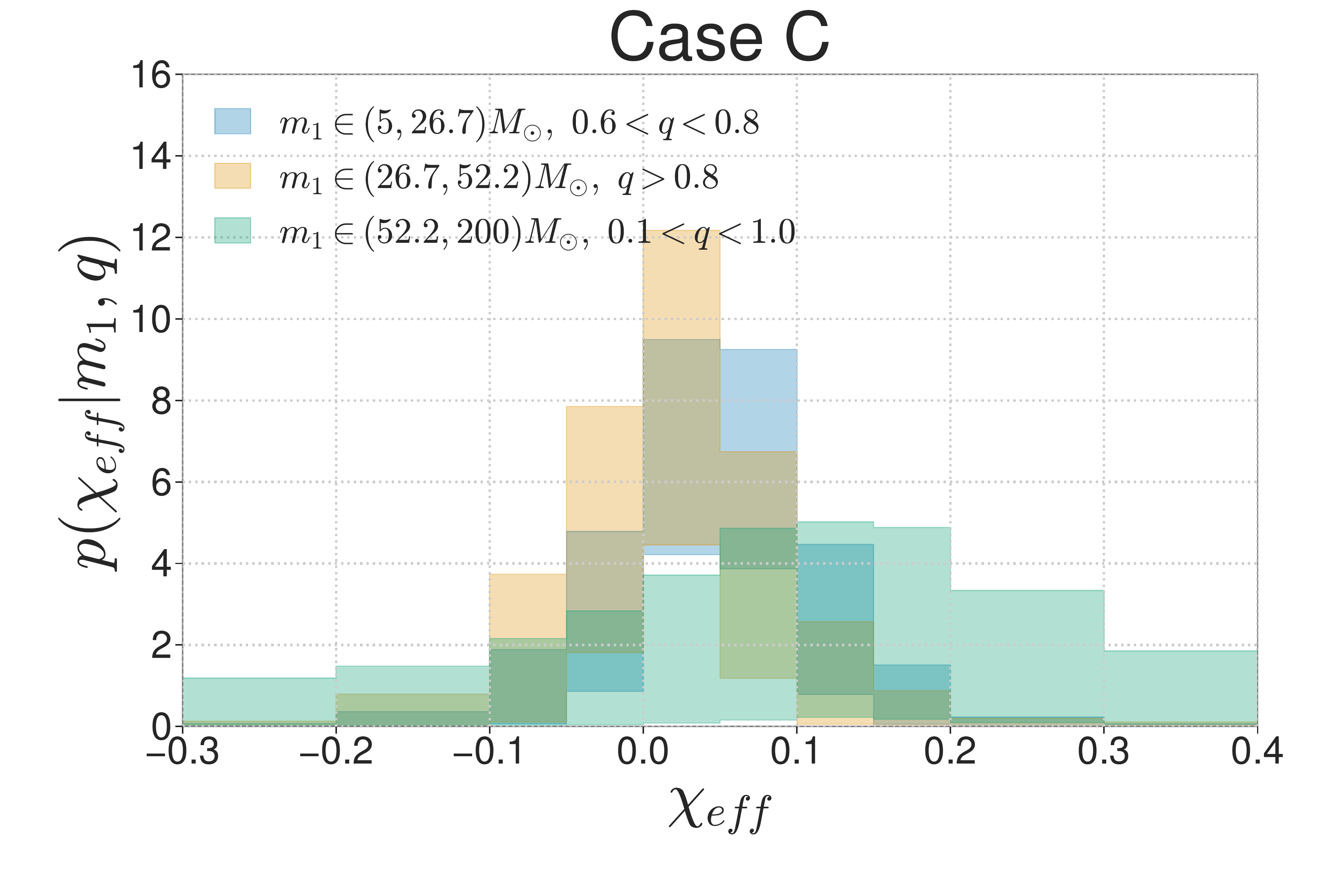}
    \caption{\label{fig:subpops-diffcase} The mass-ratio~(\textit{ledt}) and effective spin~(\textit{right}) subpopulations for the three subpopulations corresponding to the alternative choices of mass-boundaries, namely, cases A~(\textit{top}), B~(\textit{center}) and C~(\textit{bottom}) of Table~\ref{tab:diffcase}, respectively.}
\end{figure*}
\begin{figure*}
\centering
    \includegraphics[width=0.48\textwidth]{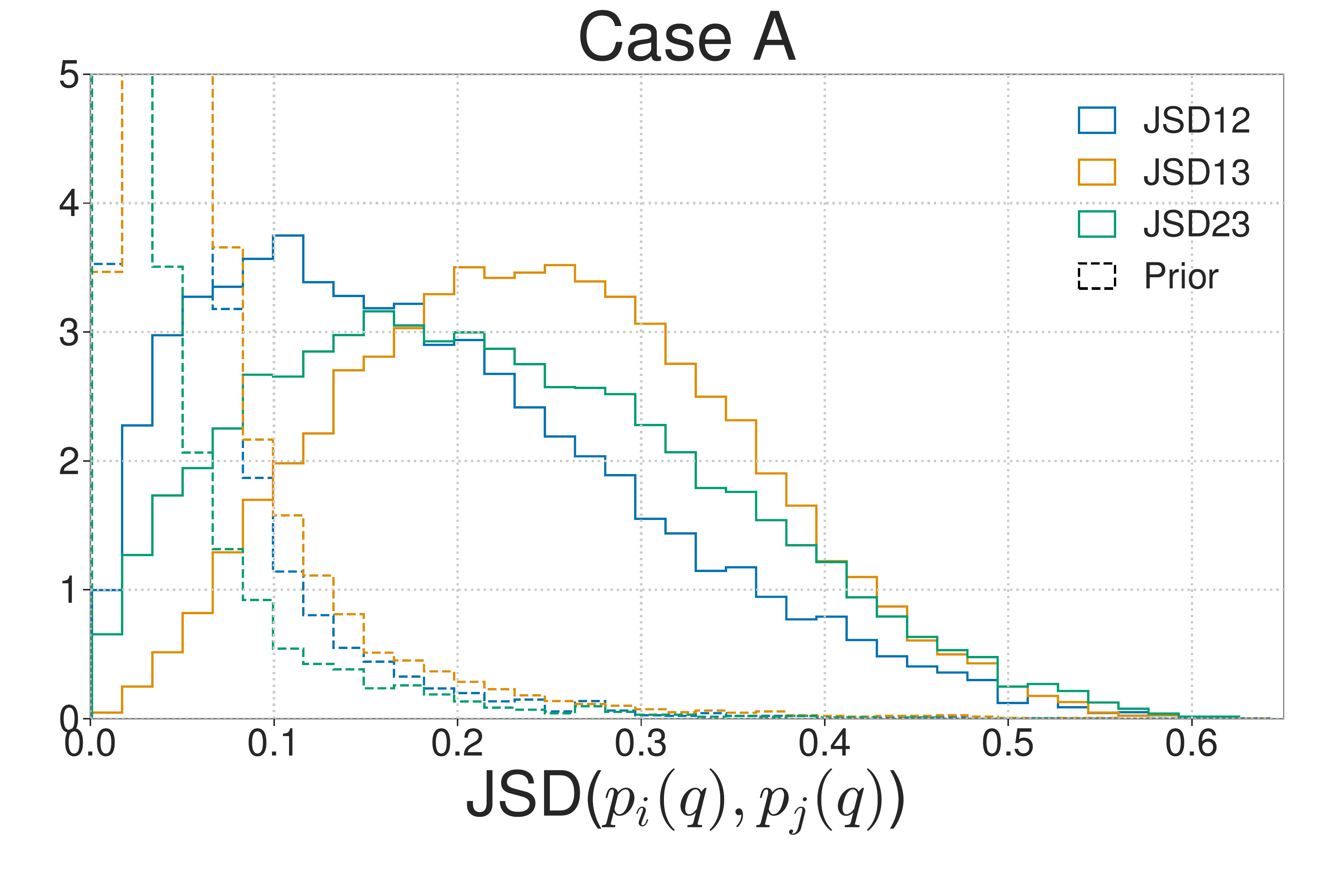}
    \includegraphics[width=0.48\textwidth]{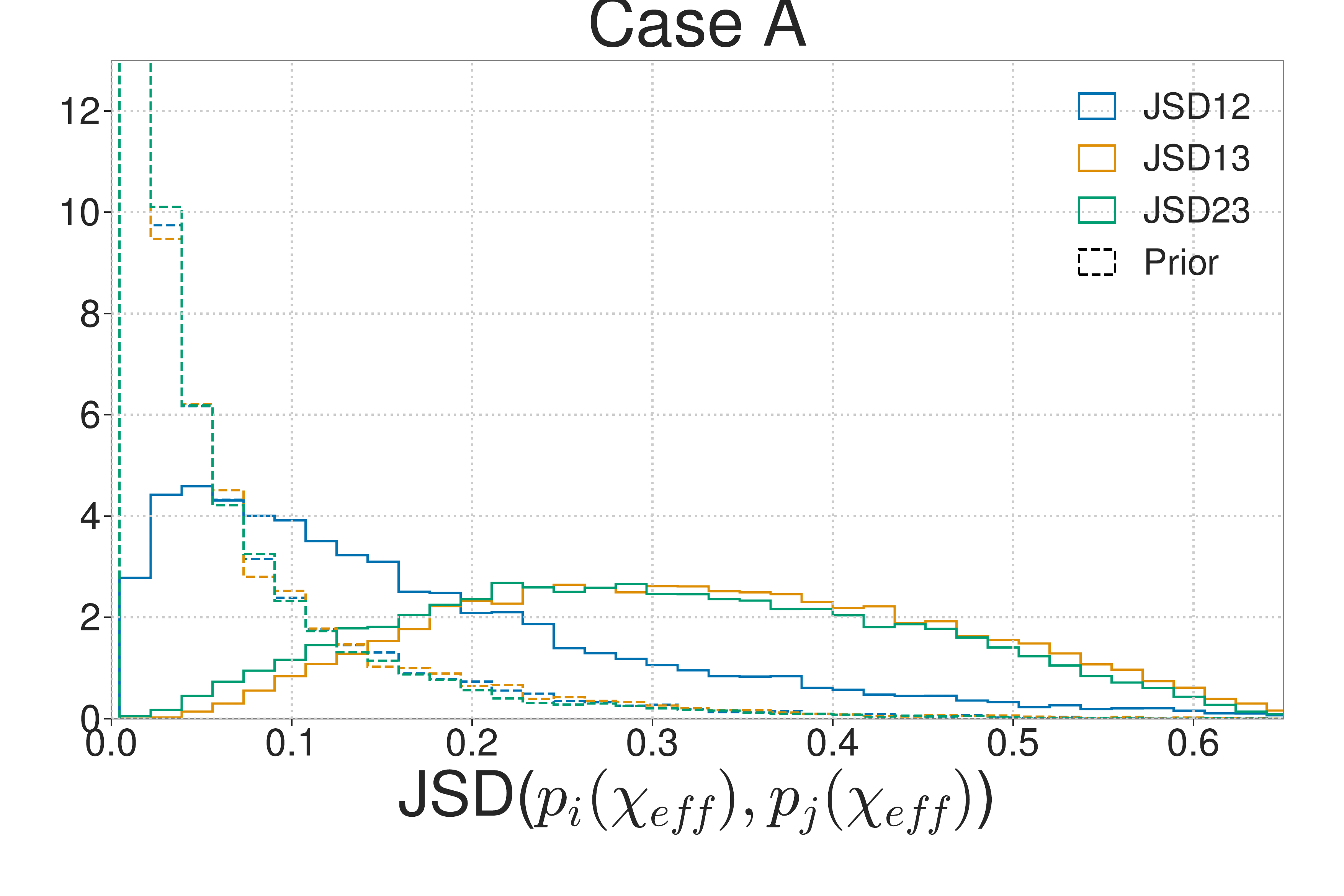}
    \includegraphics[width=0.48\textwidth]{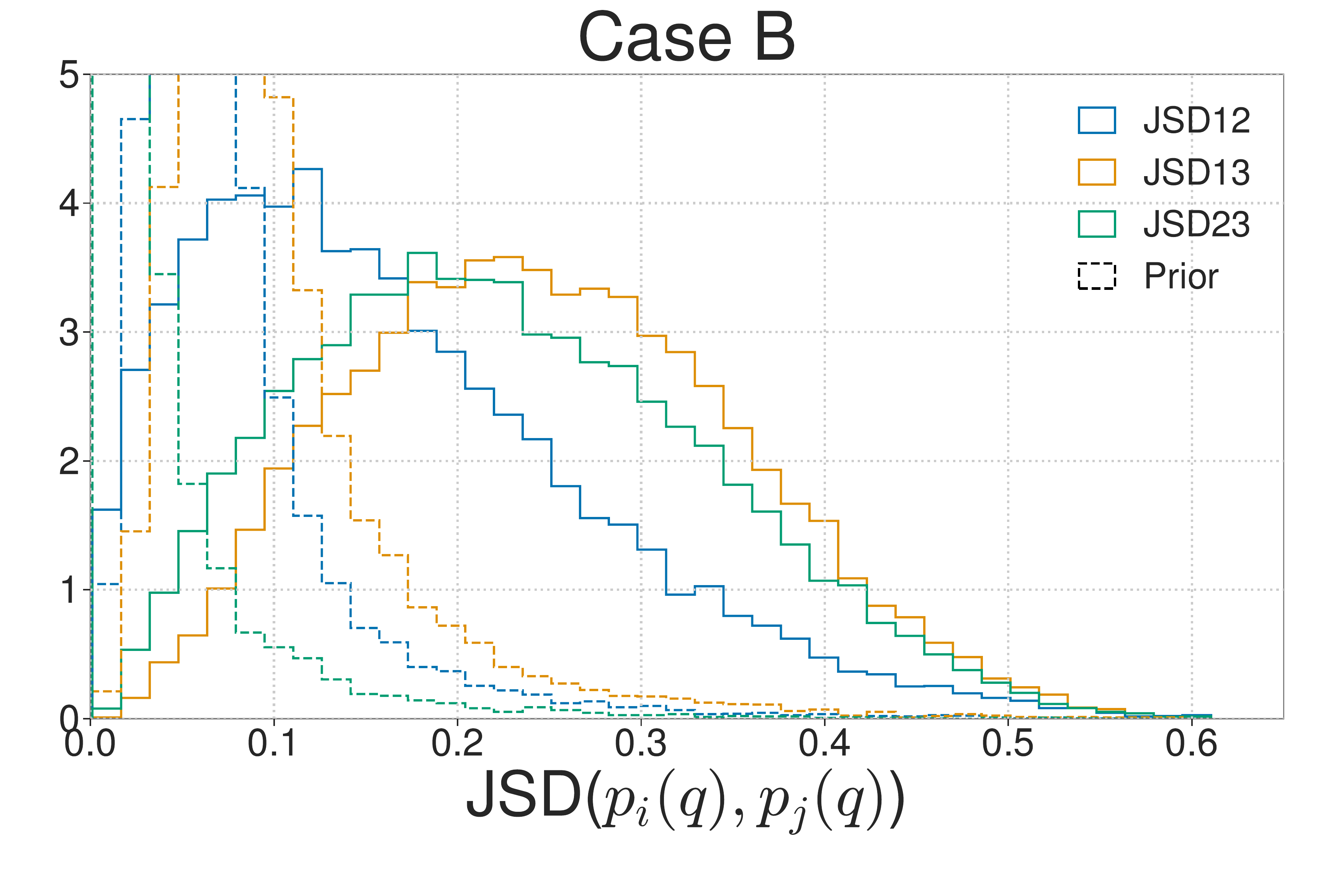}
    \includegraphics[width=0.48\textwidth]{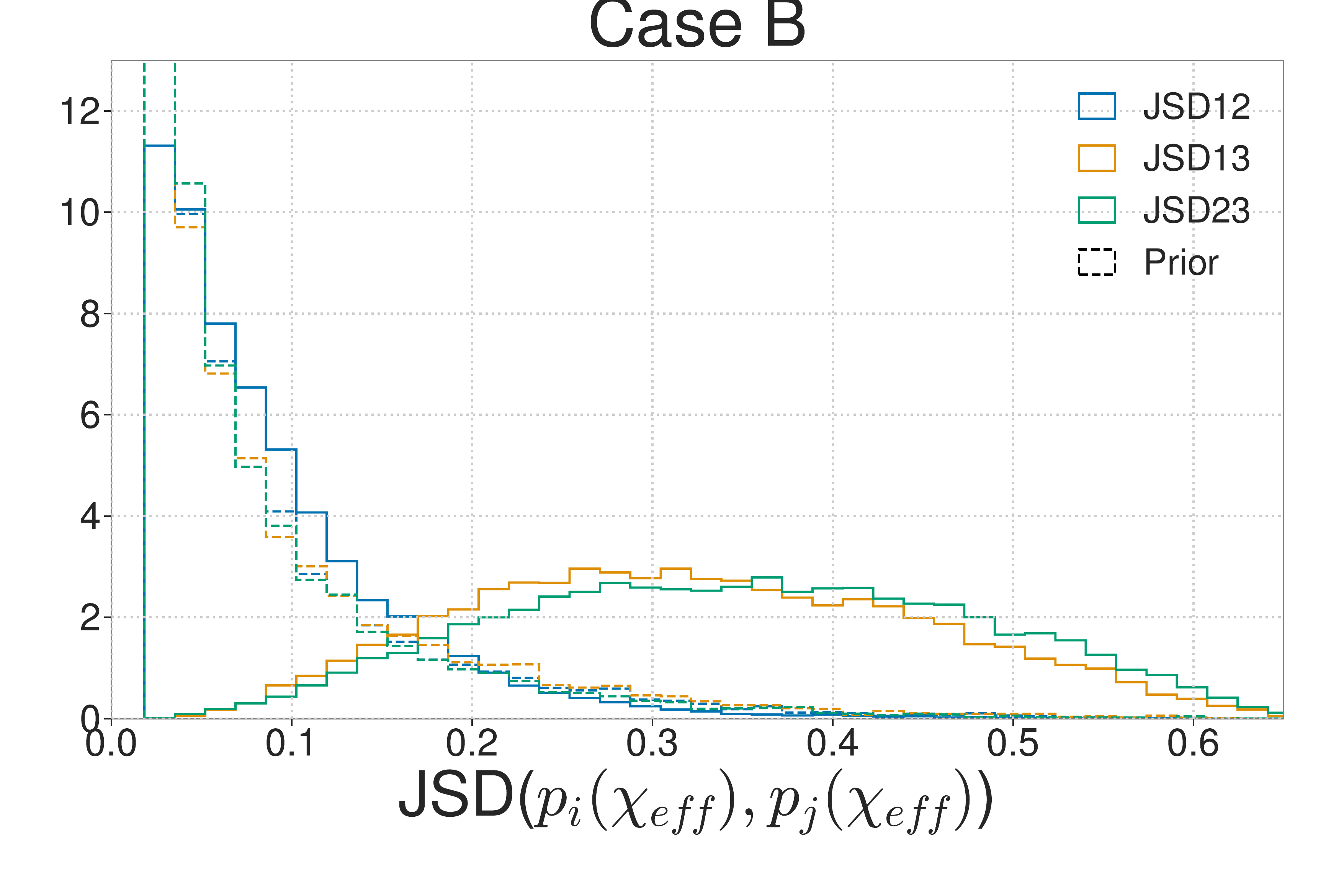}
    \includegraphics[width=0.48\textwidth]{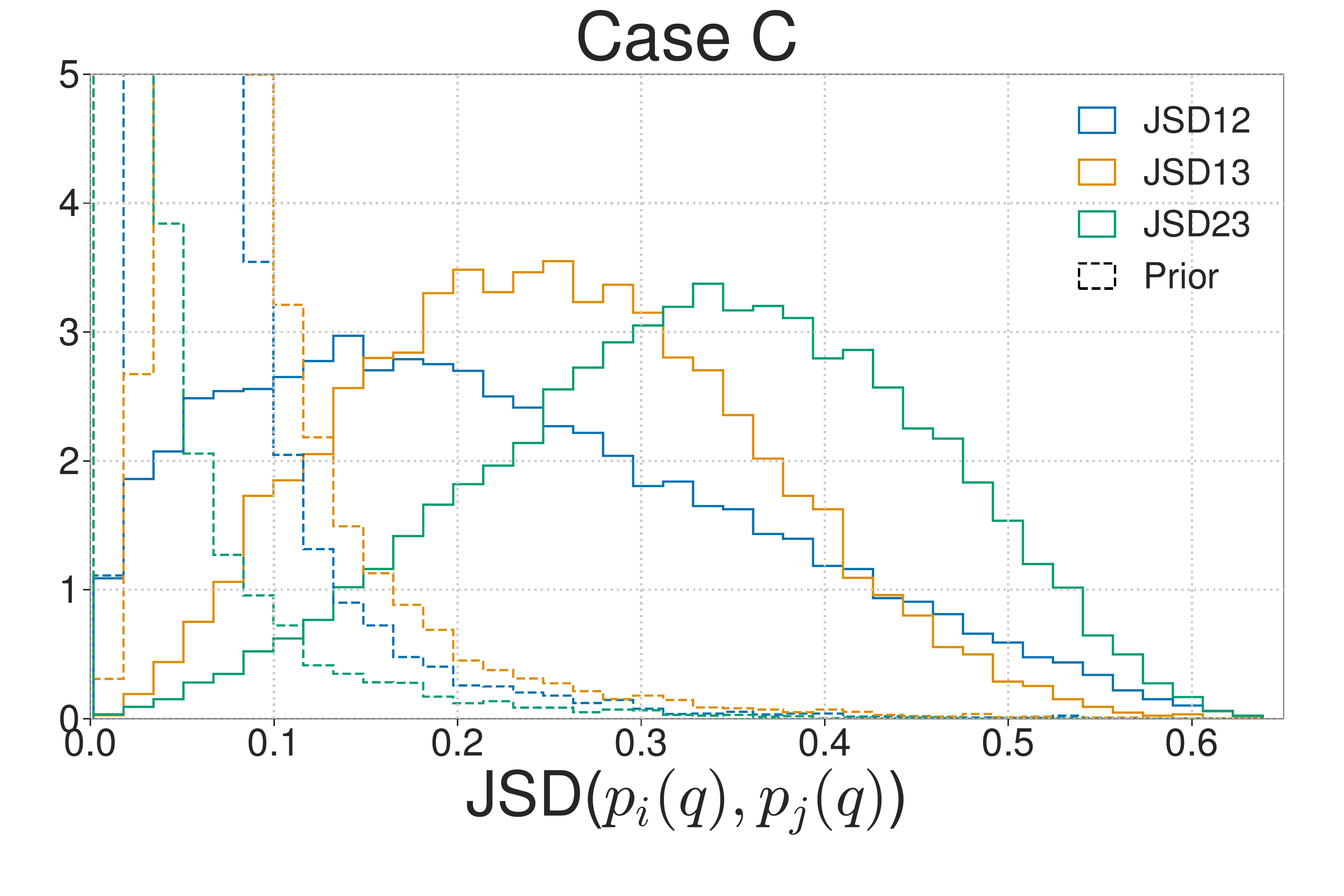}
    \includegraphics[width=0.48\textwidth]{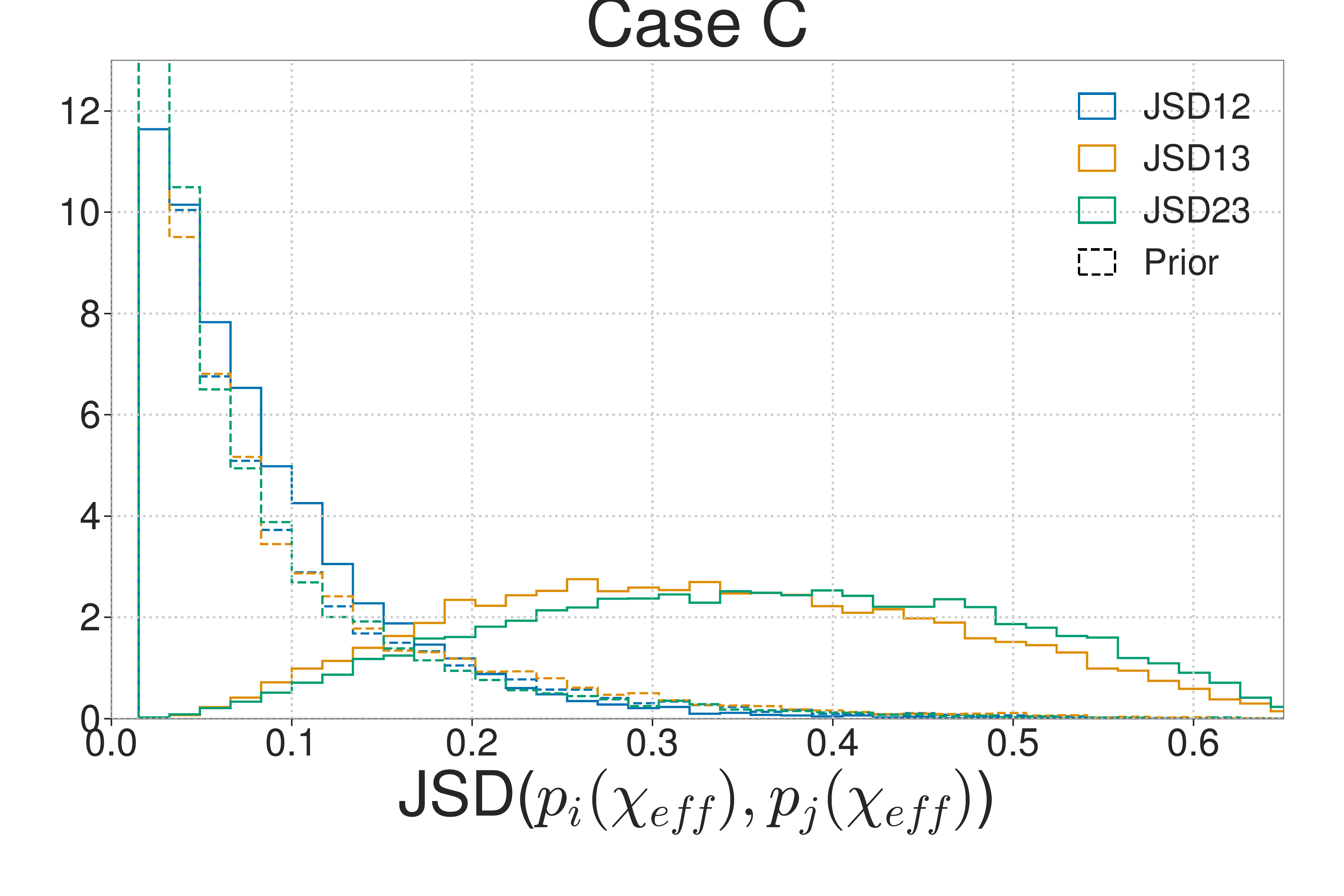}
    \caption{\label{fig:JSD-diffcase} Same as Figure~\ref{fig:JSD} but for the alternative mass boundaries, namely cases A~(\textit{top}), B~(\textit{center}) and C~(\textit{bottom}) of Table~\ref{tab:diffcase}, respectively.}
\end{figure*}

In Figure~\ref{fig:subpops-diffcase}, we show the conditional $q$ and $\chi_{eff}$ for each alternative case and show the corresponding JS divergences in Figure~\ref{fig:JSD-diffcase}. It can be seen that the existence of three distinct $q$ and $\chi_{eff}$ distributions across different mass ranges and their corresponding trends remain the same across reasonable variations of the mass-boundaries chosen to separate the three subpopulations.

\section{Variation of binning choice} \label{app: Case_28_bins}

\begin{figure*}
\centering

\includegraphics[width = 0.48\linewidth]{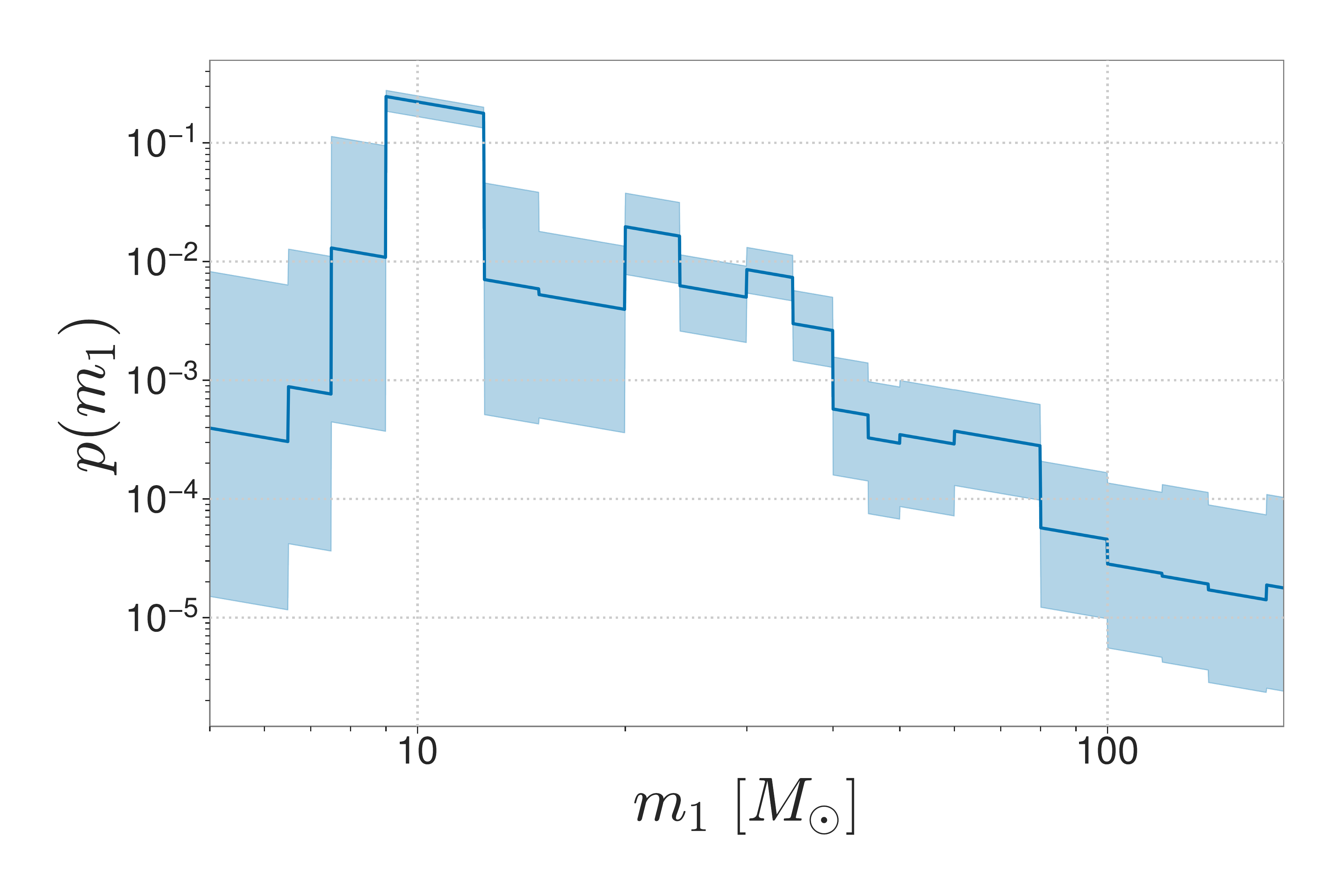}
\includegraphics[width = 0.48\linewidth]{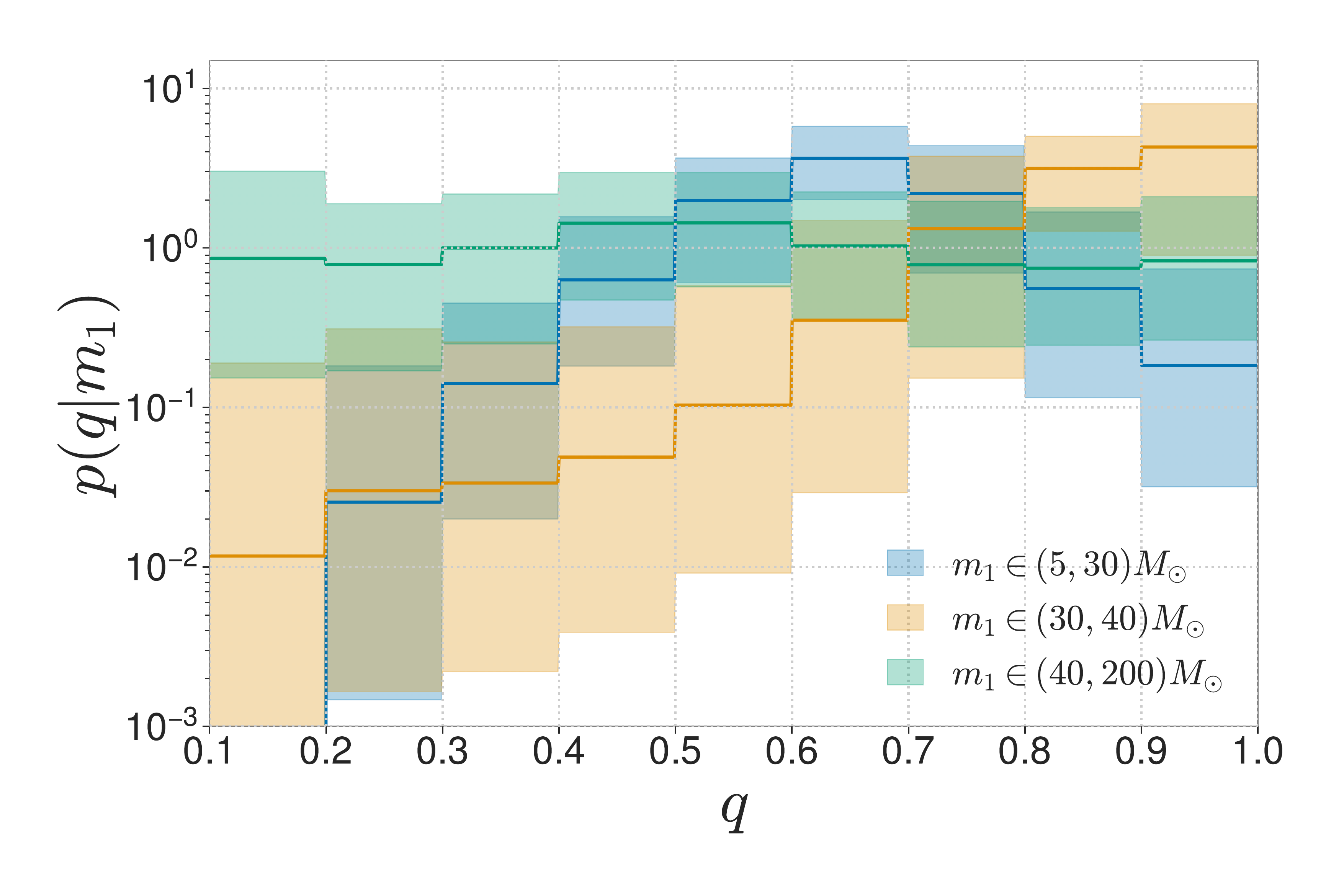}

\includegraphics[width = 0.48\linewidth]{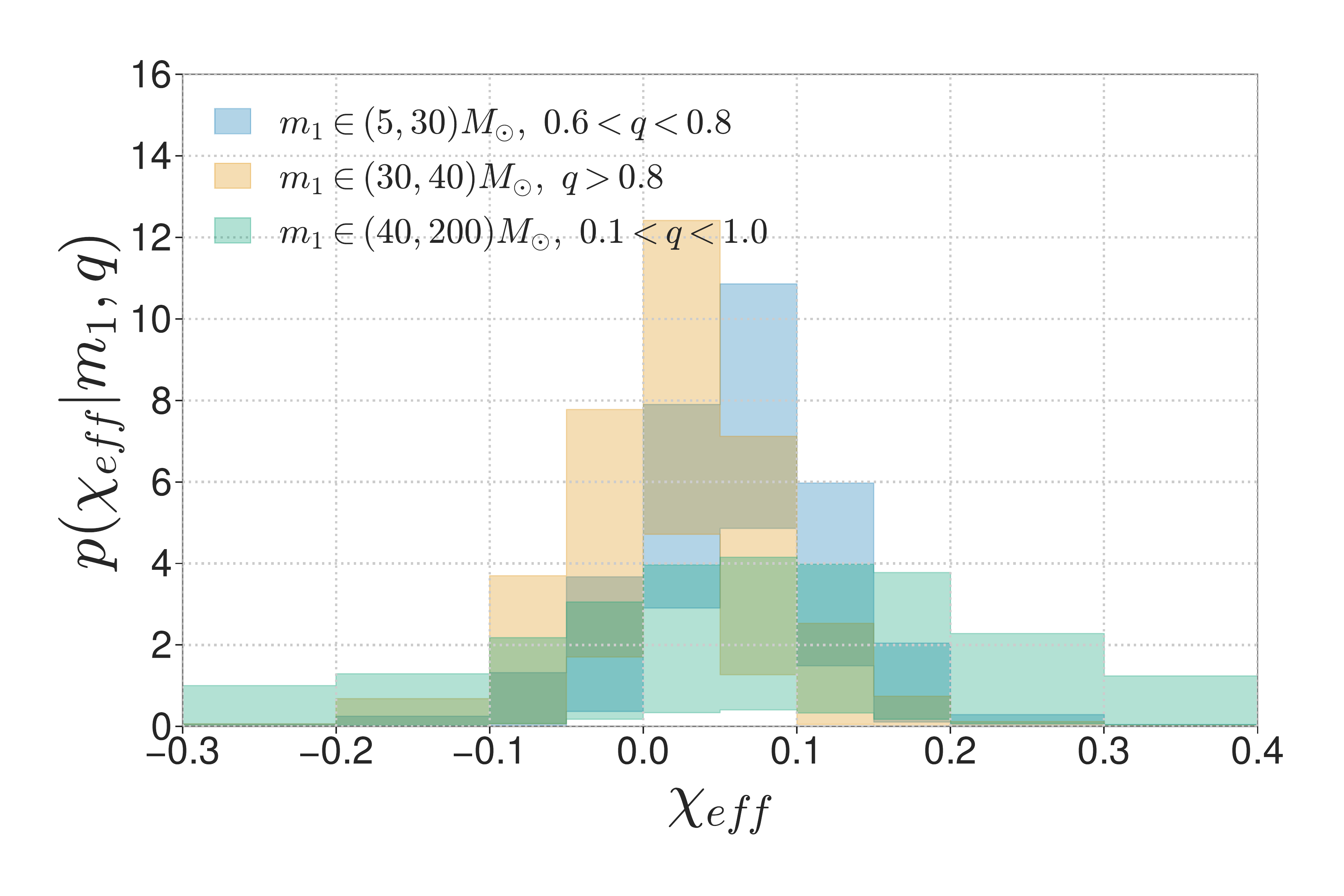}
\includegraphics[width = 0.48\linewidth]{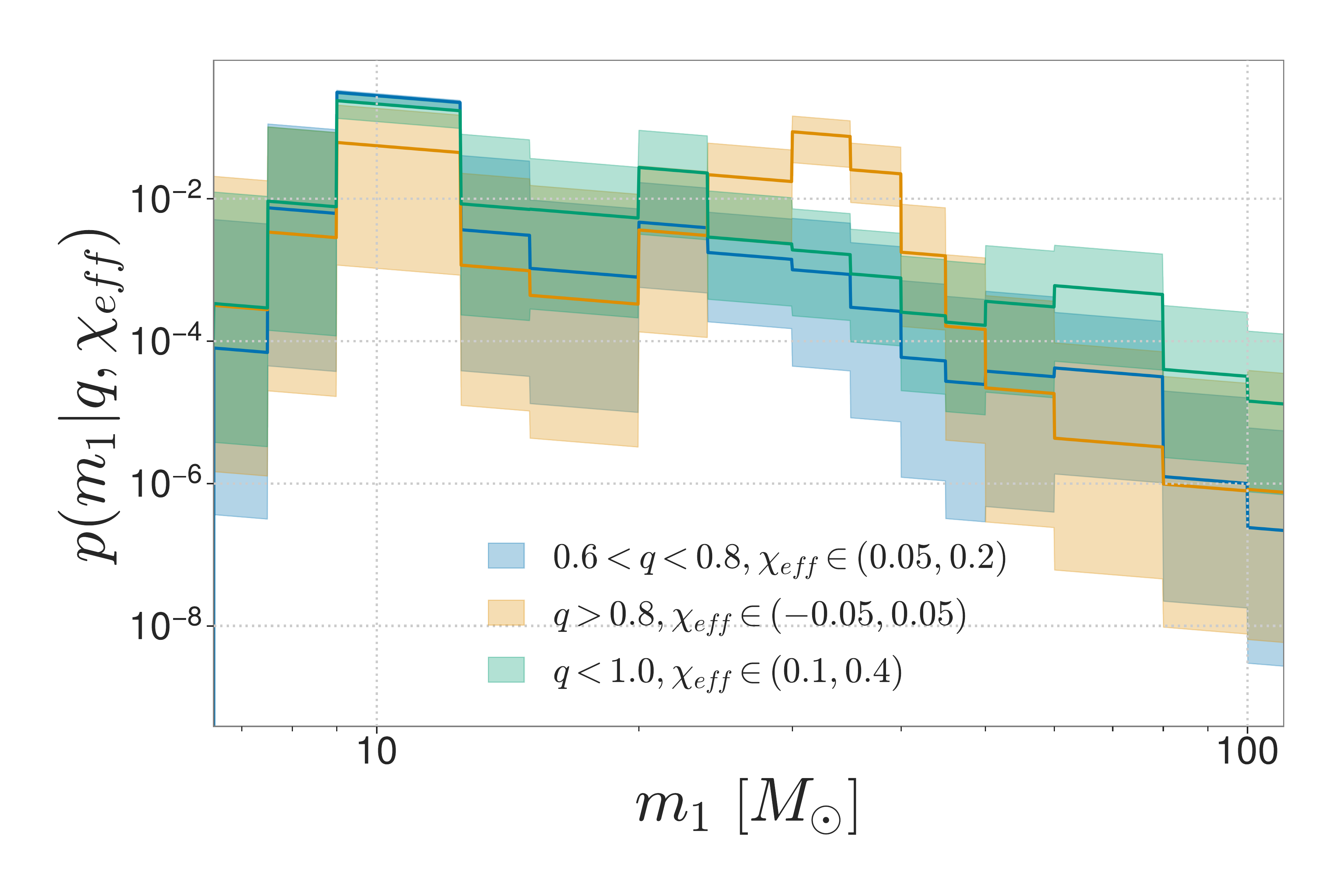}

\caption{
Marginal and conditional distributions of primary mass, mass-ratio and effective spin inferred from GWTC-4 for an alternate bin choice with non-uniform primary mass bins.
}
\label{fig: Subpop_case_28}
\end{figure*}

We show in Figure \ref{fig: Subpop_case_28} the inference trends we find for GWTC-4 for a different bin choice with non-uniform primary mass bins (see Table \ref{tab: Bin choices non uniform}). We recover results consistent with those presented in Sec. \ref{subsec: Subpopulations} and Sec. \ref{sec:astro imps}, thereby confirming that the subpopulations are robust against bin variations.

\begin{table}[H] 
\begin{adjustwidth}{0cm}{}
\adjustbox{margin=1cm 0 0 0}{%
\begin{tabular}{cc}
\hline
\hline
Parameter & Bin edges \\
\hline
$m_1 (M_{\odot})$           & \hspace{1em} 5, 6.5, 7.5, 9, 12.5, 15, 20, 24, 30, 35, 40, 45, 50, 60, 80, 100, 120, 140, 170, 200                                                                              \\

\hline
$\chi_{\rm eff}$ &   -0.7, -0.6 , -0.4 , -0.3 , -0.2 ,    -0.1 , -0.05, 0.0, 0.05,  0.1 ,  0.15, 0.2 ,   0.3 ,  0.4 ,  0.6, 0.7
                                                                                     \\
\hline
$q$   & \hspace{-2em} 0.1, 0.2, 0.3, 0.4, 0.5,   0.6,0.7, 0.8, 0.9, 1.0   \\
\hline
\end{tabular}}
\end{adjustwidth}
\caption{The non uniform mass bin choice used in Figure \ref{fig: Subpop_case_28}}\label{tab: Bin choices non uniform}
\end{table}

\section{Validation of results with uncorrelated simulation} \label{app: uncorr_sim}

To ensure that the correlations we determine by inferring the joint hyperposterior are not manifesting out of artifacts built into our model, we validate our results using a simulated population that does not contain any intrinsic correlations in the ($m_1, q, \chi_{\rm eff}$) space. This is achieved by independently drawing binary parameters from the uncorrelated population distribution and thereafter injecting these signals into synthetic detector noise, and finally passing them through the same selection and analysis pipeline as the real data. In order to perform event level parameter estimation on these injections, we employ Bayesian inference using the \textsc{Bilby} package \citep{Ashton_2019, RomeroShaw2020}, its multi-purpose nested sampler \textsc{Dynesty} \citep{Speagle2020}, and the IMRphenomD aligned spin waveform approximant \citep{Husa2016, Khan2016}. By comparing the inferred correlations (or lack thereof) recovered from this control population to those observed in the actual LVK data, we assess whether the empirical correlations are driven by model systematics or are statistically significant features of the underlying astrophysical population.

\begin{figure*}
\centering

\includegraphics[width = 0.32\linewidth]{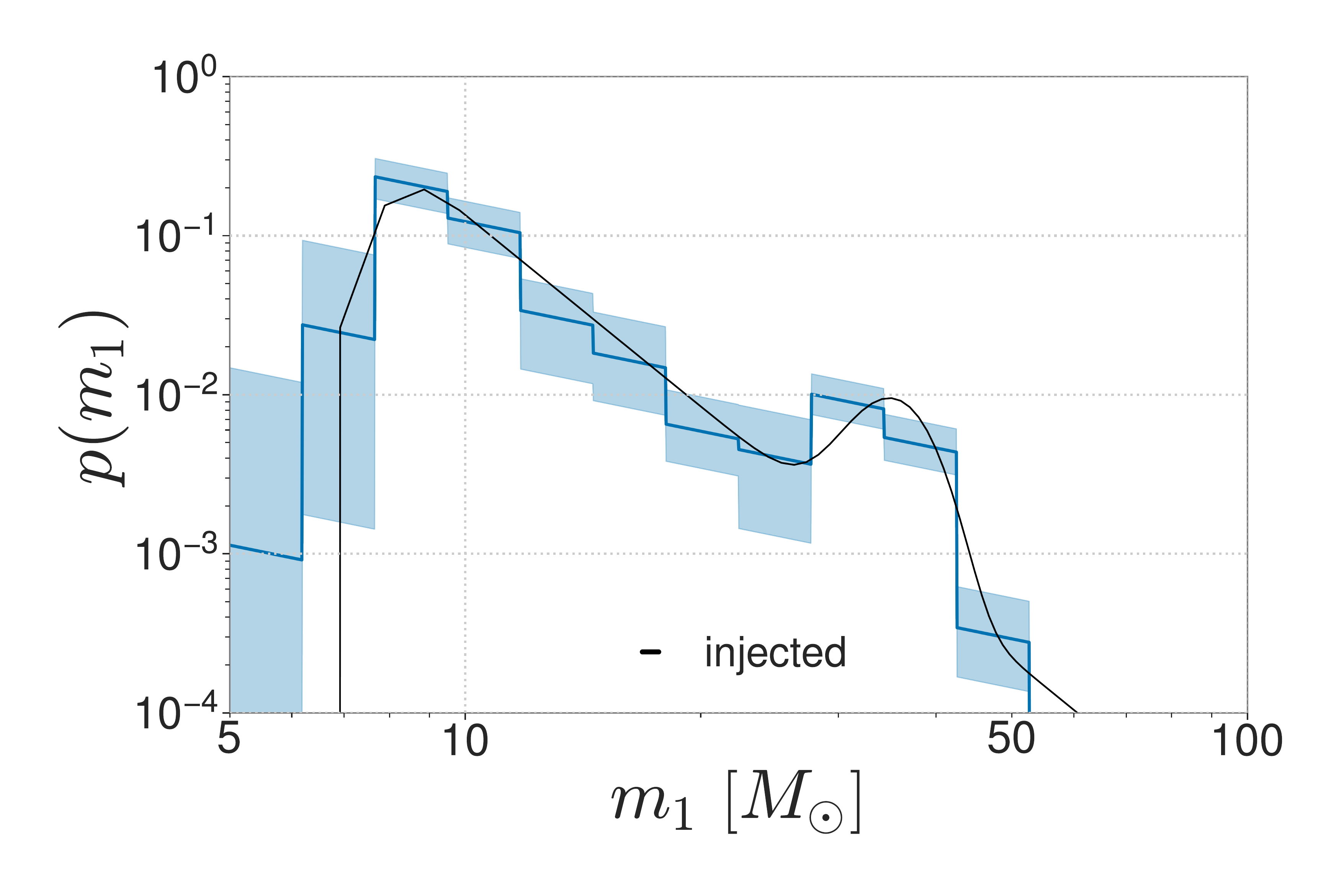}\vspace{-1mm}
\includegraphics[width = 0.32\linewidth]{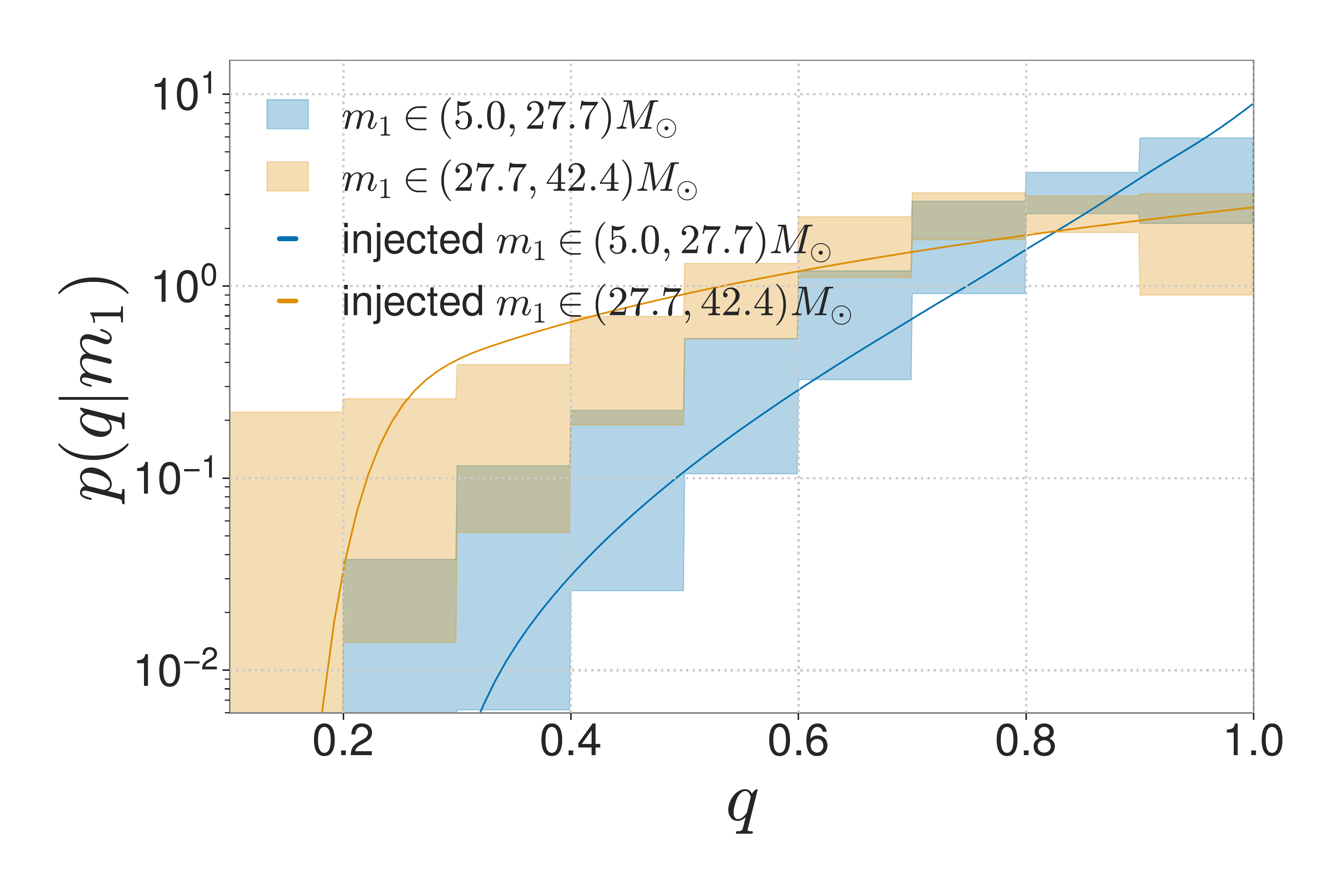}
\includegraphics[width = 0.32\linewidth]{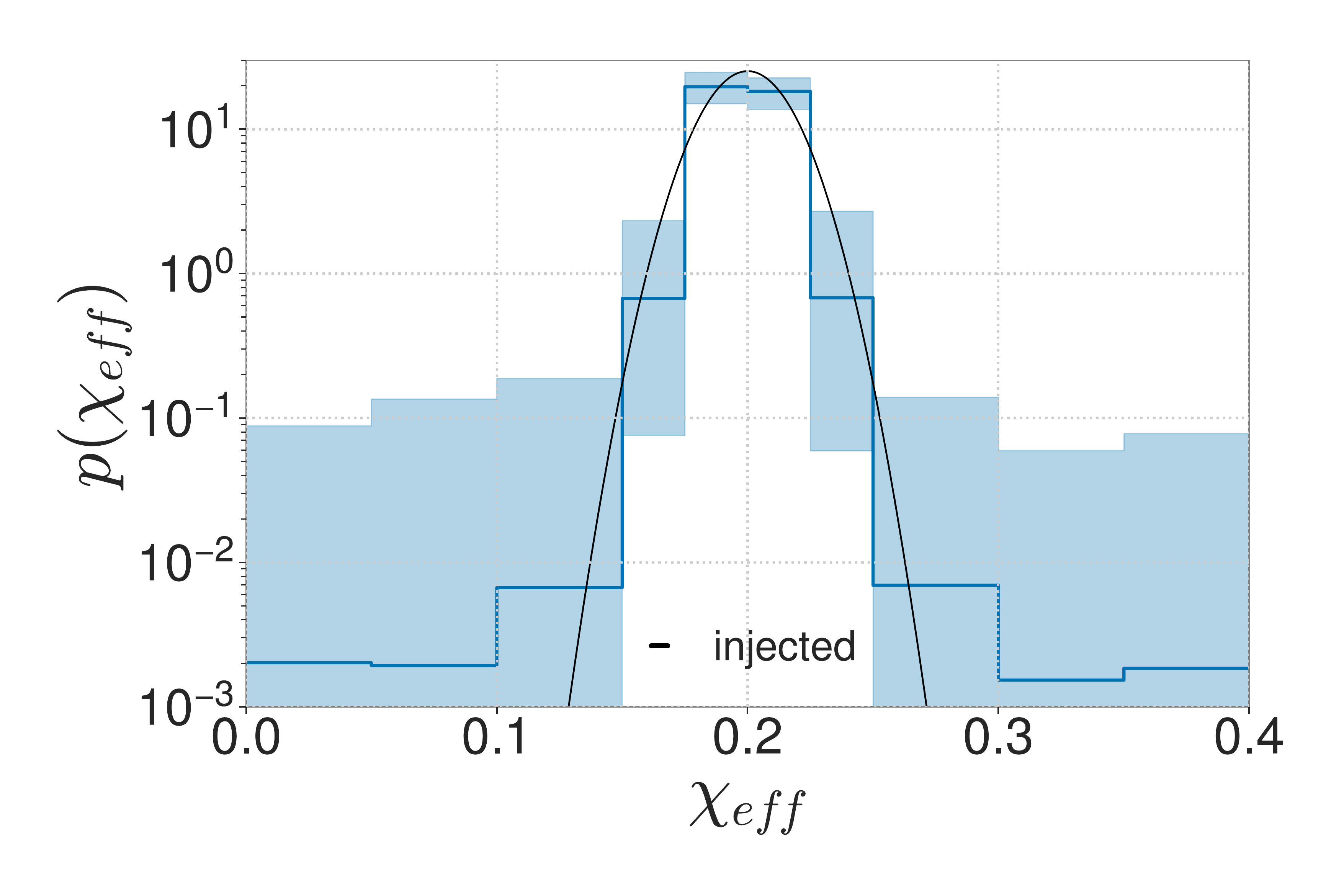}

\includegraphics[width = 0.32\linewidth]{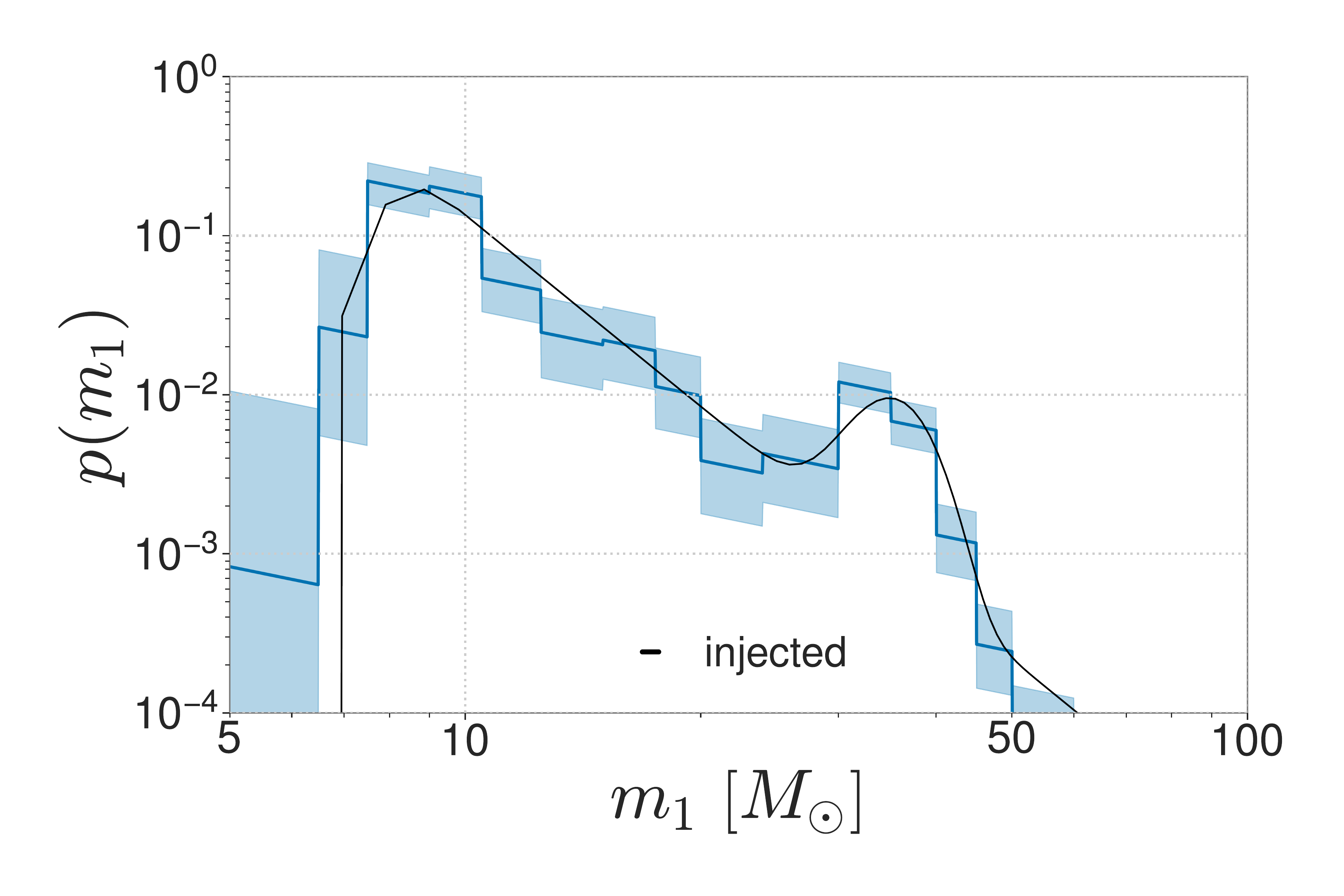}\vspace{-1mm}
\includegraphics[width = 0.32\linewidth]{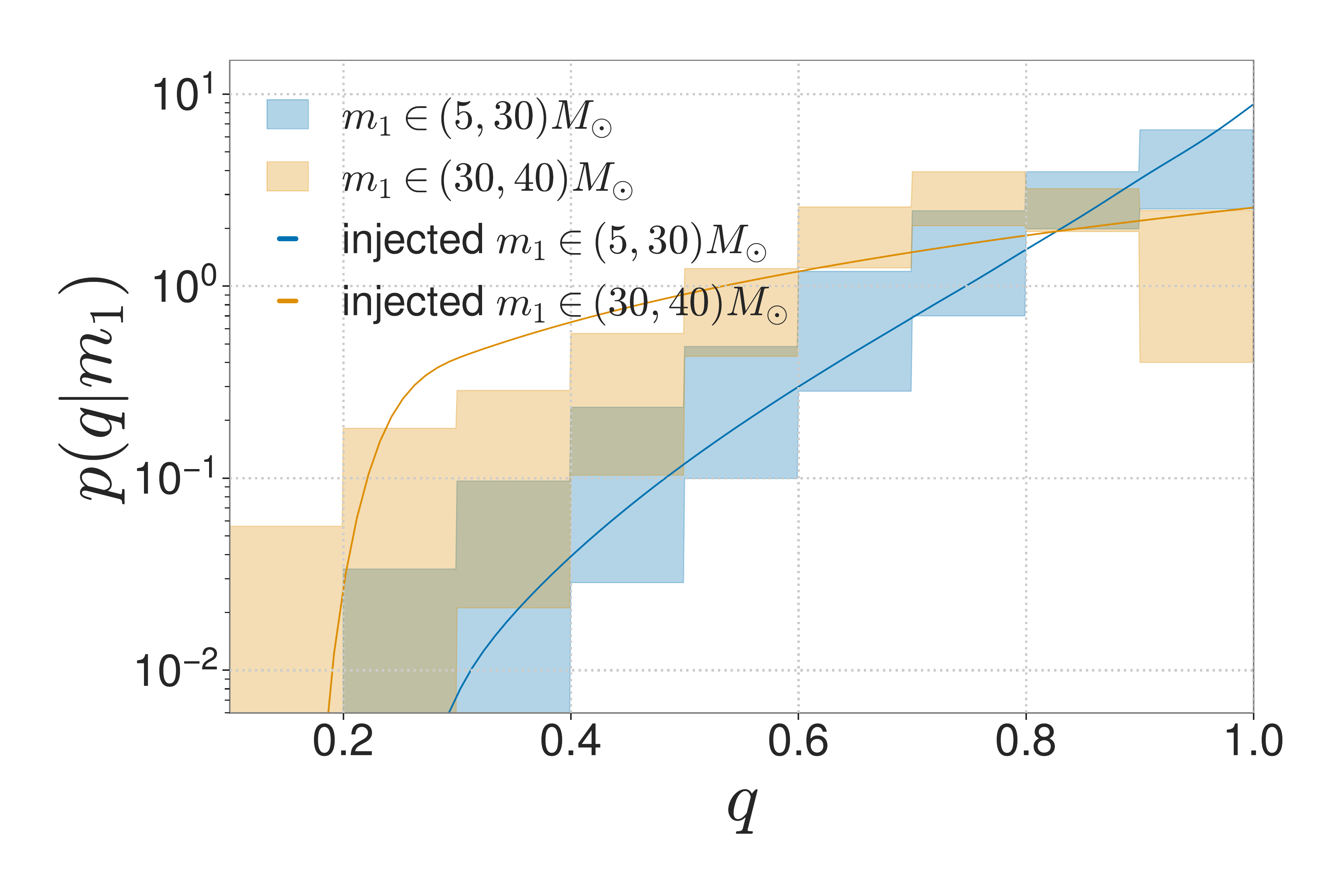}
\includegraphics[width = 0.32\linewidth]{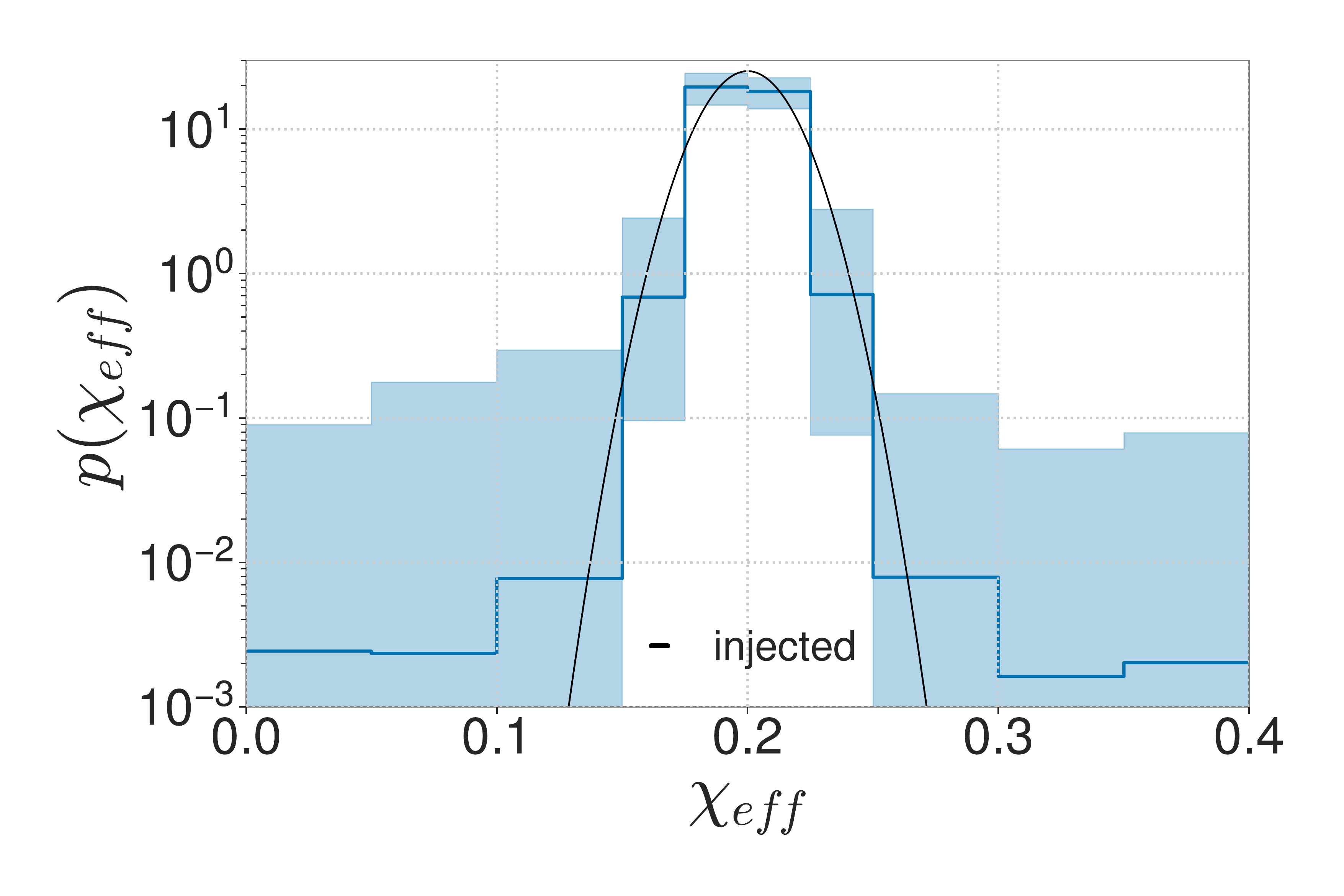}

\caption{
Marginal and conditional distributions of BBH primary mass, mass ratio and effective spin for the simulated catalog with no intrinsic correlations between the three parameters, for two different $m_1$ bin resolutions (log uniform on top and a non-uniform choice at the bottom).
}
\label{fig: Prob dist plots uncorr}
\end{figure*}

Upon conducting population inference using our framework on this simulated dataset, we find that the subpopulation trends recovered from the GWTC-4 event set are not a result of biases built into the model. The marginal primary mass, conditional mass-ratio with respect to primary mass, and marginal effective spin distributions (Figure \ref{fig: Prob dist plots uncorr}) recovered by our model are consistent with the true injected distributions. We have therefore verified that the predictions we make regarding the GWTC-4 data are not spuriously arising due to model artifacts.

\section{Convergence of Monte Carlo integrals} \label{app: MC convergence}

In population inference, the hierarchical likelihood as a function of hyperparameters is computed from Monte Carlo approximations of integrals of the population prior over single-event posterior samples and detectable injections. The accuracy of these Monte Carlo estimators is a function of the population hyperpriors and can therefore lead to biases in the inferred distribution. To mitigate this issue, previous studies have shown that hyper-parameters that lead to large variance in the Monte Carlo estimators and hence the hierarchical likelihood should be penalized for reliable population inference. 

For our BGP model, the hierarchical likelihood takes the following form:
\begin{equation}
    \log p(\vec{d}|\vec{n}) =  \log p(\vec{d}|\vec{n}) = -\sum_\gamma n^\gamma \left<VT\right>^\gamma + \sum_i \log \biggl(\sum_\gamma n^\gamma w^\gamma (d_i)\biggr),
\end{equation}
where, $\left<VT\right>^\gamma$ and $w^\gamma (d_i)$ are Monte Carlo estimators of the detectable hypervolume event-specific posterior supports in the ith bin respectively. The means~($\mu_{VT}^{\gamma},\mu_{w}^{\gamma}(d_i)$) and variances~($(\sigma_{VT}^{\gamma})^2,(\sigma_{w}^{\gamma}(d_i))^2$) of these estimators can be computed~\citep[see, for example,][]{Ray:2023, Ray2024, MaganaHernandez:2024} to evaluate both the log-likelihood and its variance due to Monte Carlo uncertainties~\citep{Essick2022, Talbot:2023pex}, as follows:
\begin{equation}
    \log p(\vec{d}|\vec{n}) =  \log p(\vec{d}|\vec{n}) = -\sum_\gamma n^\gamma \mu_{VT}^{\gamma} + \sum_i \log \biggl(\sum_\gamma n^\gamma \mu_{w}^{\gamma}(d_i)\biggr),
\end{equation}
\begin{equation}
    \sigma^2_{  \log p(\vec{d}|\vec{n})} = \sum_{\gamma}n_{\gamma}^2(\sigma_{VT}^{\gamma})^2+ \sum_{i}\frac{\sum_{\gamma}n_{\gamma}^2(\sigma_{w}^{\gamma}(d_i))^2}{\left(\sum_{\gamma}n_{\gamma}\mu_{w}^{\gamma}(d_i)\right)^2}.
\end{equation}
Here, the variance in the log-likelihood estimator $( \sigma^2_{  \log p(\vec{d}|\vec{n})})$ due to the finite number of Monte Carlo samples can become large for certain values of hyperparameters which should be rejected to avoid biases in the inferred population. We penalize the log-likelihood for hyperparameters that lead to $\sigma^2_{  \log p(\vec{d}|\vec{n})}$ larger than 1. To ensure finite gradients (as required by HMC) we impose this variance-cut asymptotically and modify the log-likelihood function as follows:
\begin{equation}
    \log p(\vec{d}|\vec{n}) =  \log p(\vec{d}|\vec{n}) = -\sum_\gamma n^\gamma \mu_{VT}^{\gamma} + \sum_i \log \biggl(\sum_\gamma n^\gamma \mu_{w}^{\gamma}(d_i)\biggr) - \log\{1+\left(\frac{1}{\sigma_{\log p(\vec{d}|\vec{n})}}\right)^{-n}\},
\end{equation}
where $n$ is a large positive integer which we choose to be 30~\citep{Callister_Farr_2024}. We have verified that all values of $n$ within the range $[30,75]$ yield comparable answers with very few hypersamples that correspond to large log-likelihood variance, ensuring unbiased population inference.

\bibliographystyle{aasjournal}
\bibliography{ref-list}

\end{document}